 \documentclass[final,3p,euclid]{elsarticle}

\usepackage[colorlinks,linkcolor=red,anchorcolor=blue,citecolor=green]{hyperref} 
\usepackage{amssymb}
\usepackage{amsmath}
\usepackage{amsthm}
\usepackage{setspace}
\usepackage{graphicx}
\usepackage{epstopdf}
\usepackage{pgfplots}
\usepackage{url}
\graphicspath{{figure/}}
\usepackage{subfigure}
\usepackage{multirow}
\usepackage[flushleft]{threeparttable}
\usepackage{makecell}
\usepackage{url}
\usepackage{color}
\usepackage{pifont}
\usepackage{xcolor}
\usepackage{float}
\usepackage{xcolor}
\usepackage{colortbl}
\usepackage{threeparttable}
\usepackage{booktabs}
\usepackage{array}
\newcolumntype{M}{>{\centering\arraybackslash}m{2cm}}
\usepackage[ruled]{algorithm2e}

\newcounter{magicrownumbers}

\journal{}

\begin{document}

\begin{frontmatter}



\title{Bayesian Active Learning for Bayesian Model Updating: the Art of Acquisition Functions and Beyond}

\author[Address1]{Jingwen Song}
\author[Address2,Address3]{Pengfei Wei\corref{cor1}}  
\ead{pengfeiwei@nwpu.edu.cn}
\address[Address1]{State IJR Center of Aerospace Design and Additive Manufacturing, School of Mechanical Engineering, Northwestern Polytechnical University, Xi’an 710072, China}
\address[Address2]{School of Power and Energy, Northwestern Polytechnical University, Xi’an 710072, China}
\address[Address3]{Science and Technology on Altitude Simulation Laboratory, Mianyang 621000, China}

\cortext[cor1]{Corresponding author at School of Power and Energy, Northwestern Polytechnical University, Xi’an 710072, China} 

\begin{abstract}
	\begin{spacing}{1.2}
Estimating posteriors and the associated model evidences, with desired accuracy and affordable computational cost, is a core issue of Bayesian model updating, and can be of great challenge given expensive-to-evaluate models and  posteriors with complex features such as multi-modalities of unequal importance, nonlinear dependencies and high sharpness. Bayesian Quadrature (BQ) equipped with active learning has emerged as a competitive framework for tackling this challenge, as it provides flexible balance between computational cost and accuracy. The performance of a BQ scheme is fundamentally dictated by the acquisition function as it exclusively governs the active generation of integration points. After reexamining one of the most advanced acquisition function from a prospective inference perspective and reformulating the quadrature rules for prediction, four new acquisition functions, inspired by distinct intuitions on expected rewards, are primarily developed, all of which are accompanied by elegant interpretations and highly efficient numerical estimators. Mathematically, these four acquisition functions measure, respectively, the prediction uncertainty of posterior, the contribution to prediction uncertainty of evidence, as well as the expected reduction of prediction uncertainties concerning posterior and evidence, and thus provide flexibility for highly effective design of integration points. These acquisition functions are further extended to the transitional BQ scheme, along with several specific  refinements, to tackle the above-mentioned challenges with high efficiency and robustness. Effectiveness of the developments is ultimately demonstrated with extensive benchmark studies and application to an engineering example. 

    \end{spacing}
\end{abstract}




\end{frontmatter}


\section{Introduction}
\begin{spacing}{1.5}
Many ill-conditioned inverse problems in computational science and engineering, such as the calibration of multi-physical simulation models \cite{kennedy2001bayesian, bi2023stochastic}, the identification of structural damages \cite{huang2019state, zeng2025data}, the uncertainty quantification of computational and AI models \cite{yin2022practical, psaros2023uncertainty}, and the identification of contamination sources \cite{jerez2021contaminant}, can be uniformly and elegantly treated using a Bayesian framework, giving rise to the Bayesian model inference problems with diverse features and challenges \cite{bingham2024inverse}. Estimation of posteriors and the associated model evidences, with desired accuracy and acceptable computational cost, is the primary focus of these updating problems, as posteriors are used for informing the subjective probability distribution of possible values of model parameters (or damage states), and, in case multiple models are available, the model evidences can be used for model selection and model averaging \cite{friel2012estimating}. It has been repeatedly shown that, resulted the ill-posedness of these inverse problems, the (unknown) nonlinear physical constraints on parameters, the complex behavior of the model functions, and the varying quality of available measurement data, the posteriors may exhibit tricky features such as multi-modalities, nonlinear dependencies, and high sharpness, even for the simple linear dynamic systems (see e.g., Ref. \cite{bryutkin2025canonical}). These complexities have rendered most of the state-of-the-art developments for estimating posteriors and model evidences either inapplicable or unable to meet the requirements in terms of numerical efficiency and accuracy.   

In the context of Bayesian model inference, one of the most renowned categories of methods is the Markov Chain Monte Carlo (MCMC) simulation, which simulates one or a group of discrete Markov chains, with the stationary distribution well-designed to converge to the target posterior distribution \cite{van2018simple, jones2022markov}, this way also to estimate the model evidence and other related posterior quantities with Monte Carlo (MC) estimators. Starting from Metropolis' original proposal-and-decision sampling scheme \cite{metropolis1953}, and Hastings' extension to asymmetric proposal distributions \cite{hastings1970monte}, known as the Metropolis-Hastings (MH) algorithm, the past half century has witnessed a rapid growth of the MCMC algorithms. Typical developments including the Hamiltonian MC \cite{neal2011mcmc,betancourt2017geometric}, which constructs the MCMC with Hamiltonian dynamics to achieve higher acceptance rate; the transitional MCMC (TMCMC) scheme \cite{ching2007transitional,betz2016transitional}, of which the fundamental principle is to approach the sharp and multi-modal posteriors using a set of intermediate distributions, each of which is simulated by resampling and MH algorithm; the preconditioned Crank-Nicolson algorithm \cite{cotter2013mcmc, hairer2014spectral}, which is dimension robust and allows even for reliable inference in infinite-dimensional Hilbert space; diverse parallel tempering MCMC strategies toward efficient simulation of high-dimensional posterior with high efficiency \cite{syed2022non}; various convergence diagnostic schemes for burn-in removal and convergence judgment \cite{roy2020convergence,south2022postprocessing}, etc. Despite those extensive developments, as many computational models are extremely expensive to evaluate and the computational resources are limited, acquiring thousands of Markov states for approaching the highly nonlinear, multi-modal and sharp posteriors with desired accuracy is impractical, let alone the necessity, for users, to develop expertise for burn-in removal, sample correlation alleviation, etc.  

Variational Bayesian Inference (VBI) is arguably the second most extensively studied and widely used class of algorithms after MCMC. It is based on assuming a parameterized density, such as a mixture Gaussian distribution, for the unknown posterior, this way to transform the problem as optimizing the variational parameters by minimizing the Kullback–Leibler (KL) divergence between the variational density and the target posterior \cite{tzikas2008variational, zhang2019advances}. This has been proved to be equivalent to maximizing a function over the design parameters, which is called Evidence Lower Bound (ELBO) or variational free energy \cite{fox2012tutorial,tran2021practical}. Under this framework, numerous numerical VBI schemes have been developed, such as the mean-field VBI which assumes independence of the model parameters for simplifying the derivations \cite{opper2001advanced},  the black-box VBI based on maximizing the ELBO using a stochastic optimization algorithm \cite{ranganath2014black}, the automatic differential VBI scheme \cite{kucukelbir2017automatic}, the VBI scheme which approximates the posterior with a normalizing flow \cite{rezende2015variational}, the Bayesian optimization scheme which avoids the computation of derivatives \cite{hong2025efficient}, and the one combining stochastic gradient descent adaptive Gaussian process model \cite{ni2021probabilistic}. Despite those developments, practitioners may still find it not straightforward to apply these methods \cite{zhang2019advances}, due to, for example, the necessity of expertise on pre-specifying a reasonable variational density for balancing the tractability and approximation accuracy, the intractability of acquiring accurate gradient information for some multi-physical simulators, etc.  

Other branches of methods include Bayesian filter algorithms \cite{yoshida2023bayesian}, transport map theory \cite{baptista2024representation}, Bayesian Updating with Structural reliability methods (BUS) \cite{straub2015bayesian}, normalizing flow \cite{papamakarios2021normalizing}, deep generative models \cite{wang2026stochastic,sun2025generative},  etc., but we don't go into further details as they are less relevant to the developments of this work. In many application scenarios with expensive-to-evaluate computational models, estimating posterior and model evidence, with desired accuracy but as few model calls as possible, has emerged as an urgent requirement. In this context, combination of Gaussian process regression with active learning has received much attention. Several branches of methods have been developed following this scheme. These include the combination of active learning method for structural reliability analysis with the BUS framework (see e.g., Refs.\cite{song2022buak, kitahara2023bayesian}), those combined with VBI for efficiently estimating the parameters of the pre-assumed density for approximating the posteriors (see e.g., Refs. \cite{hong2025efficient, ni2021probabilistic}), as well as the Bayesian Quadrature (BQ) methods (see e.g, \cite{hennig2022probabilistic}). The former two groups commonly require extra parameters to be estimated through expensive numerical procedures, which is commonly not required in a BQ scheme. Further, the BQ schemes usually provide reasonable quantification of prediction uncertainties for both posteriors and model evidences, and thus are more appealing.

Fundamentally, the BQ methods formulate the estimation of posterior density and model evidence as a statistical inference issue, and use active learning to sequentially generate probabilistic descriptions for predicting these two items and meanwhile, to quantify the prediction uncertainties. They are realized by training a Gaussian Process (GP) model for approximating a ``proxy function'', which is related to the likelihood, this way to derive probabilistic descriptions for both targets with an approximate scheme. Most of the state-of-art developments uses approximate quadrature rules. For example, with the logarithm of the likelihood being approximate by a GP model, a linearization scheme has been developed for approximately inferring Gaussian distribution for both posterior and model evidences \cite{osborne2012active}; by modeling the square-root of the likelihood with a GP model, Gaussian distributions for both posteriors and evidence have been formulated using either linearization or moment-matching \cite{gunter2014sampling}; by approximating the simulator \cite{song2025sampling} or the logarithm of likelihood \cite{kitahara2025sequential, wei2025bayesian} with a GP, MC simulation schemes based on sampling the GP model have been proposed for realizing unbiased inference. The surprising performance of these methods for capturing some complex features of the posteriors, like multi-modality and nonlinear dependencies, have been explained and boosted in Ref. \cite{wei2025bayesian} using the exponential impact, resulting in a robust and efficient algorithm called Transitional BQ (TBQ). Except the approximate quadrature rules, acquisition function also plays a critical role as it determines the production of the integration points, and thus the efficiency and accuracy. Well-developed acquisition functions toward this issue include the Uncertainty Sampling (US) \cite{acerbi2019exploration}, the Posterior Variance Contribution (PVC) function \cite{song2025sampling, kitahara2025sequential}, and the prediction uncertainty functions formulated with credible bounds \cite{wei2025bayesian}. 

The BQ framework is comprehensively reconstructed and enhanced in this work with the following contributions. First, the classical PVC function for vanilla BQ is reexamined from a prospective view, endowing it with new interpretations and more sound way for use. Second, semi-analytical quadrature rules are presented for model evidence, which enable for more efficient and precise prediction while also facilitating the quantification of prediction uncertainty. Third, as the most notable innovation, four new acquisition functions, all accompanied with mathematically sound interpretations and efficient computational methods, are devised for either non-prospective or prospective design of integration points. Fourth, together with several trivial specific improvements, both the developed quadrature rules and acquisition functions are combined with the TBQ algorithm, allowing for efficient and robust estimation of both posteriors and model evidences, even for problems with multiple disconnected modes of unequal importance, highly nonlinear dependencies and high sharpness presented in the posteriors. 

The remaining of the work is organized as follows. Section 2 presents the mathematical formulation of the Bayesian model updating problem, and a brief review of the vanilla BQ method. Section 3 first presents a new interpretation to the classical PVC function, then the new BQ rules, together with four new acquisition functions, and ultimately ends with a summary of the reformulated BQ algorithm. Improvements of TBQ algorithm with the above-mentioned developments are presented in section 4, followed by extensive benchmark studies and application to an engineering case in section 5. Section 6 gives conclusions and prospects.

\section{Problem Statement and Methods Review}
\subsection{Formulation of Bayesian model updating}
A typical Bayesian inverse problem can be formulated by the Bayes' theorem as:
\begin{equation}\label{eq:DefPosterior}
	p\left( \boldsymbol{\theta }|\mathcal{D} _{\mathrm{obs}} \right) =Z^{-1}p\left( \mathcal{D} _{\mathrm{obs}}|\boldsymbol{\theta } \right) p\left( \boldsymbol{\theta } \right) 
\end{equation} 
, where $p\left( \boldsymbol{\theta } \right) $ refers to the prior density with support $\mathbb{T} \subseteq \mathbb{R} ^d$, $p\left( \mathcal{D} _{\mathrm{obs}}|\boldsymbol{\theta } \right) $ is the likelihood function, conditioning on the observations $\mathcal{D} _{\mathrm{obs}}$, and 
\begin{equation}\label{eq:DefModelEvidence}
	Z=\int_{\mathbb{T}}{p\left( \mathcal{D} _{\mathrm{obs}}|\boldsymbol{\theta } \right) p\left( \boldsymbol{\theta } \right) \mathrm{d}\boldsymbol{\theta }}
\end{equation}
is called ``model evidence'', mathematically for normalizing the posterior density  $	p\left( \boldsymbol{\theta }|\mathcal{D} _{\mathrm{obs}} \right)$, but physically, it implies the credibility of the model or fitness of the model to the observations. The likelihood $p\left( \mathcal{D} _{\mathrm{obs}}|\boldsymbol{\theta } \right) $ can be formulated as the form:
\begin{equation}\label{eq:DefLikelihood}
	p\left( \mathcal{D} _{\mathrm{obs}}|\boldsymbol{\theta } \right) =\exp \left( -\mathcal{U} \left( \boldsymbol{\theta } \right) \right) 
\end{equation}
, where $\mathcal{U} \left( \boldsymbol{\theta } \right) $ is called ``energy function'', with its name borrowed from variational Bayesian inference and thermodynamic free energy theory  \cite{rezende2015variational,chopin2012free,friston2023free}. Commonly, the energy function, and thus the likelihood, is expensive to evaluate. The computational target of a Bayesian inverse problem is to estimate the model evidence $Z$ and posterior $	p\left( \boldsymbol{\theta }|\mathcal{D} _{\mathrm{obs}} \right)$ with desired accuracy, but with as few calls to the energy function as possible. However, given the multi-modalities, sharpness, sparsity and highly nonlinear dependencies of the posteriors, resulted from high nonlinearity of models and high-quality observations, this numerical task poses a great challenge, in terms of both prediction accuracy and efficiency.

For the Bayesian model updating problem concerned in this work, denote the simulation model as $y=\mathcal{M} \left( \boldsymbol{x},\boldsymbol{\theta } \right) $, with $\boldsymbol{x}$ being the $n$-dimensional column vector of controllable inputs, and $\boldsymbol{\theta }$ indicating the $d$-dimensional vector of deterministic-but-unknown parameters to be identified. Further, assume a set of $N_\mathrm{obs}$ observations are allocated, which is collectively implied by $\mathcal{D} _{\mathrm{obs}}=\left\{ \left( \boldsymbol{x}_{\mathrm{obs}}^{\left( l \right)},y_{\mathrm{obs}}^{\left( l \right)} \right) \right\} _{l=1}^{N_{\mathrm{obs}}}$, with the random noise of the $j$-th observation being denoted by $\epsilon ^{\left( l \right)}$. With the assumption of Gaussian white noise, i.e., $\epsilon ^{\left( l \right)}\equiv \epsilon \sim \mathcal{N} \left( 0,\sigma _{\mathrm{n}}^{2} \right) $, the energy function is formulated as:
 \begin{equation}\label{eq:Defenergyfunction}
 	\mathcal{U} \left( \boldsymbol{\theta } \right) \propto \frac{1}{\sigma _{\mathrm{n}}^{2}}\sum_{l=1}^{N_{\mathrm{obs}}}{\left( y_{\mathrm{obs}}^{\left( l \right)}-\mathcal{M} \left( \boldsymbol{x}_{\mathrm{obs}}^{\left( l \right)},\boldsymbol{\theta } \right) \right) ^2}
 \end{equation} 
, with $\sigma _{\mathrm{n}}^{2}$ being the noise variance. One can extend the definition Eq. \eqref{eq:Defenergyfunction} to cases with any types of noise distribution, and developments in this work apply to arbitrary observation likelihood and energy function. 

\subsection{Review of Vanilla Bayesian Quadrature}
Let $\Pi \left[ \cdot \right] $ denote an integral operator over the density $p\left( \boldsymbol{\theta } \right) $. Take the following $d$-dimensional integral equation to illustrate the Vanilla BQ method:
\begin{equation}\label{eq:IntegralEq}
	\mathcal{I} =\Pi \left[ g\left( \boldsymbol{\theta } \right) \right] =\int_{\mathbb{T}}{g\left( \boldsymbol{\theta } \right)}p\left( \boldsymbol{\theta } \right) \mathrm{d}\boldsymbol{\theta }
\end{equation} 
, where $g\left( \boldsymbol{\theta } \right) $ is an expensive-to-evaluate and black-box integrand. 

The BQ method is initialized by assuming a Gaussian Process (GP) model $\hat{g}\left( \boldsymbol{\theta } \right) \sim \mathcal{G} \mathcal{P} \left( m\left( \boldsymbol{\theta } \right) ,\kappa \left( \boldsymbol{\theta },\boldsymbol{\theta }^{\prime} \right) \right) $, as a prior, for describing $g\left( \boldsymbol{\theta } \right) $, where $m:\mathbb{R} ^d\rightarrow \mathbb{R} $ and $\kappa :\mathbb{R} ^d\times \mathbb{R} ^d\rightarrow \mathbb{R} $ are the prior mean and covariance functions respectively. Then, given a set of $N$ training points $\mathcal{D} _{N,\mathrm{train}}=\left\{ \mathcal{T} ,\mathcal{Y} \right\} =\left\{ \left( \boldsymbol{\theta }^{\left( t \right)},y^{\left( t \right)} \right) \right\} _{t=1}^{N}$, with $y^{\left( t \right)}=g\left( \boldsymbol{\theta }^{\left( t \right)} \right) $, the label column vector $\mathcal{Y} $ and the prediction $\hat{g}_N\left( \boldsymbol{\theta } \right) $ at an unobserved location $\boldsymbol{\theta}$ are correlated and follow Gaussian distribution, with the joint and marginal densities denoted as $p\left( \mathcal{Y} ,\hat{g}_N\left( \boldsymbol{\theta } \right) \right) $ and $p\left( \mathcal{Y} \right) $ respectively. Then, by the conditional probability formula $p\left( \hat{g}_N\left( \boldsymbol{\theta } \right) |\mathcal{Y} \right) =p\left( \mathcal{Y} ,\hat{g}_N\left( \boldsymbol{\theta } \right) \right) /p\left( \mathcal{Y} \right) $, the posterior distribution $\hat{g}_N\left( \boldsymbol{\theta } \right) $ also follows Gaussian distribution, and is denoted as $\hat{g}_N\left( \boldsymbol{\theta } \right) \sim \mathcal{G} \mathcal{P} \left( \mu _{g,N}\left( \boldsymbol{\theta } \right) ,c_{g,N}\left( \boldsymbol{\theta },\boldsymbol{\theta }^\prime \right) \right) $, with the posterior mean $\mu _{g,N}\left( \boldsymbol{\theta } \right) $, variance $\sigma _{g,N}^{2}\left( \boldsymbol{\theta } \right) $ and covariance $c_{g,N}\left( \boldsymbol{\theta },\boldsymbol{\theta }^{\prime} \right) $ formulated as:
\begin{subequations}\label{eq:PostGP}
	\begin{equation}\label{eq:PostGPMean}
		\mu _{g,N}\left( \boldsymbol{\theta } \right) =m\left( \boldsymbol{\theta } \right) +\boldsymbol{\kappa }\left( \boldsymbol{\theta },\mathcal{T} \right) \mathcal{K} ^{-1}\left( \mathcal{Y} -\boldsymbol{m}\left( \mathcal{T} \right) \right) 
	\end{equation}
\begin{equation}\label{eq:PostGPVar}
	\sigma _{g,N}^{2}\left( \boldsymbol{\theta } \right) =\kappa \left( \boldsymbol{\theta },\boldsymbol{\theta } \right) -\boldsymbol{\kappa }\left( \boldsymbol{\theta },\mathcal{T} \right) \mathcal{K} ^{-1}\boldsymbol{\kappa }\left( \mathcal{T} ,\boldsymbol{\theta } \right) 
\end{equation}
\begin{equation}\label{eq:PostGPCov}
	c_{g,N}\left( \boldsymbol{\theta },\boldsymbol{\theta }^{\prime} \right) =\kappa \left( \boldsymbol{\theta },\boldsymbol{\theta }^{\prime} \right) -\boldsymbol{\kappa }\left( \boldsymbol{\theta },\mathcal{T} \right) \mathcal{K} ^{-1}\boldsymbol{\kappa }\left( \mathcal{T} ,\boldsymbol{\theta }^{\prime} \right) 
\end{equation}
\end{subequations}  
, where $\mathcal{K} =\kappa \left( \mathcal{T} ,\mathcal{T} \right) $ is the Gram matrix, which is symmetric positive definite. One can refer to Ref. \cite{Rasmussen2006gaussian} for the pre-selection of prior mean and covariance functions and the Maximum Likelihood Estimation (MLE) of the hyper-parameters for defining these two functions.

Given the above formulation of prior GP $\hat{g}\left( \boldsymbol{\theta } \right) $ and posterior GP $\hat{g}_N\left( \boldsymbol{\theta } \right) $, a prior and a posterior Gaussian distribution can be inferred for $\mathcal{I}$ by replacing the integrand in Eq. \eqref{eq:IntegralEq} with these two GPs. Specifically, the posterior is denoted as $\hat{\mathcal{I}}_N\sim \mathcal{N} \left( \mu _{\mathcal{I} ,N},\sigma _{\mathcal{I} ,N}^{2} \right) $, with the mean $\mu _{\mathcal{I} ,N}$ and the variance $\sigma _{\mathcal{I} ,N}^{2}$ formulated as \cite{rasmussen2003bayesian,wei2020adaptive}:
\begin{subequations}\label{eq:IPost}
	\begin{equation}\label{eq:IPostMean}
		\mu _{\mathcal{I} ,N}=\Pi \left[ \mu _{g,N}\left( \boldsymbol{\theta } \right) \right] =\Pi \left[ m\left( \boldsymbol{\theta } \right) \right] +\Pi \left[ \boldsymbol{\kappa }\left( \boldsymbol{\theta },\mathcal{T} \right) \right] \mathcal{K} ^{-1}\left( \mathcal{Y} -\boldsymbol{m}\left( \mathcal{T} \right) \right) 
	\end{equation}
\begin{equation}\label{eq:IPostVar}
	\sigma _{\mathcal{I} ,N}^{2}=\Pi \Pi ^{\prime}\left[ c_{g,N}\left( \boldsymbol{\theta },\boldsymbol{\theta }^{\prime} \right) \right] =\Pi \Pi ^{\prime}\left[ \boldsymbol{\kappa }\left( \boldsymbol{\theta },\boldsymbol{\theta }^{\prime} \right) \right] -\Pi \left[ \boldsymbol{\kappa }\left( \boldsymbol{\theta },\mathcal{T} \right) \right] \mathcal{K} ^{-1}\Pi ^{\prime}\left[ \boldsymbol{\kappa }\left( \mathcal{T} ,\boldsymbol{\theta }^{\prime} \right) \right] 
\end{equation}
\end{subequations}
, where $\Pi \Pi ^{\prime}\left[ \cdot \right] =\int_{\mathbb{T} \times \mathbb{T}}{\cdot p\left( \boldsymbol{\theta } \right) p\left( \boldsymbol{\theta }^{\prime} \right) \mathrm{d}\boldsymbol{\theta }\mathrm{d}\boldsymbol{\theta }^{\prime}}$. Closed-form expressions of $\Pi \left[ \boldsymbol{\kappa }\left( \boldsymbol{\theta },\mathcal{T} \right) \right] $ and $\Pi \Pi ^{\prime}\left[ \boldsymbol{\kappa }\left( \boldsymbol{\theta },\boldsymbol{\theta }^{\prime} \right) \right] $ are available for specific types of pair $\left( \kappa ,p \right) $, e.g., for case $\kappa $ is a squared exponential kernel and $p$ is a Gaussian density. One can refer to Ref. \cite{briol2019probabilistic} for a summary and Ref. \cite{wei2020adaptive} for the closed-form expressions.  

The posterior variance $\sigma _{\mathcal{I} ,N}^{2}$ presents a reasonable quantification of prediction uncertainty of $\mu _{\mathcal{I} ,N}$ for approximating $\mathcal{I}$, given $g$ belongs to the Reproducing Kernel Hilbert Space (RKHS) associated with the kernel $\kappa$ \cite{briol2019probabilistic}. Given a fixed $N$, the magnitude of $\sigma _{\mathcal{I} ,N}^{2}$ is primarily determined by the training point $\mathcal{T}$, and several acquisition functions, such as the uncertainty sampling (US) function \cite{gunter2014sampling} and the Posterior Variance Contribution (PVC) function \cite{wei2020adaptive}, have been proposed for sequentially designing $\mathcal{T}$. The PVC function is defined as \cite{wei2020adaptive}:
\begin{equation}\label{eq:DefPVC}
	\mathcal{A} _{\mathrm{PVC}}\left( \boldsymbol{\theta } \right) =p\left( \boldsymbol{\theta } \right) \int_{\mathbb{T}}{c_{g,N}\left( \boldsymbol{\theta },\boldsymbol{\theta }^{\prime} \right) p\left( \boldsymbol{\theta }^{\prime} \right) \mathrm{d}}\boldsymbol{\theta }^{\prime}
\end{equation}
, and it admits a closed-form expression given the expectation of the kernel $\kappa$ has analytical solution. The  explanation of the PVC function is clear. As $\sigma _{\mathcal{I} ,N}^{2}=\int_{\mathbb{T}}{\mathcal{A} _{\mathrm{PVC}}\left( \boldsymbol{\theta } \right) \mathrm{d}\theta}$, it measures the contribution of the prediction uncertainty of the GP model at the unobserved location $\boldsymbol{\theta}$ to the prediction variance $\sigma _{\mathcal{I} ,N}^{2}$, with the integration of the spatial correlation between $\hat{g}_N\left( \boldsymbol{\theta } \right) $ and $\hat{g}_N\left( \boldsymbol{\theta }^{\prime} \right) $ across the whole support $\mathbb{T} $ of $\boldsymbol{\theta }^{\prime}$. It has also been shown in Ref. \cite{hong2024parallelization} that, for any $\boldsymbol{\theta }^{\left( t \right)}\in \mathcal{T}$, $\mathcal{A} _{\mathrm{PVC}}\left( \boldsymbol{\theta }^{\left( t \right)} \right) =0$, indicating that the prediction uncertainty at a valid training point makes no contribution to the prediction variance  $\sigma _{\mathcal{I} ,N}^{2}$. The next optimal training point can then be specified as $\boldsymbol{\theta }^+=\mathrm{arg}\max _{\theta \in \mathbb{R}}\mathcal{A} _{\mathrm{PVC}}\left( \boldsymbol{\theta } \right) $. As will see later, the PVC function encompasses a more profound and intricate connotation.   

\section{Developments of Acquisition Functions}\label{sec:ProAcqFun}
\subsection{Reexamination of PVC function}\label{subsec:ReaxamPVC}

In this subsection, we reexamine the PVC function from a prospective view for the vanilla BQ, which further motivates the development of this work. To illustrate this, suppose we have a new training point, denoted as $\left( \boldsymbol{\theta }^+,y^+ \right) $, where $y^+=\hat{y}_N\left( \boldsymbol{\theta }^+ \right) $ is an Gaussian random variable before actually observing it by calling the $g$-function. Based on the Bayesian incremental learning scheme, with $\left( \boldsymbol{\theta }^+,y^+ \right) $ being added, the GP model is updated as $\hat{g}_{N+1}\left( \boldsymbol{\theta } \right) $ with its posterior mean $\mu _{g,N+1}\left( \boldsymbol{\theta }|\boldsymbol{\theta }^+,y^+ \right) $, variance $\sigma _{g,N+1}^{2}\left( \boldsymbol{\theta }|\boldsymbol{\theta }^+ \right) $ and covariance $c_{g,N+1}\left( \boldsymbol{\theta },\boldsymbol{\theta }^{\prime}|\boldsymbol{\theta }^+ \right) $ being updated as \cite{chevalier2013corrected, zhou2024look}:
\begin{subequations}\label{eq:PostUpdatedGP}
\begin{equation}\label{eq:PostMeanUpdatedGP}
\mu _{g,N+1}\left( \boldsymbol{\theta }|\boldsymbol{\theta }^+,y^+ \right) =\mu _{g,N}\left( \boldsymbol{\theta } \right) +\frac{c_{g,N}\left( \boldsymbol{\theta }^+,\boldsymbol{\theta } \right)}{\sigma _{g,N}^{2}\left( \boldsymbol{\theta }^+ \right)}\left( y^+-\mu _{g,N}\left( \boldsymbol{\theta }^+ \right) \right),  
\end{equation}
\begin{equation}\label{eq:PostVarUpdatedGP}
	\sigma _{g,N+1}^{2}\left( \boldsymbol{\theta }|\boldsymbol{\theta }^+ \right) =\sigma _{g,N}^{2}\left( \boldsymbol{\theta } \right) -\frac{c_{g,N}^{2}\left( \boldsymbol{\theta }^+,\boldsymbol{\theta } \right)}{\sigma _{g,N}^{2}\left( \boldsymbol{\theta }^+ \right)}
\end{equation}
, and
\begin{equation}\label{eq:PostCovUpdatedGP}
	c_{g,N+1}\left( \boldsymbol{\theta },\boldsymbol{\theta }^{\prime}|\boldsymbol{\theta }^+ \right) =c_{g,N}\left( \boldsymbol{\theta },\boldsymbol{\theta }^{\prime} \right) -\frac{c_{g,N}\left( \boldsymbol{\theta },\boldsymbol{\theta }^+ \right) c_{g,N}\left( \boldsymbol{\theta }^+,\boldsymbol{\theta }^{\prime} \right)}{\sigma _{g,N}^{2}\left( \boldsymbol{\theta }^+ \right)}
\end{equation}
\end{subequations}
, respectively. It can be observed from Eq. \eqref{eq:PostMeanUpdatedGP} that the posterior mean depends linearly on the Gaussian variable $y^+$, thus is still a GP over $\boldsymbol{\theta}$, with the mean equal to $\mu _{g,N}\left( \boldsymbol{\theta } \right)$, and the covariance depending on $\boldsymbol{\theta}^+$; whereas, from Eqs. \eqref{eq:PostVarUpdatedGP} and \eqref{eq:PostCovUpdatedGP}, both the posterior variance and covariance are independent of $y^+$, and thus are deterministic functions. Accordingly, the prospective vanilla BQ rule can be updated as:
\begin{subequations}\label{eq:BQruleNplus1}
\begin{equation}\label{eq:BQMeanNplus1}
		\mu _{\mathcal{I} ,N+1}\left( \boldsymbol{\theta }^+,y^+ \right) =\mu _{\mathcal{I} ,N}+\frac{\Pi \left[ c_{g,N}\left( \boldsymbol{\theta }^+,\boldsymbol{\theta } \right) \right]}{\sigma _{g,N}^{2}\left( \boldsymbol{\theta }^+ \right)}\left( y^+-\mu _{g,N}\left( \boldsymbol{\theta }^+ \right) \right) 
\end{equation}
\begin{equation}\label{eq:BQVarNplus1}
	\sigma _{\mathcal{I} ,N+1}^{2}\left( \boldsymbol{\theta }^+ \right) =\sigma _{\mathcal{I} ,N}^{2}-\frac{\left( \Pi \left[ c_{g,N}\left( \boldsymbol{\theta }^+,\boldsymbol{\theta } \right) \right] \right) ^2}{\sigma _{g,N}^{2}\left( \boldsymbol{\theta }^+ \right)}.
\end{equation}
\end{subequations} 
Still, $	\mu _{\mathcal{I} ,N+1}\left( \boldsymbol{\theta }^+,y^+ \right) $ depends linearly on $y^+$, but $\sigma _{\mathcal{I} ,N+1}^{2}\left( \boldsymbol{\theta }^+ \right)$ is independent of $y^+$.

Based on the formulation of Eq. \eqref{eq:BQruleNplus1}, the PVC function is reformulated from two perspectives. First, let's define the expected gain $\mathcal{G} _{\mu}\left( \boldsymbol{\theta }^+ \right) $, of accepting $\boldsymbol{\theta}^+$ as a new training point, as the expected squared change of the posterior mean of the vanilla BQ rule, i.e,
\begin{equation}\label{eq:GainVBQ1}
	\mathcal{G} _{\mu}\left( \boldsymbol{\theta }^+ \right) =\int_{\mathbb{R}}{\left( \mu _{\mathcal{I} ,N+1}\left( \boldsymbol{\theta }^+,y^+ \right) -\mu _{\mathcal{I} ,N} \right) ^2f\left( y^+ \right)}\mathrm{d}y^+
\end{equation}
, where $f\left( y^+ \right) $ refers to the Gaussian density of $y^+$. The optimal new point $\boldsymbol{\theta }^+$ should be the one at which this expected gain is maximized, this way to maximize the expected movement of the mean prediction. From Eqs. \eqref{eq:BQMeanNplus1} and \eqref{eq:GainVBQ1}, the expected gain $\mathcal{G} _{\mu}\left( \boldsymbol{\theta }^+ \right)$ can be further formulated as:
\begin{equation}\label{eq:GainVBQ1_PVC}
	\mathcal{G} _{\mu}\left( \boldsymbol{\theta }^+ \right) =\frac{\left( \Pi \left[ c_{g,N}\left( \boldsymbol{\theta }^+,\boldsymbol{\theta } \right) \right] \right) ^2}{\sigma _{g,N}^{2}\left( \boldsymbol{\theta }^+ \right)}=\frac{\left( \mathcal{A} _{\mathrm{PVC}}\left( \boldsymbol{\theta }^+ \right) /p\left( \boldsymbol{\theta }^+ \right) \right) ^2}{\sigma _{g,N}^{2}\left( \boldsymbol{\theta }^+ \right)}.
\end{equation} 

Further, consider another type of expected gain formulated as:
\begin{equation}\label{eq:GainVBQ2}
	\mathcal{G} _{\sigma}\left( \boldsymbol{\theta }^+ \right) =\sigma _{\mathcal{I} ,N+1}^{2}\left( \boldsymbol{\theta }^+ \right) -\sigma _{\mathcal{I} ,N}^{2}
\end{equation}
, which represents the (expected) reduction of prediction variance of the vanilla BQ rule given $(\boldsymbol{\theta}^+,y^+)$ being added to the training data. From Eqs. \eqref{eq:BQVarNplus1} and \eqref{eq:GainVBQ2}, it can be immediately observed that $\mathcal{G} _{\sigma}\left( \boldsymbol{\theta }^+ \right)$ and $\mathcal{G} _{\mu}\left( \boldsymbol{\theta }^+ \right) $ admit exactly the same closed-form mathematical formulation. It is then concluded that, for any  $\boldsymbol{\theta }^{\left( t \right)}\in \mathcal{T}$, $\mathcal{G} _{\mu}\left( \boldsymbol{\theta }^{\left( t \right)} \right) =\mathcal{G} _{\sigma}\left( \boldsymbol{\theta }^{\left( t \right)} \right) =0$, indicating no gain can be achieved from reusing a point as a training point. As a summary, the gain function $\mathcal{G} _{\mu}\left( \boldsymbol{\theta } \right) $ (or $\mathcal{G} _{\sigma}\left( \boldsymbol{\theta } \right) $)  informs the point, by accepting which as a new training point, the prediction mean is expected to move the most, and meanwhile, the prediction variance will be reduced the most, thus it definitely embodies a prospective view for vanilla BQ. This interpretation endows the PVC function with a new connotation, which is that absolute PVC function $\left| \mathcal{A} _{\mathrm{PVC}}\left( \boldsymbol{\theta } \right) \right|$ does have a prospective view. 

One can use $\mathcal{G} _{\mu}\left( \boldsymbol{\theta } \right) $ formulated by Eq.\eqref{eq:GainVBQ1_PVC} as a new acquisition function for vanilla BQ, and it will definitely result in different design of training points. Our experience through several numerical experiments show that, in most implementations, $\mathcal{G} _{\mu}\left( \boldsymbol{\theta } \right) $ shows better performance, in terms of accuracy and efficiency, than $\mathcal{A} _{\mathrm{PVC}}\left( \boldsymbol{\theta } \right) $. An interesting complementary conclusion beyond Ref. \cite{wei2020adaptive} can then be drawn as that, the point with global minimum value of PVC function also brings high expected gain, which was originally explained as invaluable since its PVC value is negative. 

Readers can conduct their own numerical experiments to compare the performance of  $\mathcal{A} _{\mathrm{PVC}}\left( \boldsymbol{\theta } \right) $, $\left| \mathcal{A} _{\mathrm{PVC}}\left( \boldsymbol{\theta } \right) \right|$ and $\mathcal{G} _{\mu}\left( \boldsymbol{\theta } \right) $ for vanilla BQ. Here, we don't give more details, as our concern is mainly on the computation of model evidence $Z$ and posterior $p(\boldsymbol{\theta}|\mathcal{D}_\mathrm{obs})$, instead of the vanilla BQ rule for solving the integral equation in Eq. \eqref{eq:IntegralEq}. However, the conclusions presented in this subsection motivate the developments of more acquisition functions for estimating the model evidence with mutated BQ rule, as will be given in the next subsection.

\subsection{Reconstruction of BQ Rules and Acquisition Functions }
The above-reviewed vanilla BQ rule, which is based on approximating the integrand with a GP model, is not suitable for estimating the model evidence $Z$, due to the inefficiency of capturing the sharply peaked, non-negative and potentially multi-modal behaviors of the likelihood functions, as explained in Ref. \cite{wei2025bayesian}. Following this work, the target to be approximated by a GP model, also called the proxy function, is the logarithm of the likelihood function or equivalently, the negative energy function. It has been explained in  Ref. \cite{wei2025bayesian} that this practice is capable of boosting the so-called `exponential impact' for more effective active learning. The developments in this work will further give full play to the potential of this positive effect in enhancing the algorithms. Without confusing with the integrand of Eq. \eqref{eq:IntegralEq}, it is denoted as $g\left( \boldsymbol{\theta } \right) =-\mathcal{U} \left( \boldsymbol{\theta } \right) =\log p\left( \mathcal{D} _{\mathrm{obs}}|\boldsymbol{\theta } \right) $. In the previous work, the credible intervals (CIs) were utilized for quantifying the prediction uncertainties and devising the acquisition functions. In this subsection, new estimators, quantification of prediction uncertainties and resultant PVC function, are presented.       

Given the proxy function $g(\boldsymbol{\theta})$ being approximated by a GP model $\hat{g}_N\left( \boldsymbol{\theta } \right) \sim \mathcal{G} \mathcal{P} \left( \mu _{g,N}\left( \boldsymbol{\theta } \right) ,c_{g,N}\left( \boldsymbol{\theta },\boldsymbol{\theta }^\prime \right) \right) $ trained with $\mathcal{D} =\left\{ \left( \boldsymbol{\theta }^{\left( t \right)},y^{\left( t \right)} \right) \right\} _{t=1}^{N}$, a resultant stochastic model $\hat{p}_N\left( \mathcal{D} _{\mathrm{obs}}|\boldsymbol{\theta } \right) =\exp \left( \hat{g}_N\left( \boldsymbol{\theta } \right) \right) $ for approximating the likelihood can be inferred. It is readily observable that $\hat{p}_N\left( \mathcal{D} _{\mathrm{obs}}|\boldsymbol{\theta } \right) $ is a logarithmic GP (LGP) model and is denoted as $\hat{p}_N\left( \mathcal{D} _{\mathrm{obs}}|\boldsymbol{\theta } \right) \sim \mathcal{L} \mathcal{G} \mathcal{P} \left( \mu _{\mathrm{like},N}\left( \boldsymbol{\theta } \right) ,c_{\mathrm{like},N}\left( \boldsymbol{\theta },\boldsymbol{\theta }^{\prime} \right) \right) $, with the posterior mean $\mu _{\mathrm{like},N}\left( \boldsymbol{\theta } \right) $, variance $\sigma _{\mathrm{like},N}^{2}\left( \boldsymbol{\theta } \right) $ and covariance $c_{\mathrm{like},N}\left( \boldsymbol{\theta },\boldsymbol{\theta }^{\prime} \right) $ formulated as:
\begin{subequations}\label{eq:PostLikelihood}
	\begin{equation}\label{eq:PostMeanLikelihood}
		\mu _{\mathrm{like},N}\left( \boldsymbol{\theta } \right) =\exp \left( \mu _{g,N}\left( \boldsymbol{\theta } \right) +\frac{\sigma _{g,N}^{2}\left( \boldsymbol{\theta } \right)}{2} \right), 
	\end{equation}
\begin{equation}\label{eq:PostVarLikelihood}
	\sigma _{\mathrm{like},N}^{2}\left( \boldsymbol{\theta } \right) =\left( \exp \left( \sigma _{g,N}^{2}\left( \boldsymbol{\theta } \right) \right) -1 \right) \mu _{\mathrm{like},N}^{2}\left( \boldsymbol{\theta } \right) 
\end{equation}
, and
\begin{equation}\label{eq:PostCOVLikelihood}
	c_{\mathrm{like},N}\left( \boldsymbol{\theta },\boldsymbol{\theta }^{\prime} \right) =\mu _{\mathrm{like},N}\left( \boldsymbol{\theta } \right)  \left( \exp \left( c_{g,N}\left( \boldsymbol{\theta },\boldsymbol{\theta }^{\prime} \right) \right) -1 \right)\mu _{\mathrm{like},N} \left( \boldsymbol{\theta }^{\prime} \right) . 
\end{equation}
\end{subequations}
Eqs. \eqref{eq:PostMeanLikelihood} and \eqref{eq:PostVarLikelihood} can be directly obtained from the moment formulas of log-normal distribution, while Eq. \eqref{eq:PostCOVLikelihood} can be derived from the joint Gaussian density of $\hat{g}_{N}(\boldsymbol{\theta})$ and $\hat{g}_{N}(\boldsymbol{\theta}^\prime)$. Here we don't give more details.   

Additionally, the resultant model evidence, formulated by $\hat{Z}_N=\int_{\mathbb{T}}{\hat{p}_N\left( \mathcal{D} _{\mathrm{obs}}|\boldsymbol{\theta } \right) p\left( \boldsymbol{\theta } \right) \mathrm{d}\boldsymbol{\theta }}$, is a random variable, commonly with no closed-form distribution, but its mean and variance can be formulated, similar to the vanilla BQ rule \cite{wei2020adaptive},  as:
\begin{subequations}\label{eq:PostEvidence}
	\begin{equation}\label{eq:PostMeanEvidence}
		\mu _{Z,N}=\int_{\mathbb{T}}{\mu _{\mathrm{like},N}\left( \boldsymbol{\theta } \right) p\left( \boldsymbol{\theta } \right) \mathrm{d}\boldsymbol{\theta }}
	\end{equation} 
\begin{equation}\label{eq:PostVarEvidence}
	\sigma _{Z,N}^{2}=\int_{\mathbb{T} \times \mathbb{T}}{c_{\mathrm{like},N}\left( \boldsymbol{\theta },\boldsymbol{\theta }^{\prime} \right)}p\left( \boldsymbol{\theta } \right) p\left( \boldsymbol{\theta }^{\prime} \right) \mathrm{d}\boldsymbol{\theta }\mathrm{d}\boldsymbol{\theta }^{\prime}.
\end{equation}
\end{subequations}
Besides, using the Cauchy-Schwarz inequality $c_{\mathrm{like},N}\left( \boldsymbol{\theta },\boldsymbol{\theta }^{\prime} \right) \leqslant \sigma _{\mathrm{like},N}\left( \boldsymbol{\theta } \right) \sigma _{\mathrm{like},N}\left( \boldsymbol{\theta }^{\prime} \right) $, an upper bound of $\sigma _{Z,N}^{2}$ can be derived, to avoid the computation of the covariance $c_{g,N}\left( \boldsymbol{\theta },\boldsymbol{\theta }^{\prime} \right) $, as:
\begin{equation}\label{eq:UppBoundVarZ}
	\sigma _{Z,N}^{2}\leqslant \left( \int_{\mathbb{T}}{\sigma _{\mathrm{like},N}\left( \boldsymbol{\theta } \right) p\left( \boldsymbol{\theta } \right) \mathrm{d}\boldsymbol{\theta }} \right) ^2\triangleq \overline{\sigma }_{Z,N}^{2}.
\end{equation}
Ordinarily, there is no closed-form solutions for $\mu _{Z,N}$, $\sigma _{Z,N}^{2}$ and $\overline{\sigma }_{Z,N}^{2}$; whereas, due to the closed-form expressions of the integrands $\mu _{\mathrm{like},N}\left( \boldsymbol{\theta } \right) $, $\sigma _{\mathrm{like},N}^{2}\left( \boldsymbol{\theta } \right)$ and $c_{\mathrm{like},N}\left( \boldsymbol{\theta },\boldsymbol{\theta }^{\prime} \right) $, as given by Eq. \eqref{eq:PostLikelihood}, these three items can be efficiently estimated by crude or advanced Monte Carlo (MC) quadrature. 

With the above formulation, the quantity $\hat{p}_N\left( \mathcal{D} _{\mathrm{obs}}|\boldsymbol{\theta } \right) p\left( \boldsymbol{\theta } \right) $ for predicting the unnormalized posterior $p\left( \mathcal{D} _{\mathrm{obs}}|\boldsymbol{\theta } \right) p\left( \boldsymbol{\theta } \right) $ also follows a log-normal distribution, which is denoted as $\mathcal{L} \mathcal{G} \mathcal{P} \left( \mu _{\mathrm{post},N}\left( \boldsymbol{\theta } \right) ,c_{\mathrm{post},N}\left( \boldsymbol{\theta },\boldsymbol{\theta }^{\prime} \right) \right) $. The mean $\mu _{\mathrm{post},N}\left( \boldsymbol{\theta } \right) $, variance $\sigma _{\mathrm{post},N}^{2}\left( \boldsymbol{\theta } \right) $ and covariance $c_{\mathrm{post},N}\left( \boldsymbol{\theta },\boldsymbol{\theta }^{\prime} \right) $ are formulated as:
 \begin{subequations}\label{eq:PostMomentsPosterior}
 	\begin{equation}\label{eq:PostMeanPosterior}
 		\mu _{\mathrm{post},N}\left( \boldsymbol{\theta } \right) =\mu _{\mathrm{like},N}\left( \boldsymbol{\theta } \right) p\left( \boldsymbol{\theta } \right), 
 	\end{equation}
    \begin{equation}\label{eq:PostVarPosterior}
    	\sigma _{\mathrm{post},N}^{2}\left( \boldsymbol{\theta } \right) =\sigma _{\mathrm{like},N}^{2}\left( \boldsymbol{\theta } \right) p^2\left( \boldsymbol{\theta } \right) 
    \end{equation}
, and
   \begin{equation}\label{eq:PostCovPosterior}
   	c_{\mathrm{post},N}\left( \boldsymbol{\theta },\boldsymbol{\theta }^{\prime} \right) =p\left( \boldsymbol{\theta } \right) c_{\mathrm{like},N}\left( \boldsymbol{\theta },\boldsymbol{\theta }^{\prime} \right) p\left( \boldsymbol{\theta }^{\prime} \right). 
   \end{equation}
 \end{subequations} 

Till now, under the BQ framework, the posterior moments $\mu_{Z,N}$ and $\sigma^2_{Z,N}$ have been formulated with Eq. \eqref{eq:PostEvidence} to provide a probabilistic prediction of the deterministic-but-unknown model evidence $Z$; and accordingly, the posterior moments, $\mu_{\mathrm{post},N}(\boldsymbol{\theta})$, $\sigma^2_{\mathrm{post},N}(\boldsymbol{\theta})$, and $c_{\mathrm{post},N}\left( \boldsymbol{\theta },\boldsymbol{\theta }^{\prime} \right)$ have been built with Eq. \eqref{eq:PostMomentsPosterior} for predicting the unnormalized posteriors $p\left( \mathcal{D} _{\mathrm{obs}}|\boldsymbol{\theta } \right) p\left( \boldsymbol{\theta } \right) $, which is a deterministic-but-unknown function. One should not confuse these posterior features with those of the model parameters $\boldsymbol{\theta}$. The posterior uncertainty of $\boldsymbol{\theta}$, termed as ``Type-A uncertainty'', refers to the degree of unknowness of the model parameters $\boldsymbol{\theta}$, which is summarized by the target posterior density $p(\boldsymbol{\theta}|\mathcal{D}_\mathrm{obs})$; while the variance $\sigma^2_{Z,N}$ and $\sigma^2_{\mathrm{post},N}(\boldsymbol{\theta})$ summarize the numerical prediction uncertainties (termed as ``Type-B uncertainty'') associated with the model evidence $Z$ and posterior density $p(\boldsymbol{\theta}|\mathcal{D}_\mathrm{obs})$, both of which are deterministic-but-unknown quantities. For the numerical computation of $Z$ and $p(\boldsymbol{\theta}|\mathcal{D}_\mathrm{obs})$, the posterior variance of $\boldsymbol{\theta}$ is a deterministic value, and is not reducible during the numerical computation process. However, the variance $\sigma^2_{Z,N}$ and $\sigma^2_{\mathrm{post},N}(\boldsymbol{\theta})$ can be reduced given more training points, and the reduction efficiency relies on the training point design strategy, i.e., the utilized acquisition function. The posterior components of $Z$ and $p(\boldsymbol{\theta}|\mathcal{D}_\mathrm{obs})$ will then be employed to formulate the convergence criteria and develop efficient acquisition functions by means efficiently reducing these prediction uncertainties. To better distinguish between the two types of uncertainty described above, the relevant notations are summarized in Table \ref{table:TwoTypesUncernt}.

\newcolumntype{C}{>{\centering\arraybackslash}m{2cm}}
\newcolumntype{D}{>{\raggedright\arraybackslash}m{5.4cm}}
\newcolumntype{E}{>{\centering\arraybackslash}m{2.2cm}}
\newcolumntype{F}{>{\centering\arraybackslash}m{5.2cm}}

\begin{table}[H]
	\caption{Summary and illustration of the two types of uncertainty }
	\label{table:TwoTypesUncernt}
	\centering
	\begin{threeparttable}
		\begin{tabular}{C D E F}
			\hline 
			Uncertainty types  & Explanations & Quantified by & Roles\\ 
			\hline
			Type-A & Posterior uncertainty of model parameters $\boldsymbol{\theta}$.  & $p(\boldsymbol{\theta}|\mathcal{D}_\mathrm{obs})$& Deterministic-but-unknown function to be estimated by TBQ.\\
			Type-B & Numerical uncertainties associated with estimating $Z$ and $p(\boldsymbol{\theta}|\mathcal{D}_\mathrm{obs})$. & $\sigma^2_{Z,N}$ and $\sigma^2_{\mathrm{post},N}(\boldsymbol{\theta})$& Used for designing acquisition functions and indicating the convergence of TBQ.  \\
			\hline
		\end{tabular}
	\end{threeparttable}
\end{table}

Motivated by the above explanation, two non-prospective acquisition functions are proposed. The first one, named as \textbf{Prediction Uncertainty Quantification (PUQ)} function, is defined as:
\begin{equation}\label{eq:PUQ}
	\mathcal{A} _{\mathrm{PUQ}}\left( \boldsymbol{\theta } \right) =\sigma _{\mathrm{like},N}\left( \boldsymbol{\theta } \right) p\left( \boldsymbol{\theta } \right) 
\end{equation}     
, which admits a closed-form expression, and thus its global maximum point can be efficiently identified by e.g., Particle Swarm Optimization (PSO) and Genetic Algorithm (GA). This newly obtained training point is the location at which the prediction of the posterior density has the largest uncertainty, and is also the point of which the prediction uncertainty contributes the most to the upper bound $\overline{\sigma }_{Z,N}^{2}$, as indicated by Eq. \eqref{eq:UppBoundVarZ}. It is believed that, by updating the GP model, the prediction uncertainties associated with the posterior can be reduced to a great extent.

The second acquisition function, inspired by the \textbf{PVC function} for vanilla BQ \cite{wei2020adaptive}, is formulated as:
\begin{equation}\label{eq:PVCmodifeid}
	\mathcal{A} _{\mathrm{PVC}}\left( \boldsymbol{\theta } \right) =p\left( \boldsymbol{\theta } \right) \int_{\mathbb{T}}{c_{\mathrm{like},N}\left( \boldsymbol{\theta },\boldsymbol{\theta }^{\prime} \right)}p\left( \boldsymbol{\theta }^{\prime} \right) \mathrm{d}\boldsymbol{\theta }^{\prime}
\end{equation}
, which does not admit a closed-form expression, but can still be efficiently estimated by the MC quadrature rule. We use the same notation $\mathcal{A} _{\mathrm{PVC}}\left( \boldsymbol{\theta } \right) $ with Eq. \eqref{eq:DefPVC}, but one should not confuse these two. In what follows, without specific statement, the PVC function refers to the one defined by Eq. \eqref{eq:PVCmodifeid}. Similarly, this PVC function measures the contribution of the prediction uncertainty of $\hat{p}_N\left( \mathcal{D} _{\mathrm{obs}}|\boldsymbol{\theta } \right) $ to the posterior variance of $\hat{Z}_N$, with the integration of the spatial correlation information at $\boldsymbol{\theta}$ over the whole support $\mathbb{T}$. As $\exp \left( c_{g,N}\left( \boldsymbol{\theta },\boldsymbol{\theta }^{\prime} \right) \right)$ may be either larger or smaller than one, this PVC function can be either negative or positive, and both the global maximum and minimum points contributes substantially to $\sigma _{Z,N}^{2}$. Thus, one can also search the new training point by maximizing the absolute PVC function. Naturally, updating the GP model using the point with the maximum PVC value, the variance, and thus the prediction uncertainty, of the model evidence, can achieve a great reduction.  

\subsection{Prospective Acquisition Functions}
It has been shown, e.g., in our previous work for reliability analysis \cite{wei2023expected}, that  an acquisition function with a prospective view usually has the potential of substantially reducing the required number of function calls for a pre-specified accuracy. This motivates the development of prospective acquisition functions for learning the model evidence and posterior in this section. To achieve this, denote the LGP model induced from $\exp \left( \hat{g}_{N+1}\left( \boldsymbol{\theta } \right) \right) $ as $\hat{p}_{N+1}\left( \mathcal{D} _{\mathrm{obs}}|\boldsymbol{\theta } \right) \sim \mathcal{L} \mathcal{G} \mathcal{P} \left( \mu _{\mathrm{like},N+1}\left( \boldsymbol{\theta }|\boldsymbol{\theta }^+,y^+ \right) ,c_{\mathrm{like},N+1}\left( \boldsymbol{\theta },\boldsymbol{\theta }^{\prime}|\boldsymbol{\theta }^+ \right) \right) $, of which the posterior mean $\mu _{\mathrm{like},N+1}\left( \boldsymbol{\theta }|\boldsymbol{\theta }^+,y^+ \right) $, variance $\sigma _{\mathrm{like},N+1}^{2}\left( \boldsymbol{\theta }|\boldsymbol{\theta }^+,y^+ \right) $ and covariance $c_{\mathrm{like},N+1}\left( \boldsymbol{\theta },\boldsymbol{\theta }^{\prime}|\boldsymbol{\theta }^+,y^+ \right) $ can be obtained by simply replacing all the subscripts ``$N$'' in Eq. \eqref{eq:PostLikelihood} with ``$N+1$''. Further, the posterior mean $\mu _{Z,N+1}(\boldsymbol{\theta}^+,y^+)$ and variance $\sigma _{Z,N+1}^{2}(\boldsymbol{\theta}^+,y^+)$ of the model evidence, resulted from the update of the GP model with $\left( \boldsymbol{\theta }^+,y^+ \right) $, can be formulated by replacing all the subscripts ``$N$'' in Eq. \eqref{eq:PostEvidence} with ``$N+1$''. 

The first prospective acquisition function, called\textbf{ Prospective Likelihood Uncertainty Reduction (PLUR)} function, is formulated as:
\begin{equation}\label{eq:DefPLUR}
	\mathcal{A} _{\mathrm{PLUR}}\left( \boldsymbol{\theta }^+ \right) =\int_{\mathbb{T}}{\mu _{\mathrm{like},N}^{2}\left( \boldsymbol{\theta } \right) \left( \exp \left( \frac{c_{g,N}^{2}\left( \boldsymbol{\theta }^+,\boldsymbol{\theta } \right)}{\sigma _{g,N}^{2}\left( \boldsymbol{\theta }^+ \right)} \right) -1 \right) p(\boldsymbol{\theta })\mathrm{d}\boldsymbol{\theta }}.
\end{equation}
As its name reveals, the PLUR function measures the expected reduction of prediction uncertainty for the likelihood. To illustrate this, the following two equations are presented: 
\begin{equation}\label{eq:PLURInterp}
	\begin{split}
		\mathcal{A} _{\mathrm{PLUR}}\left( \boldsymbol{\theta }^+ \right) &=\int_{\mathbb{R}}{\int_{\mathbb{T}}{\left( \mu _{\mathrm{like},N+1}\left( \boldsymbol{\theta }|\boldsymbol{\theta }^+,y^+ \right) -\mu _{\mathrm{like},N}\left( \boldsymbol{\theta } \right) \right) ^2p\left( \boldsymbol{\theta } \right) f\left( y^+ \right) \mathrm{d}\boldsymbol{\theta }\mathrm{d}y^+}}
		\\
		&=\int_{\mathbb{R}}{\int_{\mathbb{T}}{\left( \sigma _{\mathrm{like},N}^{2}\left( \boldsymbol{\theta } \right) -\sigma _{\mathrm{like},N+1}^{2}\left( \boldsymbol{\theta }|\boldsymbol{\theta }^+,y^+ \right) \right) p\left( \boldsymbol{\theta } \right) f\left( y^+ \right) \mathrm{d}\boldsymbol{\theta }}\mathrm{d}y^+}.
	\end{split}
\end{equation}  
The mathematical derivations of both equations in Eq. \eqref{eq:PLURInterp} are reported in \ref{Append:Proof20}. The first row of Eq. \eqref{eq:PLURInterp} implies that $	\mathcal{A} _{\mathrm{PLUR}}\left( \boldsymbol{\theta }^+ \right)$ can be explained as the expected cumulative change of the posterior mean of the likelihood (and also the unnormalized posterior), if the GP model is updated with a new training point $\boldsymbol{\theta }^+ $; while the second row reveals that $	\mathcal{A} _{\mathrm{PLUR}}\left( \boldsymbol{\theta }^+ \right)$ can be interpreted as the expected cumulative reduction of the prediction variance of the likelihood (and thus also the unnormalized posterior), given $\boldsymbol{\theta }^+ $ is specified as a new training point for updating. Thereof, the acquisition function $	\mathcal{A} _{\mathrm{PLUR}}\left( \boldsymbol{\theta }^+ \right)$ measures the prospective gain on improving the prediction accuracy of the likelihood and the posterior. Intuitively, by updating the GP model, and then the resultant results, using the point $\boldsymbol{\theta}^+$ with the maximum PLUR value, it is expected to improve the accuracy for predicting the posterior the most. Taking the above interpretations into account, the training points are sequentially determined as $\boldsymbol{\theta }^+=\mathrm{arg}\max _{\theta \in \mathbb{T}}\mathcal{A} _{\mathrm{PLUR}}\left( \boldsymbol{\theta } \right) $.       
 
Concerning the expected gain of predicting the model evidence, the second prospective acquisition function, called \textbf{Prospective Evidence Uncertainty Reduction (PEUR)} function, is defined as:
\begin{equation}\label{eq:DefPEUR}
\mathcal{A} _{\mathrm{PEUR}}\left( \boldsymbol{\theta }^+ \right) =\int_{\mathbb{T} \times \mathbb{T}}{\mu _{\mathrm{like},N}\left( \boldsymbol{\theta } \right) \mu _{\mathrm{like},N}\left( \boldsymbol{\theta }^{\prime} \right) \left( \exp \left( \frac{c_{g,N}\left( \boldsymbol{\theta },\boldsymbol{\theta }^+ \right) c_{g,N}\left( \boldsymbol{\theta }^+,\boldsymbol{\theta }^{\prime} \right)}{\sigma _{g,N}^{2}\left( \boldsymbol{\theta }^+ \right)} \right) -1 \right) p\left( \boldsymbol{\theta } \right) p\left( \boldsymbol{\theta }^{\prime} \right) \mathrm{d}\boldsymbol{\theta }\mathrm{d}\boldsymbol{\theta }^{\prime}}.
\end{equation} 
The name ``PEUR'' stems from the fact that it measures the expected reduction of prediction uncertainty associated with the model evidence $Z$. To illustrate this, the following two equations are developed:
\begin{equation}\label{eq:PEURInterp}
	\begin{split}
		\mathcal{A} _{\mathrm{PEUR}}\left( \boldsymbol{\theta }^+ \right) &=\int_{\mathbb{R}}{\left( \mu _{Z,N+1}\left( \boldsymbol{\theta }^+,y^+ \right) -\mu _{Z,N} \right) ^2f\left( y^+ \right) \mathrm{d}y^+}
		\\
		&=\sigma _{Z,N}^{2}-\int_{\mathbb{R}}{\sigma _{Z,N+1}^{2}(\boldsymbol{\theta }^+,y^+)f\left( y^+ \right) \mathrm{d}y^+}.
	\end{split}
\end{equation}
The mathematical proof of Eq. \eqref{eq:PEURInterp} is presented in \ref{Append:Proof22}. The right-hand term of the first line of Eq. \eqref{eq:PEURInterp} is explained the expected amount of change of the mean prediction of $Z$, and that of the second row is interpreted as the expected reduction of the prediction variance of $Z$, both resulted from updating the GP model with a new training point $\left( \boldsymbol{\theta }^+,y^+ \right) $. Thus, specifying the new training point as the point at which the PEUR function is maximized, i.e., $\boldsymbol{\theta }^+=\mathrm{arg}\max _{\theta \in \mathbb{T}}\mathcal{A} _{\mathrm{PEUR}}\left( \boldsymbol{\theta } \right) $, it is expected to improve the prediction accuracy of the model evidence to a maximum amount.

Up to now, four acquisition functions, i.e., the PUQ function defined by Eq. \eqref{eq:PUQ}, the PVC function defined in Eq. \eqref{eq:PVCmodifeid}, the PLUR function defined by Eq. \eqref{eq:DefPLUR} and the PEUR function defined with Eq. \eqref{eq:DefPEUR}, have all been proposed with clear intuitions and explanations. They are collectively summarized in Table \ref{table:SummaryAcqFun}, where the last column illustrates the computational complexity level of each acquisition function. One notes that the PVC function has also been used in Ref. \cite{kitahara2025sequential}, but the computation is quite demanding as it depends on sampling of both GP model and $\boldsymbol{\theta}$. In case the acquisition functions do not admit closed-form expressions, they can be estimated with the MC quadrature rule, based on a sample matrix $\mathcal{T} _{\mathrm{MC}}=\left( \boldsymbol{\theta }_{\mathrm{MC}}^{\left( k \right)} \right) _{k=1}^{N_{\mathrm{MC}}}$ (for $\mathcal{A} _{\mathrm{PVC}} $ and $\mathcal{A} _{\mathrm{PLUR}} $) or two independent sample matrices $\mathcal{T} _{\mathrm{MC}}$ and $\mathcal{T} _{\mathrm{MC}}^{\prime}$ (for $\mathcal{A} _{\mathrm{PEUR}}$), where $\mathcal{T} _{\mathrm{MC}}^{\prime}$ can be generated by randomly permuting the rows of $\mathcal{T} _{\mathrm{MC}}$. This also applies to the estimation of the posterior mean $\mu_{Z,N}$ and variance $\sigma_{Z,N}^2$, as formulated by Eq. \eqref{eq:PostEvidence}. Compared with those prospective acquisition functions for structural reliability analysis, such as those reported in Ref. \cite{zhou2024look, wei2023expected}, an appealing feature of $\mathcal{A}_\mathrm{PLUR}$ and $\mathcal{A}_\mathrm{PEUR}$ is that, although being initially defined as expectations over the probability distribution of $y^+$, it is not required to estimate the integrals over $y^+$. This feature is highly beneficial: while achieving better expected rewards, the computational cost of the acquisition function does not increase significantly. Indeed, the cost for computing the PLUR function is almost the same as for computing the PVC function, and that for PEUR is a little bit higher, as illustrated in Table \ref{table:SummaryAcqFun}. This is distinct yet encouraging property for both the PLUR and PEUR functions that sets them apart from prospective acquisition functions in other Bayesian numerical analysis, like the Expected Integrated Error Reduction (EIER) developed for Bayesian active learning of failure probability \cite{wei2023expected}, and the Knowledge Gradient (KG) function developed for Bayesian optimization \cite{frazier2008knowledge}. 

\newcolumntype{C}{>{\centering\arraybackslash}m{3.2cm}}
\newcolumntype{D}{>{\raggedright\arraybackslash}m{10.4cm}}
\newcolumntype{E}{>{\centering\arraybackslash}m{2.2cm}}

\begin{table}[H]
	\caption{Summary of acquisition functions for estimating model evidence and posteriors with BQ rules}
	\label{table:SummaryAcqFun}
	\centering
	\begin{threeparttable}
		\begin{tabular}{C D E}
			\hline 
			Notations  & Explanations & Complexities \tnote{1}\\ 
			\hline
			$\mathcal{A} _{\mathrm{PUQ}}\left( \boldsymbol{\theta } \right) $, Eq. \eqref{eq:PUQ}& Prediction variance of posterior at $\boldsymbol{\theta}$.  & I\\
		$\mathcal{A} _{\mathrm{PVC}}\left( \boldsymbol{\theta } \right) $, Eq. \eqref{eq:PVCmodifeid}& Contribution of prediction uncertainty at $\boldsymbol{\theta}$ to that of $Z$. &II\\
			$\mathcal{A} _{\mathrm{PLUR}}\left( \boldsymbol{\theta } \right) $, Eq. \eqref{eq:DefPLUR}&Expected change of mean prediction, or expected reduction of prediction variance, of likelihood/posterior, attained by adopting $\boldsymbol{\theta}$.   &II\\
			$\mathcal{A} _{\mathrm{PEUR}}\left( \boldsymbol{\theta } \right) $,  Eq. \eqref{eq:DefPEUR}&Expected change of mean prediction, or expected reduction of prediction variance, of model evidence, achieved by adopting $\boldsymbol{\theta}$.  & $\mathrm{II}^+$\\
			
			\hline
		\end{tabular}
		\begin{spacing}{1}
			\begin{tablenotes}
				{\small\item [1] Complexity level I indicates that the acquisition function admits a closed-form expression; Level II implies that the acquisition function is defined as a $d$-dimensional integral over $\mathbb{T}$ with a closed-form integrand; Level $\mathrm{II}^+$ reveals that the acquisition function is defined by a $2d$-dimensional integral of over $\mathbb{T}\times \mathbb{T}$ with a closed-form integrand; Thus, with higher level of complexity, the computational cost gets more demanding, but level  $\mathrm{II}^+$ is a little bit higher than level  $\mathrm{II}$ as both are estimated with single-loop MC estimators.} 
			\end{tablenotes}
		\end{spacing}
		
	\end{threeparttable}
\end{table}

\subsection{Bayesian Quadrature Algorithm}
With any one of the four developed acquisition functions as an ``engine'' for active learning, the BQ algorithm for learning the model evidence and posterior is summarized  Algorithm \ref{Alg: BQ}. Two stopping conditions are developed as:
\begin{subequations}\label{eq:StopCond}
	\begin{equation}\label{eq:StopCond1}
		\frac{\overline{\sigma }_{Z,N}}{\mu _{Z,N}}\leqslant \epsilon _1
	\end{equation}
	and 
	\begin{equation}\label{eq:StopCond2}
		\frac{\sigma _{Z,N}}{\mu _{Z,N}}\leqslant \epsilon _2
	\end{equation}
\end{subequations}
, where $\overline{\sigma }_{Z,N}$ is the upper bound of the STandard Deviation (STD) of $Z$, and is computed by MC estimator based on Eq. \eqref{eq:UppBoundVarZ}, $\sigma _{Z,N}$ is the exact STD of $Z$, which can be computed by a MC estimator based on Eq. \eqref{eq:PostVarEvidence}. The computation of $\overline{\sigma }_{Z,N}$ is based on a $(N_\mathrm{MC}\times d)$-dimensional sample matrix $\mathcal{T} _{\mathrm{MC}}$, while $\sigma _{Z,N}$ is based on two, i.e.,  $\mathcal{T} _{\mathrm{MC}}$ and $\mathcal{T} _{\mathrm{MC}}^\prime$. If $\mathcal{A} _{\mathrm{PVC}}\left( \boldsymbol{\theta } \right) $ or $\mathcal{A} _{\mathrm{PLUR}}\left( \boldsymbol{\theta } \right) $ is utilized as the acquisition function, it is suggested to use Eq. \eqref{eq:StopCond1} as a  rough judgment, and for iteration steps where it is satisfied, use the second one. For this scheme, $\epsilon_1$ can be set as, e.g., 0.05$\sim$0.1, and $\epsilon_2$ as, e.g., 0.01$\sim$0.05, based on the users' tolerance to prediction error. However, if $\mathcal{A} _{\mathrm{PEUR}}\left( \boldsymbol{\theta } \right) $ is selected as the acquisition function, itself needs to be computed with the two sample matrices, one can simply use the second stopping condition. In Algorithm \ref{Alg: BQ}, only one condition is described for simplicity. 

One notes that, although the stopping conditions are defined using the prediction uncertainty associated with the model evidence, meeting the stopping criterion also implies the convergence of the posterior estimate. For example, if the condition of Eq. \eqref{eq:StopCond1} is satisfied, it means $\bar{\sigma}_{Z,N}$ is sufficiently small, which further indicates $\sigma _{\mathrm{like},N}\left( \boldsymbol{\theta} \right)$ is also negligibly small, as $\bar{\sigma}_{Z,N}$ is formulated as the integral of the non-negative function $\sigma _{\mathrm{like},N}\left( \boldsymbol{\theta } \right) $ over the prior, as revealed by Eq. \eqref{eq:UppBoundVarZ}. Further, this indicates that the likelihood, and thus the unnormalized posterior, is estimated with high accuracy. Given both the unnormalized posterior and the model evidence are estimated with controllable error tolerance, the normalized posterior is also estimated with desired accuracy. One can also formulate stopping criterion using the posterior variance $\sigma_{\mathrm{post},j}^2(\boldsymbol{\theta})$ of the unnormalized posterior, but no in-depth discussion is provided herein due to space limitation. 

\vspace{0.5cm}

\begin{spacing}{1.3}
	\begin{algorithm}[H]
		\label{Alg: BQ}
		\caption{The BQ algorithm}
		\LinesNumbered 
		\KwIn{$g(\boldsymbol{\theta})$, $\epsilon$, $N_0$, $N_\mathrm{MC}$}
		\KwOut{$\mu _{Z,N}$, $\sigma _{Z,N}$ (or $\overline{\sigma }_{Z,N}$), $\mu _{\mathrm{post},N}\left( \boldsymbol{\theta } \right) $, $\sigma _{\mathrm{post},N}\left( \boldsymbol{\theta } \right) $, $N$ }
		Initialization: $\mathrm{StopFlag}=0$, $N=N_0$;\\
		$\mathcal{T} _{\mathrm{MC}}=\left\{ \boldsymbol{\theta }_{\mathrm{MC}}^{\left( k \right)} \right\} _{k=1}^{N_{\mathrm{MC}}} \gets$ LSH sampling from $p(\boldsymbol{\theta})$;\\
		$\mathcal{T} _{\mathrm{MC}}^{\prime}=\left\{ \boldsymbol{\theta }_{\mathrm{MC}}^{\prime \left( k \right)} \right\} _{k=1}^{N_{\mathrm{MC}}}\gets $ Randomly permute the rows of $\mathcal{T} _{\mathrm{MC}}$;\\
		$\mathcal{T} =\left\{ \boldsymbol{\theta }^{\left( t \right)} \right\} _{t=1}^{N}\gets$ LSH sampling from $p(\boldsymbol{\theta})$;\\
		$\mathcal{Y} \gets g(\mathcal{T})$;\\
		\While{$\mathrm{StopFlag}=0$\,}{
			Train $\hat{g}_N\left( \boldsymbol{\theta} \right) $ with $\mathcal{D} _{N,\mathrm{train}}=\left\{ \mathcal{T} ,\mathcal{Y} \right\} $;\\
			$\mu _{Z,N} \gets$ MC estimate of Eq. \eqref{eq:PostMeanEvidence} based on $\mathcal{T}_\mathrm{MC}$;\\
			$\overline{\sigma }_{Z,N} \gets$ MC estimate of Eq. \eqref{eq:UppBoundVarZ}  based on $\mathcal{T}_\mathrm{MC}$; or\\
			$\sigma _{Z,N}\gets$ MC estimate of Eq. \eqref{eq:PostVarEvidence} based on $\mathcal{T}_\mathrm{MC}$ and $\mathcal{T}_\mathrm{MC}^\prime$;\\
			
			\eIf{\textup{Eq. \eqref{eq:StopCond} satisfied}}{
				$\mu _{\mathrm{like},N}\left( \boldsymbol{\theta } \right)$ and $\sigma _{\mathrm{like},N}\left( \boldsymbol{\theta } \right) \gets$ Eq. \eqref{eq:PostLikelihood};\\
				$\text{StopFlag}=1$;
			}{
				$\boldsymbol{\theta }^+\gets $ $\mathrm{arg}\max _{\boldsymbol{\theta }\in \mathbb{T}}\mathcal{A} \left( \boldsymbol{\theta } \right) $;\\ 
				$y^+\gets g\left( \boldsymbol{\theta }^+ \right) $; \\
				$\mathcal{T} \gets \mathcal{T} \cup \left\{ \boldsymbol{\theta }^+ \right\} $;\\ $\mathcal{Y} \gets \mathcal{Y} \cup \left\{ y^+ \right\} $;\\
				$N\gets N+1$;\\
			}
		}	
		$\mu _{\mathrm{post},N}\left( \boldsymbol{\theta } \right)$ and $\sigma _{\mathrm{post},N}\left( \boldsymbol{\theta } \right) \gets$ Eq. \eqref{eq:PostMomentsPosterior};\\
	\end{algorithm}
\end{spacing}

\vspace{0.5cm}

The inputs to Algorithm \ref{Alg: BQ} consist of, the proxy function (logarithm of likelihood, or negative energy function) $g(\boldsymbol{\theta})$, the prior density $p(\boldsymbol{\theta})$, the stopping threshold $\epsilon$ (either $\epsilon_1$ or $\epsilon_2$), the initial training sample size $N_0$ (e.g., 12), and the MC sample size $N_\mathrm{MC}$ (suggested to be $5\times 10^3\sim 1\times10^4$). The outputs of Algorithm \ref{Alg: BQ} include the mean prediction $\mu _{Z,N}$, STD $\sigma _{Z,N}$ (or $\overline{\sigma }_{Z,N}$) for indicating the prediction uncertainty for $Z$, the posterior mean $\mu _{\mathrm{post},N}\left( \boldsymbol{\theta } \right) $ and STD $\sigma _{\mathrm{post},N}\left( \boldsymbol{\theta } \right) $ of the unnormalized posterior.  

\section{Improvements for Transitional Bayesian Quadrature}
The developments in section \ref{sec:ProAcqFun} are then further utilized to reform the TBQ algorithm for problems with posteriors being extremely divergent from the priors. The improvements are mainly in three aspects, i.e., stopping conditions for each tempering stage, acquisition functions for querying new training points, as well as estimators together with their uncertainty measures for predicting posteriors and evidences. Before the proposition of these specific improvements, the TBQ algorithm framework is briefly reviewed. One can refer to Ref. \cite{wei2025bayesian} for Bayesian model updating and Ref. \cite{wei2025transitional} for structural reliability analysis.

\subsection{Framework of Transitional Bayesian Quadrature}
The TBQ algorithm, inspired by the TMCMC algorithm \cite{ching2007transitional}, is based on approaching the target posterior by a set of intermediate densities scaled by a tempering parameter $\gamma_j$ varying from zero to one. Specifically, given $\gamma_j\in [0, 1]$, the intermediate posterior density at the $j$-th tempering stage is defined as:
\begin{equation}\label{eq:DefIntermediateDensity}
p_j\left( \boldsymbol{\theta }|\mathcal{D} _{\mathrm{obs}} \right) =Z_{j}^{-1}\left[ p\left( \mathcal{D} _{\mathrm{obs}}|\boldsymbol{\theta } \right) \right] ^{\gamma _j}p\left( \boldsymbol{\theta } \right) 
\end{equation}
, where 
\begin{equation}
	Z_j=\int_{\mathbb{T}}{\left[ p\left( \mathcal{D} _{\mathrm{obs}}|\boldsymbol{\theta } \right) \right] ^{\gamma _j}p\left( \boldsymbol{\theta } \right) \mathrm{d}\boldsymbol{\theta }}
\end{equation}
is the normalizing constant (or called intermediate model evidence) of the $j$-th stage. With $\gamma_j=0$, $p_j\left( \boldsymbol{\theta }|\mathcal{D} _{\mathrm{obs}} \right)$ equals exactly to the prior density and $Z_j=1$; while with $\gamma_j$ approaching one, $p_j\left( \boldsymbol{\theta }|\mathcal{D} _{\mathrm{obs}} \right)$ is prone to the target posterior $p\left( \boldsymbol{\theta }|\mathcal{D} _{\mathrm{obs}} \right) $, and $Z_j$ approaches to the target model evidence $Z$.

With $p_{j-1}\left( \boldsymbol{\theta }|\mathcal{D} _{\mathrm{obs}} \right)$ as the weight density, the evidence ratio $Z_j/Z_{j-1}$ of the two consecutive tempering stages can be formulated as:
\begin{equation}\label{eq:RatioDefinition}
	\frac{Z_{j}}{Z_{j-1}}=\int_{\mathbb{T}}{\left[ p\left( \mathcal{D} _{\mathrm{obs}}|\boldsymbol{\theta } \right) \right] ^{\gamma _{j}-\gamma _{j-1}}p_{j-1}\left( \boldsymbol{\theta }|\mathcal{D}_\mathrm{obs} \right) \mathrm{d}\boldsymbol{\theta }}.
\end{equation}
The above formulation aligns with the Bayesian incremental learning scheme, i.e., one can take the posterior $p_{j-1}\left( \boldsymbol{\theta }|\mathcal{D}_\mathrm{obs} \right)$ of the $(j-1)$-th tempering stage as a prior, and then the incremental quantity $\left[ p\left( \mathcal{D} _{\mathrm{obs}}|\boldsymbol{\theta } \right) \right] ^{\gamma _{j}-\gamma _{j-1}}$ as a likelihood, to infer a posterior $p_j\left( \boldsymbol{\theta }|\mathcal{D} _{\mathrm{obs}} \right)$ and the associated model evidence $Z_j$, for the $j$-th stage. 

\begin{figure}[H]
	\centering
	\setlength{\abovecaptionskip}{-0.0cm}
	\includegraphics[scale=0.5]{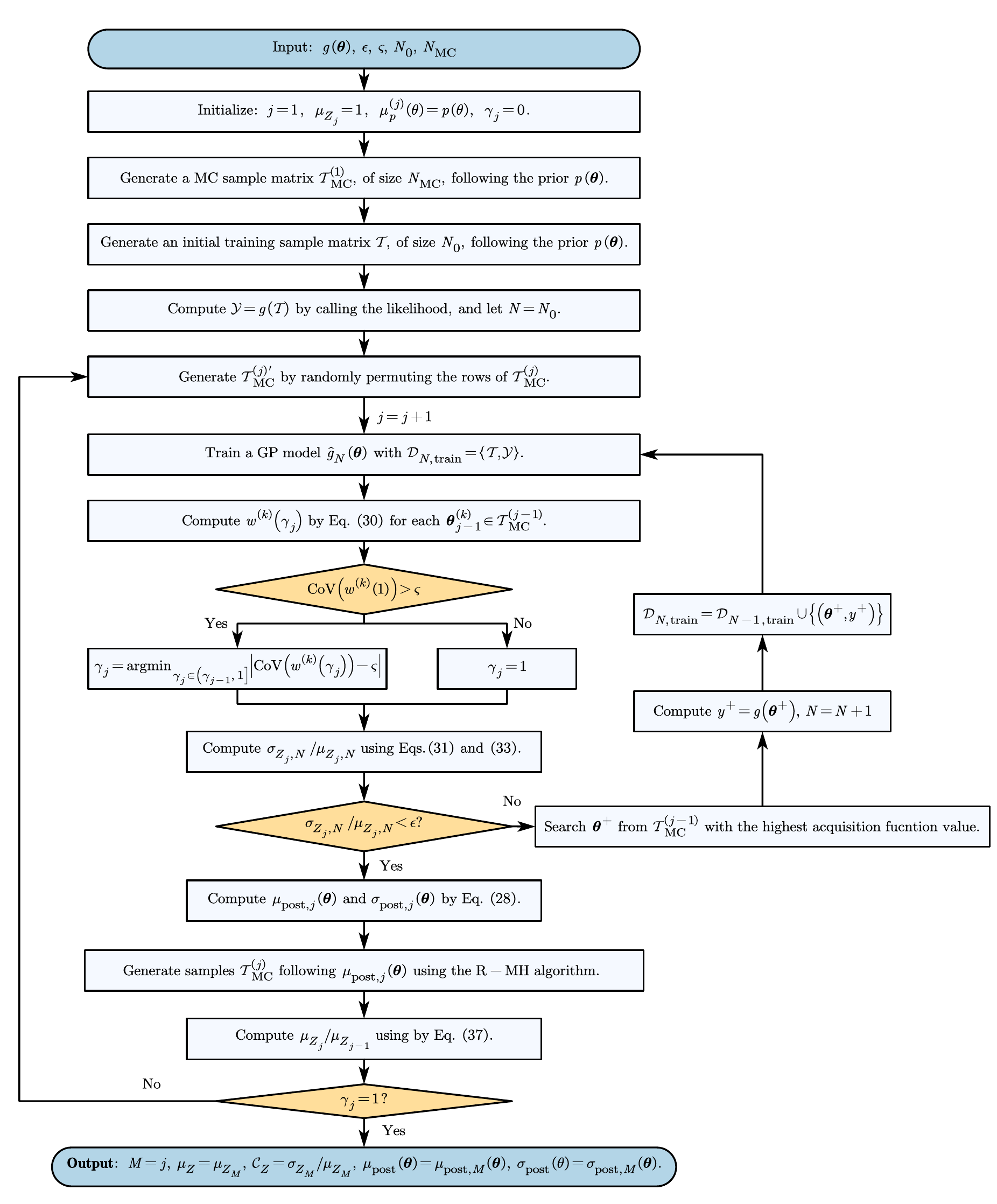}
	\caption{\centering{Flowchart of the TBQ algorithm.}}
	\label{fig:TBQflowchart}
\end{figure}

The fundamental principle of TBQ is to learn a sequence of values, denoted as $0=\gamma _1<\gamma _2<\cdots <\gamma _M=1$, for the tempering parameter, and meanwhile, to estimate $p_j\left( \boldsymbol{\theta }|\mathcal{D} _{\mathrm{obs}} \right)$ and $Z_j/Z_{j-1}$ for each tempering stage with desired accuracy, both using the BQ algorithm, this way to sequentially approach the target posterior $p\left( \boldsymbol{\theta }|\mathcal{D} _{\mathrm{obs}} \right) $ and model evidence $Z$ using the estimates of $p_M\left( \boldsymbol{\theta }|\mathcal{D} _{\mathrm{obs}} \right)$ and $Z_M$ respectively. One can refer to Ref. \cite{wei2025bayesian} for the original TBQ algorithm. The TBQ algorithm reformed with the developments in section \ref{sec:ProAcqFun} is conceptually summarized in Algorithm \ref{Alg:TBQ}, of which the five critical procedures, labeled as \textcolor{blue}{\ding{192}-\ding{196}}, are described in the subsequent subsections, together with corresponding parameters settings. To facilitate readers in coding the algorithm, a more detailed algorithmic flowchart is also presented in Figure \ref{fig:TBQflowchart}, accompanied with a source code for implementing the TBQ algorithm (see Section \ref{sec:conclusion} for the link).

\vspace{0.5 cm}
\begin{spacing}{1.25}
	\begin{algorithm}[H]
		\label{Alg:TBQ}
		\caption{The TBQ algorithm}
		\LinesNumbered 
		\KwIn{$g(\boldsymbol{\theta})$, $\epsilon$, $\varsigma$, $N_0$, $N_\mathrm{MC}$, $N_\mathrm{lim}$}
		\KwOut{$\mu _Z$, $C _Z$, $\mu _{\mathrm{post}}\left( \boldsymbol{\theta } \right) $,$\sigma _{\mathrm{post}}\left( \boldsymbol{\theta } \right) $, $N$ }
		Initialization: $N=N_0$, $j=1$, $\mu _{Z_j}=1$,$\mu _{p}^{\left( j \right)}\left( \boldsymbol{\theta } \right) =p\left( \boldsymbol{\theta } \right) $, $\gamma_j=0$;\\
		$\mathcal{T}^{(1)} _{\mathrm{MC}} \gets$ Generate $N_\mathrm{MC}$ samples following  $p(\boldsymbol{\theta})$;\\
		$\mathcal{T} =\left\{ \boldsymbol{\theta }^{\left( t \right)} \right\} _{t=1}^{N}\gets$ Generate $N_0$ initial training samples following $p(\boldsymbol{\theta})$;\\
		$\mathcal{Y} \gets g(\mathcal{T})$;\\
		\While{$\gamma_j<1$}{
		    $\mathcal{T} _{\mathrm{MC}}^{(j)\prime}\gets $ Randomly permute the rows of $\mathcal{T} _{\mathrm{MC}}^{(j)}$;\\
		    			$j\gets j+1$;\\
		   	\While{ $N<N_\mathrm{lim}$ }{
		   			$\hat{g}_N\left( \boldsymbol{\theta} \right) \gets$ Train a GP model with $\mathcal{D} _{N,\mathrm{train}}=\left\{ \mathcal{T} ,\mathcal{Y} \right\} $;\\
		   			\eIf{$\mathrm{CoV}\left( w^{\left( k \right)}\left( 1 \right) \right) <\varsigma$}{$\gamma_{j}$ $\gets$ 1}{$\gamma_{j}$ $\gets$ $\mathrm{arg}\min _{\gamma _j\in \left( \gamma _{j-1},1 \right]}\left| \mathrm{CoV}\left( w^{\left( k \right)}\left( \gamma _j \right) \right) -\varsigma \right|$ \textcolor{blue}{\ding{192}};
		   			}
		   	    	$\mu_{Z_j,N}$ and $\sigma_{Z_j,N}$ $\gets$ compute the mean and STD predictions \textcolor{blue}{\ding{193}};\\  	    	
		   	    	\eIf{$\mathrm{Stopping \,\,\, condition}$ \textcolor{blue}{\ding{193}} satisfied }{Break the inner \textbf{while} loop;}{
		   	    	$\boldsymbol{\theta}^+ \gets$ Search a new training point from $\mathcal{T}_\mathrm{MC}^{(j-1)}$ or the support of the prior density by maximizing an acquisition function \textcolor{blue}{\ding{194}};\\
		   	    	$y^+\gets g(\boldsymbol{\theta}^+)$;\\
		   	    	$\mathcal{D}_{N+1,\mathrm{train}}\leftarrow\mathcal{D}_{N,\mathrm{train}}\cup\{\left(\boldsymbol{\theta}^{+},y^{+}\right)\}$; \\
		   	    	$N\gets N+1$;
		   	    }
		   	}   
	   	$\mu _{\mathrm{post},j}\left( \boldsymbol{\theta } \right)$ and $\sigma _{\mathrm{post},j}\left( \boldsymbol{\theta } \right) \gets$ Eq. \eqref{eq:PostMomentsLike_jStage};\\
	   	 		
	   	$\mathcal{T}^{(j)} _{\mathrm{MC}} \gets$ Generate samples following the mean estimate $\mu _{\mathrm{post},j}\left( \boldsymbol{\theta } \right) $  \textcolor{blue}{\ding{195}}; \\
	   	$\mu _{Z_j}/\mu _{Z_{j-1}}$ $\gets$ Mean prediction \textcolor{blue}{\ding{196}};\\
	    }
        $M\gets j$;\\
        $\mu _Z$, $\mathcal{C} _Z \gets$ Predict $Z$ and estimate the CoV \textcolor{blue}{\ding{196}};\\
    	 $\mu _{\mathrm{post}}\left( \boldsymbol{\theta } \right) \gets \mu _{\mathrm{post},M}\left( \boldsymbol{\theta } \right) $, $\sigma _{\mathrm{post}}\left( \boldsymbol{\theta } \right) \gets \sigma _{\mathrm{post},M}\left( \boldsymbol{\theta } \right) $;\\	
	\end{algorithm}
\end{spacing}
\vspace{0.5 cm}

The transitional learning scheme by adaptive increment of $\gamma_j$ also presents a flexible trade-off between exploration and exploitation. Specifically, in the early tempering stages, this scheme encourages exploration as $p_j\left( \boldsymbol{\theta }|\mathcal{D} _{\mathrm{obs}} \right)$ with small $\gamma_j$ has a richer support; while in the later stage as $\gamma_j$ approaching one, $p_j\left( \boldsymbol{\theta }|\mathcal{D} _{\mathrm{obs}} \right)$ tends to the target posterior, and the algorithm is more inclined to exploit the local modes where the target posterior admits high values. The movement from one stage to the subsequent one is controlled by an adaptive updating scheme for specifying $\gamma_j$, as will be introduced in subsection \ref{subsec: SpecifGamma}. A more refined trade-off between exploration and exploitation within one fixed stage is controlled by the acquisition function, as will be presented in subsection \ref{subsec: AcqFunGamma}. 

In what follows, the posterior moments of the LGP $\hat{p}_{j,N}\left( \mathcal{D} _{\mathrm{obs}}|\boldsymbol{\theta } \right) $ for describing the intermediate likelihood $p_j\left( \mathcal{D} _{\mathrm{obs}}|\boldsymbol{\theta } \right) $,  and those of the LGP $\hat{p}_{j,N}\left( \mathcal{D} _{\mathrm{obs}}|\boldsymbol{\theta } \right) p\left( \boldsymbol{\theta } \right) $ for approximating the unnormalized intermediate posterior $p_j\left( \mathcal{D} _{\mathrm{obs}}|\boldsymbol{\theta } \right) p\left( \boldsymbol{\theta } \right) $, will be frequently utilized.  To make a clear discrimination, the posterior mean, variance and covariance of the likelihood $\hat{p}_{j,N}\left( \mathcal{D} _{\mathrm{obs}}|\boldsymbol{\theta } \right) $, at the $j$-th tempering stage, are denoted as $\mu _{\mathrm{like},j,N}\left( \boldsymbol{\theta } \right) $, $\sigma _{\mathrm{like},j,N}^{2}\left( \boldsymbol{\theta } \right) $ and $c_{\mathrm{like},j,N}\left( \boldsymbol{\theta },\boldsymbol{\theta }^{\prime} \right) $, respectively, and are formulated by:
\begin{subequations}\label{eq:PostMomentsLike_jStage}
	\begin{equation}\label{eq:PostMeanLike_jStage}
		\mu _{\mathrm{like},j,N}\left( \boldsymbol{\theta } \right) =\exp \left( \gamma _j\mu _{g,N}+\frac{\gamma _{j}^{2}\sigma _{g,N}^{2}}{2} \right), 
	\end{equation}
	\begin{equation}\label{eq:PostVarLike_jStage}
		\sigma _{\mathrm{like},j,N}^{2}\left( \boldsymbol{\theta } \right) =\left( \exp \left( \gamma _{j}^{2}\sigma _{g,N}^{2}\left( \boldsymbol{\theta } \right) \right) -1 \right) \mu _{\mathrm{like},j,N}^{2}\left( \boldsymbol{\theta } \right),  
	\end{equation}
	and
	\begin{equation}\label{eq:PostCOVLike_jStage}
		c_{\mathrm{like},j,N}\left( \boldsymbol{\theta },\boldsymbol{\theta }^{\prime} \right) =\mu _{\mathrm{like},j,N}\left( \boldsymbol{\theta } \right) \left( \exp \left( \gamma _{j}^{2}c_{g,N}\left( \boldsymbol{\theta },\boldsymbol{\theta }^{\prime} \right) \right) -1 \right) \mu _{\mathrm{like},j,N}\left( \boldsymbol{\theta }^{\prime} \right) 
	\end{equation}    
\end{subequations}
, which are generated following Eq. \eqref{eq:PostLikelihood}. Our experience show that, in the early training stage where $\sigma_{g,N}^2$ is high, the mean estimate of Eq. \eqref{eq:PostMeanLike_jStage} as well as its resultant estimate of model evidence may exhibit high bias, which may mislead the active learning. To alleviate this, it is suggested to use the biased estimate $\mu _{\mathrm{like},j,N}\left( \cdot \right) =\exp \left( \gamma _j\mu _{g,N}(\cdot) \right)$  for estimating both posterior mean of likelihood and model evidence. Further, the posterior mean, variance and covariance of the posterior $\hat{p}_{j,N}\left( \mathcal{D} _{\mathrm{obs}}|\boldsymbol{\theta } \right) p\left( \boldsymbol{\theta } \right) $ are respectively indicated by $\mu _{\mathrm{post},j,N}\left( \boldsymbol{\theta } \right) $, $\sigma _{\mathrm{post},j,N}^{2}\left( \boldsymbol{\theta } \right) $ and $c_{\mathrm{post},j,N}\left( \boldsymbol{\theta },\boldsymbol{\theta }^{\prime} \right) $, of which the closed-form expressions can be easily obtained.

\subsection{Estimation of tempering parameter} \label{subsec: SpecifGamma}
While inferring for the $j$-th stage, it can be concluded that both $Z_{j-1}$ and $p_{j-1}\left( \mathcal{D} _{\mathrm{obs}}|\boldsymbol{\theta } \right) p\left( \boldsymbol{\theta } \right) $ have been estimated with desired accuracy, and are fixed at their mean estimates $\mu _{Z_{j-1}}$ and  $\mu _{\mathrm{post},j-1}\left( \boldsymbol{\theta } \right) $ respectively. One notes that, the ``$N$'' is dropped out from the subscripts of these estimators to indicate that they are fixed, and are no longer updated after the training for the $(j-1)$-th stages being completed.  Then, Eq. \eqref{eq:RatioDefinition} can be reformulated as:
\begin{equation}\label{eq:Ratio}
	\frac{Z_j}{\mu _{Z_{j-1}}}=\int_{\mathbb{T}}{\frac{p_j\left( \mathcal{D} _{\mathrm{obs}}|\boldsymbol{\theta } \right)}{\mu _{\mathrm{like},j-1}\left( \boldsymbol{\theta } \right)}\frac{\mu _{\mathrm{post},j-1}\left( \boldsymbol{\theta } \right)}{\mu _{Z_{j-1}}}\mathrm{d}\boldsymbol{\theta }}.
\end{equation}
Given $\gamma_{j-1}$ fixed, $\gamma_j$ can be specified by constraining the variation of the MC samples of the integrand of Eq. \eqref{eq:Ratio}, as suggested by the original TMCMC \cite{ching2007transitional} and TBQ \cite{wei2025bayesian}. Specifically, suppose a sample population $\mathcal{T} _{\mathrm{MC}}^{\left( j-1 \right)}=\left\{ \boldsymbol{\theta }_{j-1}^{\left( k \right)} \right\} _{k=1}^{N_{\mathrm{MC}}}$ following the distribution $\mu _{\mathrm{post},j-1}\left( \boldsymbol{\theta } \right) /\mu _{Z_{j-1}}$ has been readily generated (see subsection \ref{subsec: Sampling}), a weight $w^{(k)}(\gamma_j)$, depending on the value of $\gamma_{j}$, can be computed for each sample, as:

\begin{equation}\label{eq:weightSamp}
	w^{\left( k \right)}(\gamma_{j})=\frac{\mu _{\mathrm{like},j,N}\left( \boldsymbol{\theta }_{j-1}^{\left( k \right)} \right)}{\mu _{\mathrm{like},j-1}\left( \boldsymbol{\theta }_{j-1}^{\left( k \right)} \right)}
\end{equation} 
, where $\mu _{\mathrm{like},j,N}\left( \cdot \right) =\exp \left( \gamma _j\mu _{g,N}(\cdot) \right)$ takes the biased estimation. Intuitively, the variation of the estimator for the ratio $\mu _{Z_{j,N}}/\mu _{Z_{j-1}}$ can be controlled by constraining the variation of $w^{\left( k \right)}(\gamma_{j})$. Indeed, the CoV of the samples $w^{\left( k \right)}(\gamma_j)$ across $k=1,2,\cdots,N_\mathrm{MC}$, denoted as $\mathrm{CoV}\left( w^{\left( k \right)}\left( \gamma_{j} \right) \right)$, tends to increase with the value of $\gamma_{j}$. Thus, $\gamma _{j}$ can be specified by forcing the CoV of $w^{\left( k \right)}(\gamma_j)$ equal to a user-specified value $\varsigma$. In case $\mathrm{CoV}\left( w^{\left( k \right)}\left( 1 \right) \right)$ is smaller than $\varsigma$, $\gamma_j$ can be simply updated as 1. For practical implementation in this work, the Matlab function \textit{fminbnd} is utilized with default setting, which means to specify $\gamma_{j}$ by minimizing the difference between $\mathrm{CoV}\left( w^{\left( k \right)}\left( \gamma_{j} \right) \right)$ and $\varsigma$ within the support $(\gamma_{j-1},1]$,  using the algorithm combining golden section search and parabolic interpolation. This procedure is the same as that suggested in TMCMC \cite{ching2007transitional} and TBQ \cite{wei2025bayesian}. Specifically, $\varsigma$ is suggested to take value between 0.5 and 1.  

One notes that, following each update of the GP model, the value of the tempering parameter $\gamma_{j}$ needs to be updated accordingly by forcing the CoV of $w^{\left( k \right)}(\gamma_j)$ equal to or not exceed $\varsigma$, not only across tempering stages, but also within each specific stage, as suggested in Algorithm \ref{Alg:TBQ}. This active updating procedure is computationally negligible, but highly positive for precisely controlling the divergence between intermediate posteriors of two consecutive stages, and thus constructive in improving the robustness of the algorithm.

\subsection{Stopping Condition} \label{subsec: StopCond}
Following each update of $\gamma_{j}$, it is then required to judge whether the loop for the $j$-th tempering stage should be broken or not. This can be realized by checking the prediction uncertainty of the ratio $Z_{j}/Z_{j-1}$. Similar to Eq. \eqref{eq:PostEvidence}, the posterior mean of Eq. \eqref{eq:Ratio} can be formulated and estimated by:
\begin{equation}\label{eq:PostMeanRatio}
	\frac{\mu _{Z_j,N}}{\mu _{Z_{j-1}}}=\int_{\mathbb{T}}{\frac{\mu _{\mathrm{like},j,N}\left( \boldsymbol{\theta } \right)}{\mu _{\mathrm{like},j-1}\left( \boldsymbol{\theta } \right)}\frac{\mu _{\mathrm{post},j-1}\left( \boldsymbol{\theta } \right)}{\mu _{Z_{j-1}}}\mathrm{d}\boldsymbol{\theta }}\cong \frac{1}{N_{\mathrm{MC}}}\sum_{k=1}^{N_{\mathrm{MC}}}{\frac{\mu _{\mathrm{like},j,N}\left( \boldsymbol{\theta }_{j-1}^{\left( k \right)} \right)}{\mu _{\mathrm{like},j-1}\left( \boldsymbol{\theta }_{j-1}^{\left( k \right)} \right)}}
\end{equation}
where $\mu _{\mathrm{like},j,N}\left( \cdot \right) =\exp \left( \gamma _j\mu _{g,N}(\cdot) \right)$ takes the biased estimation and $\boldsymbol{\theta }_{j-1}^{\left( k \right)}\in \mathcal{T} _{\mathrm{MC}}^{\left( j-1 \right)}$, with $\mathcal{T}_\mathrm{MC}^{(j-1)}$ following the probability distribution $\mu _{\mathrm{post},j-1}\left( \boldsymbol{\theta } \right) /\mu _{Z_{j-1}}$. 

Further, the posterior variance of the ratio is formulated by:
\begin{equation}\label{eq:PostVarRatio}
	\frac{\sigma _{Z_j,N}^{2}}{\mu _{Z_{j-1}}^{2}}=\int_{\mathbb{T} \times \mathbb{T}}{\frac{c_{\mathrm{like},j,N}\left( \boldsymbol{\theta },\boldsymbol{\theta }^{\prime} \right)}{\mu _{\mathrm{like},j-1}\left( \boldsymbol{\theta } \right) \mu _{\mathrm{like},j-1}\left( \boldsymbol{\theta }^{\prime} \right)}\frac{\mu _{\mathrm{post},j-1}\left( \boldsymbol{\theta } \right) \mu _{\mathrm{post},j-1}\left( \boldsymbol{\theta }^{\prime} \right)}{\mu _{Z_{j-1}}^{2}}\mathrm{d}\boldsymbol{\theta }\mathrm{d}\boldsymbol{\theta }^{\prime}}
\end{equation}
, which is evaluated by the MC estimator formulated with the sample population $\mathcal{T}_\mathrm{MC}^{(j-1)}$ following $\mu _{\mathrm{post},j-1}\left( \boldsymbol{\theta } \right) /\mu _{Z_{j-1}}$ and $\mathcal{T}_\mathrm{MC}^{\prime (j-1)}$ following $\mu _{\mathrm{post},j-1}\left( \boldsymbol{\theta }^{\prime} \right) /\mu _{Z_{j-1}}$, i.e.,
\begin{equation}\label{eq:PostVarRatioMCestimate}
	\frac{\sigma _{Z_j,N}^{2}}{\mu _{Z_{j-1}}^{2}}\cong \frac{1}{N_{\mathrm{MC}}}\sum_{k=1}^{N_{\mathrm{MC}}}{\frac{c_{\mathrm{like},j,N}\left( \boldsymbol{\theta }_{j-1}^{\left( k \right)},\boldsymbol{\theta }_{j-1}^{\left( k \right) \prime} \right)}{\mu _{\mathrm{like},j-1}\left( \boldsymbol{\theta }_{j-1}^{\left( k \right)} \right) \mu _{\mathrm{like},j-1}\left( \boldsymbol{\theta }_{j-1}^{\left( k \right) \prime} \right)}}
\end{equation}
, where $\boldsymbol{\theta }_{j-1}^{\left( k \right) \prime}\in \mathcal{T} _{\mathrm{MC}}^{\prime \left( j-1 \right)}$, with $\mathcal{T} _{\mathrm{MC}}^{\prime \left( j-1 \right)}$ being generated by randomly permuting the rows of $\mathcal{T} _{\mathrm{MC}}^{\left( j-1 \right)}$, and $c_{\mathrm{like},j,N}(\cdot,\cdot)$ is formulated by Eq. \eqref{eq:PostCOVLike_jStage}.

Analogous to Eq. \eqref{eq:UppBoundVarZ}, an upper bound of the STD is formulated as:  
\begin{equation}\label{eq:PostVarBoundRatio}
	\frac{\bar{\sigma}_{Z_j,N}}{\mu _{Z_{j-1}}}=\int_{\mathbb{T}}{\frac{\sigma _{\mathrm{like},j,N}\left( \boldsymbol{\theta } \right)}{\mu _{\mathrm{like},j-1}\left( \boldsymbol{\theta } \right)}\frac{\mu _{\mathrm{post},j-1}\left( \boldsymbol{\theta } \right)}{\mu _{Z_{j-1}}}\mathrm{d}\boldsymbol{\theta }}\cong \frac{1}{N_{\mathrm{MC}}}\sum_{k=1}^{N_{\mathrm{MC}}}{\frac{\sigma _{\mathrm{like},j,N}\left( \boldsymbol{\theta }_{j-1}^{\left( k \right)} \right)}{\mu _{\mathrm{like},j-1}\left( \boldsymbol{\theta }_{j-1}^{\left( k \right)} \right)}}
\end{equation}
, and can be numerically estimated with $\mathcal{T}_\mathrm{MC}^{(j-1)}$, where $\sigma _{\mathrm{like},j,N}(\cdot)$ is formulated by Eq. \eqref{eq:PostVarLike_jStage}.

The CoV $\sigma _{Z_j,N}/\mu _{Z_j,N}$, which is computed by Eqs. \eqref{eq:PostVarRatioMCestimate} and \eqref{eq:PostMeanRatio}, or its upper bound $\bar{\sigma}_{Z_j,N}/\mu _{Z_j,N}$, which is estimated from Eqs. \eqref{eq:PostVarBoundRatio} and \eqref{eq:PostMeanRatio}, can then be served as stopping conditions, in analogy with Eq. \eqref{eq:StopCond}.  
\subsection{Acquisition functions} \label{subsec: AcqFunGamma}
Following the stopping condition for the $j$-th stage not satisfied, it is then required to specify a new training point to achieve a considerable reduction of the prediction uncertainties on the likelihood $p_j(\boldsymbol{\theta}|\mathcal{D}_\mathrm{obs})$ and the model evidence $Z_j$. This can be achieved by querying the sample population $\mathcal{T}_\mathrm{MC}^{(j-1)}$ or performing an optimization search within the support of the prior, both with an acquisition function. One note that there will be minor difference between the acquisition functions used for the above two schemes, as the candidates are different.  

As the target is to estimate $\exp \left( \gamma _{j}g\left( \boldsymbol{\theta } \right) \right)p(\boldsymbol{\theta}) $ and $Z_j$ with desired accuracy, the four acquisition functions are respectively modified as:
\begin{subequations}\label{eq:AcqFunModifeid}
	\begin{equation}\label{eq:PUQModifeid}
		\mathcal{A} _{\mathrm{PUQ},j}\left( \boldsymbol{\theta } \right) =\sigma _{\mathrm{like},j,N}\left( \boldsymbol{\theta } \right) p\left( \boldsymbol{\theta } \right) ,  
	\end{equation}
	\begin{equation}\label{eq:PVCModifeid}
		\mathcal{A} _{\mathrm{PVC},j}\left( \boldsymbol{\theta } \right) =p\left( \boldsymbol{\theta } \right) \int_{\mathbb{T}}{\frac{c_{\mathrm{like},j,N}\left( \boldsymbol{\theta },\boldsymbol{\theta }^{\prime} \right)}{\mu _{\mathrm{like},j-1}\left( \boldsymbol{\theta }^{\prime} \right)}\frac{\mu _{\mathrm{post},j-1}\left( \boldsymbol{\theta }^{\prime} \right)}{\mu _{Z_{j-1}}}\mathrm{d}\boldsymbol{\theta }^{\prime}},
	\end{equation}
	\begin{equation}\label{eq:PLURModifeid}
		\mathcal{A} _{\mathrm{PLUR},j}\left( \boldsymbol{\theta }^+ \right) =\int_{\mathbb{T}}{\frac{\mu _{\mathrm{like},j,N}^{2}\left( \boldsymbol{\theta } \right)}{\mu _{\mathrm{like},j-1}\left( \boldsymbol{\theta } \right)}\left( \exp \left( \frac{\gamma _{j}^{2}c_{g,N}^{2}\left( \boldsymbol{\theta }^+,\boldsymbol{\theta } \right)}{\sigma _{g,N}^{2}\left( \boldsymbol{\theta }^+ \right)} \right) -1 \right) \frac{\mu _{\mathrm{post},j-1}\left( \boldsymbol{\theta } \right)}{\mu _{Z_{j-1}}}\mathrm{d}\boldsymbol{\theta }}
	\end{equation}
, and 
	\begin{equation}\label{eq:PVRModifeid}
		\begin{split}
			\mathcal{A} _{\mathrm{PEUR},j}\left( \boldsymbol{\theta }^+ \right) =\int_{\mathbb{T} \times \mathbb{T}}&{\frac{\mu _{\mathrm{like},j,N}\left( \boldsymbol{\theta } \right) \mu _{\mathrm{like},j,N}\left( \boldsymbol{\theta }^{\prime} \right)}{\mu _{\mathrm{like},j-1}\left( \boldsymbol{\theta } \right) \mu _{\mathrm{like},j-1}\left( \boldsymbol{\theta }^{\prime} \right)}\left( \exp \left( \frac{\gamma _{j}^{2}c_{g,N}\left( \boldsymbol{\theta },\boldsymbol{\theta }^+ \right) c_{g,N}\left( \boldsymbol{\theta }^+,\boldsymbol{\theta }^{\prime} \right)}{\sigma _{g,N}^{2}\left( \boldsymbol{\theta }^+ \right)} \right) -1 \right)}
			\\
			&\times \frac{\mu _{\mathrm{post},j-1}\left( \boldsymbol{\theta } \right) \mu _{\mathrm{post},j-1}\left( \boldsymbol{\theta }^{\prime} \right)}{\mu _{Z_{j-1}}^{2}}\mathrm{d}\boldsymbol{\theta }\mathrm{d}\boldsymbol{\theta }^{\prime}.
		\end{split}
	\end{equation}
\end{subequations}
The above four acquisition functions are applied to search for the optimal training point from the candidate sample pool $\mathcal{T}_\mathrm{MC}^{(j-1)}$; whereas, to search from the prior support $\mathbb{T}$ using an optimization algorithm, the above acquisition functions can be weighted by $\mu_{\mathrm{post},j-1}(\boldsymbol{\theta})$. The mathematical derivations for the above acquisition function are similar to those presented in Table \ref{table:SummaryAcqFun}, and one can implement the derivation following the same procedures presented in  \ref{Append:Proof20} and \ref{Append:Proof22}. Analogous to those presented in Table \ref{table:SummaryAcqFun}, $\mathcal{A} _{\mathrm{PUQ},j}\left( \boldsymbol{\theta } \right)$ is formulated in a closed form; $\mathcal{A} _{\mathrm{PVC},j}\left( \boldsymbol{\theta } \right)$ and $\mathcal{A} _{\mathrm{PIVR},j}\left( \boldsymbol{\theta }^+ \right)$ are defined as integrals over the density $\mu _{\mathrm{post},j-1}\left( \boldsymbol{\theta } \right) /\mu _{Z_{j-1}}$, and thus can be computed by MC estimators formulated with  $\mathcal{T}^{ (j-1)} _{\mathrm{MC}}$; and $\mathcal{A} _{\mathrm{PEUR},j}\left( \boldsymbol{\theta }^+ \right)$ is defined by the integral over $\mu _{\mathrm{post},j-1}\left( \boldsymbol{\theta } \right) /\mu _{Z_{j-1}}$ and $\mu _{\mathrm{post},j-1}\left( \boldsymbol{\theta }^\prime \right) /\mu _{Z_{j-1}}$, and thus can be estimated by a MC estimator formulated with $\mathcal{T}^{ (j-1)} _{\mathrm{MC}}$ and $\mathcal{T}^{\prime (j-1)} _{\mathrm{MC}}$. 

\subsection{Sampling scheme}\label{subsec: Sampling}
Following the stopping condition for the $j$-th stage being satisfied, the mean estimate of the unnormalized posterior $p_j\left( \mathcal{D} _{\mathrm{obs}}|\boldsymbol{\theta } \right) p\left( \boldsymbol{\theta } \right) $ will be fixed at $\mu _{\mathrm{post},j}\left( \boldsymbol{\theta } \right) =\mu _{\mathrm{like},j,N}\left( \boldsymbol{\theta } \right) p\left( \boldsymbol{\theta } \right) $. It is then required to generate a sample population $\mathcal{T} _{\mathrm{MC}}^{\left( j \right)}$ of size $N_\mathrm{MC}$, following $\mu _{\mathrm{post},j}\left( \boldsymbol{\theta } \right) $ to proceed with the training for the next stage, and for estimating $Z_j$ (see subsection \ref{subsec: PredUQ}). For this sub-task, a resampling-then-Metropolis-Hastings (R-MH) sampling have been recommended in the previous work \cite{wei2025bayesian} and the original TMCMC algorithm \cite{ching2007transitional}, which is also adopted in this work. It is based on resampling from $\mathcal{T}_\mathrm{MC}^{(j-1)}$ with replacement and then with each generated sample as a seed to grow a Markov chain of equal length using the MH algorithm, this way to collect the final state of each chain as the sample population $\mathcal{T}_\mathrm{MC}^{(j)}$. One can refer to Algorithm 2 of Ref. \cite{wei2025bayesian} for details. 

\subsection{Predictions and their uncertainty quantification} \label{subsec: PredUQ}
After the stopping condition being satisfied for the $j$-th tempering stage, it is required to estimate the corresponding model evidence $Z_{j}$ and the unnormalized posterior $p\left( \mathcal{D} _{\mathrm{obs}}|\boldsymbol{\theta } \right) p\left( \boldsymbol{\theta } \right) $, and also to quantify the prediction uncertainties. For the unnormalized posterior, its mean and STD can be trivially computed by multiplying $\mu _{\mathrm{like},j}\left( \boldsymbol{\theta } \right) $ generated by Eq. \eqref{eq:PostMeanLike_jStage} and  $\sigma _{\mathrm{like},j}\left( \boldsymbol{\theta } \right) $ generated by Eq. \eqref{eq:PostVarLike_jStage}, respectively, with $p\left( \boldsymbol{\theta } \right)$. For $Z_j/\mu _{Z_{j-1}}$, its posterior mean can be computed by the crude MC estimator (mean of integrand samples) of Eq. \eqref{eq:PostMeanRatio} formulated with $\mathcal{T} _{\mathrm{MC}}^{\left( j-1 \right)}$, and its posterior variance, resulted from the variation of the GP prediction, can be computed by the crude MC estimator of Eq. \eqref{eq:PostVarRatio} formulated with $\mathcal{T} _{\mathrm{MC}}^{\left( j-1 \right)}$ and $\mathcal{T} _{\mathrm{MC}}^{\prime \left( j-1 \right)}$. For the mean estimate of $\mu _{Z_j}/\mu _{Z_{j-1}}$, a more robust estimator can be formulated with the bridging scheme, as suggested in Refs. \cite{bennett1976efficient, katafygiotis2007estimation}. Specifically, in case $\mu _{\mathrm{like},j}\left( \boldsymbol{\theta } \right) $ and $\mu _{\mathrm{like},j-1}\left( \boldsymbol{\theta } \right) $ show high divergence, an artificial likelihood, for bridging these two, can be defined, for example, as their geometric average \cite{katafygiotis2007estimation}:
\begin{equation}\label{eq:BridgingLike}
	\mu _{\mathrm{like},j-1/2}\left( \boldsymbol{\theta } \right) =\sqrt{\mu _{\mathrm{like},j}\left( \boldsymbol{\theta } \right) \mu _{\mathrm{like},j-1}\left( \boldsymbol{\theta } \right)}.
\end{equation}   
Given the normalizing constant $\mu _{Z_{j-1/2}}=\int_{\mathbb{T}}{\mu _{\mathrm{like},j-1/2}\left( \boldsymbol{\theta } \right) p\left( \boldsymbol{\theta } \right)}\mathrm{d}\boldsymbol{\theta }$, Eq. \eqref{eq:PostMeanRatio} can be equivalently formulated and estimated as:
\begin{equation}\label{eq:bridgeestimate}
	\frac{\mu _{Z_j}}{\mu _{Z_{j-1}}}=\frac{\mu _{Z_j}/\mu _{Z_{j-1/2}}}{\mu _{Z_{j-1}}/\mu _{Z_{j-1/2}}}=\frac{\int_{\mathbb{T}}{\frac{\mu _{\mathrm{like},j-1/2}\left( \boldsymbol{\theta } \right)}{\mu _{\mathrm{like},j-1}\left( \boldsymbol{\theta } \right)}\frac{\mu _{\mathrm{post},j-1}\left( \boldsymbol{\theta } \right)}{\mu _{Z_{j-1}}}\mathrm{d}\boldsymbol{\theta }}}{\int_{\mathbb{T}}{\frac{\mu _{\mathrm{like},j-1/2}\left( \boldsymbol{\theta } \right)}{\mu _{\mathrm{like},j}\left( \boldsymbol{\theta } \right)}\frac{\mu _{\mathrm{post},j}\left( \boldsymbol{\theta } \right)}{\mu _{Z_j}}\mathrm{d}\boldsymbol{\theta }}}\approx \frac{\sum\nolimits_{k=1}^{N_{\mathrm{MC}}}{\frac{\mu _{\mathrm{like},j-1/2}\left( \boldsymbol{\theta }_{j-1}^{\left( k \right)} \right)}{\mu _{\mathrm{like},j-1}\left( \boldsymbol{\theta }_{j-1}^{\left( k \right)} \right)}}}{\sum\nolimits_{k=1}^{N_{\mathrm{MC}}}{\frac{\mu _{\mathrm{like},j-1/2}\left( \boldsymbol{\theta }_{j}^{\left( k \right)} \right)}{\mu _{\mathrm{like},j}\left( \boldsymbol{\theta }_{j}^{\left( k \right)} \right)}}}.
\end{equation}
One notes that the above estimator is formulated with both sample populations $\mathcal{T} _{\mathrm{MC}}^{\left( j-1 \right)}$ and $\mathcal{T} _{\mathrm{MC}}^{\left( j \right)}$, and thus can only be implemented following the generation of the sample population $\mathcal{T} _{\mathrm{MC}}^{\left( j \right)}$ from $\mu _{\mathrm{post},j}\left( \boldsymbol{\theta } \right) $. The posterior variance given by Eq. \eqref{eq:PostVarRatio} can also be reformulated with a bridging scheme, but it is not encouraged as it is not worth the cost. The one formulated with Eq. \eqref{eq:PostVarRatio} is sufficient for indicating the prediction uncertainty.

Following the convergence of the stage with $\gamma_j=1$ being reached, let $M=j$, then the unnormalized target posterior $p\left( \mathcal{D} _{\mathrm{obs}}|\boldsymbol{\theta } \right) p\left( \boldsymbol{\theta } \right) $ is readily estimated with the mean estimate $\mu _{\mathrm{post},M}\left( \boldsymbol{\theta } \right) $, and the prediction variance is also readily quantified by $\sigma _{\mathrm{post},M}\left( \boldsymbol{\theta } \right) $. The mean estimate of $Z$ is ultimately computed by the products of ratios as:
 \begin{equation}\label{eq:MeanEstimateEvidence}
 	\mu _Z=\mu _{Z_M}=\prod_{j=2}^M{\frac{\mu _{Z_j}}{\mu _{Z_{j-1}}}}.
 \end{equation} 
Note that the variation of all MC estimators are constrained by the parameter $\varsigma$, and thus for simplicity, the variations of the estimators caused by the randomness of the sample population, are neglected. In this case, the $\mu _{Z_j}$ is regarded as the precise estimation of the normalizing constant defined by $\mu _{Z_j}=\int_{\mathbb{T}}{\mu _{\mathrm{like},j}\left( \boldsymbol{\theta } \right) p\left( \boldsymbol{\theta } \right)}\mathrm{d}\boldsymbol{\theta }$ for $j=2,3,\cdots,M$. Thus, the variation of the estimator $\mu _Z$, solely caused by the prediction uncertainty of the GP model, is uniquely reflected by the variation of the last stage. Further, the posterior CoV of the estimate $\mu _Z$, denoted as $\mathcal{C} _Z$, can be evaluated by:
\begin{equation}
	\mathcal{C} _Z=\frac{\frac{\sigma _{Z_M}}{\mu _{Z_{M-1}}}\prod\nolimits_{j=2}^{M-1}{\frac{\mu _{Z_j}}{\mu _{Z_{j-1}}}}}{\prod\nolimits_{j=2}^M{\frac{\mu _{Z_j}}{\mu _{Z_{j-1}}}}}=\frac{\sigma _{Z_M}/\mu _{Z_{M-1}}}{\mu _{Z_M}/\mu _{Z_{M-1}}}
\end{equation}    
, which is exactly the posterior CoV of the evidence ratio of the last stage.   

\section{Numerical Experiments and Engineering Applications}
In this section, a number of numerical benchmarks and engineering examples will be addressed with the BQ and/or TBQ algorithms equipped each of the four acquisition functions. The efficiency of each implementation is indicated using the total number of likelihood function calls, and the accuracy is demonstrated by comparing with the ground truths and/or the reference results generated by TMCMC. All computations are implemented on a personal laptop with an 11-th Gen Intel Core i7-11800H (2.30 GHz) CPU, 32 GB RAM memory and a NVIDIA RTX A2000 GPU with 4 GB memory. 

\subsection{Four two-dimensional benchmark examples}\label{subsec:four2Dexamples}
Four two-dimensional (2D) examples with increasing complexities of posteriors are first used for demonstrating and illustrating the proposed methods, all of which are adapted from Ref. \cite{rezende2015variational}. The energy functions are formulated as:
\begin{subequations}\label{eq:EnergyFunExample1}
\begin{equation}
		\mathcal{U} _1\left( \boldsymbol{\theta } \right) =\frac{1}{2}\left( \frac{\left\| \boldsymbol{\theta } \right\| -2}{0.4} \right) ^2-\log \left( \exp \left( -\frac{1}{2}\left( \frac{\theta _1-2}{0.6} \right) ^2 \right) +\exp \left( -\frac{1}{2}\left( \frac{\theta _1+2}{0.6} \right) ^2 \right) \right) 
\end{equation}
\begin{equation}
	\mathcal{U} _2\left( \boldsymbol{\theta } \right) =\frac{1}{2}\left( \frac{\theta _2+\omega _1\left( \theta _1 \right)}{0.4} \right) ^2
\end{equation}
\begin{equation}
	\mathcal{U} _3\left( \boldsymbol{\theta } \right) =-\log \left( \exp \left( -\frac{1}{2}\left( \frac{\theta _2+\omega _1\left( \theta _1 \right)}{0.35} \right) ^2 \right) +\exp \left( -\frac{1}{2}\left( \frac{\theta _2+\omega _1\left( \theta _1 \right) -\omega _2\left( \theta _1 \right)}{0.35} \right) ^2 \right) \right) 
\end{equation}
\begin{equation}
	\mathcal{U} _4\left( \boldsymbol{\theta } \right) =-\log \left( \exp \left( -\frac{1}{2}\left( \frac{\theta _2+\omega _1\left( \theta _1 \right)}{0.4} \right) ^2 \right) +\exp \left( -\frac{1}{2}\left( \frac{\theta _2+\omega _1\left( \theta _1 \right) -\omega _3\left( \theta _1 \right)}{0.35} \right) ^2 \right) \right) 
\end{equation}
\end{subequations}
, where $\omega _1\left( \theta \right) =\sin \left( 0.5\pi \theta \right) $, $\omega _2\left( \theta \right) =3\exp \left( -0.5\left( \theta -1 \right) ^2/0.6^2 \right) $ and $\omega _3\left( \theta \right) =3/\left( 1+\exp \left( -\left( \theta -1 \right) /0.2 \right) \right) $. The prior distribution of $\theta_1$ and $\theta_2$ are assumed to be uniform with support [-4, 4].  

\begin{figure}[H]
	\centering
	\setlength{\abovecaptionskip}{-0.2cm}
	\includegraphics[scale=0.65]{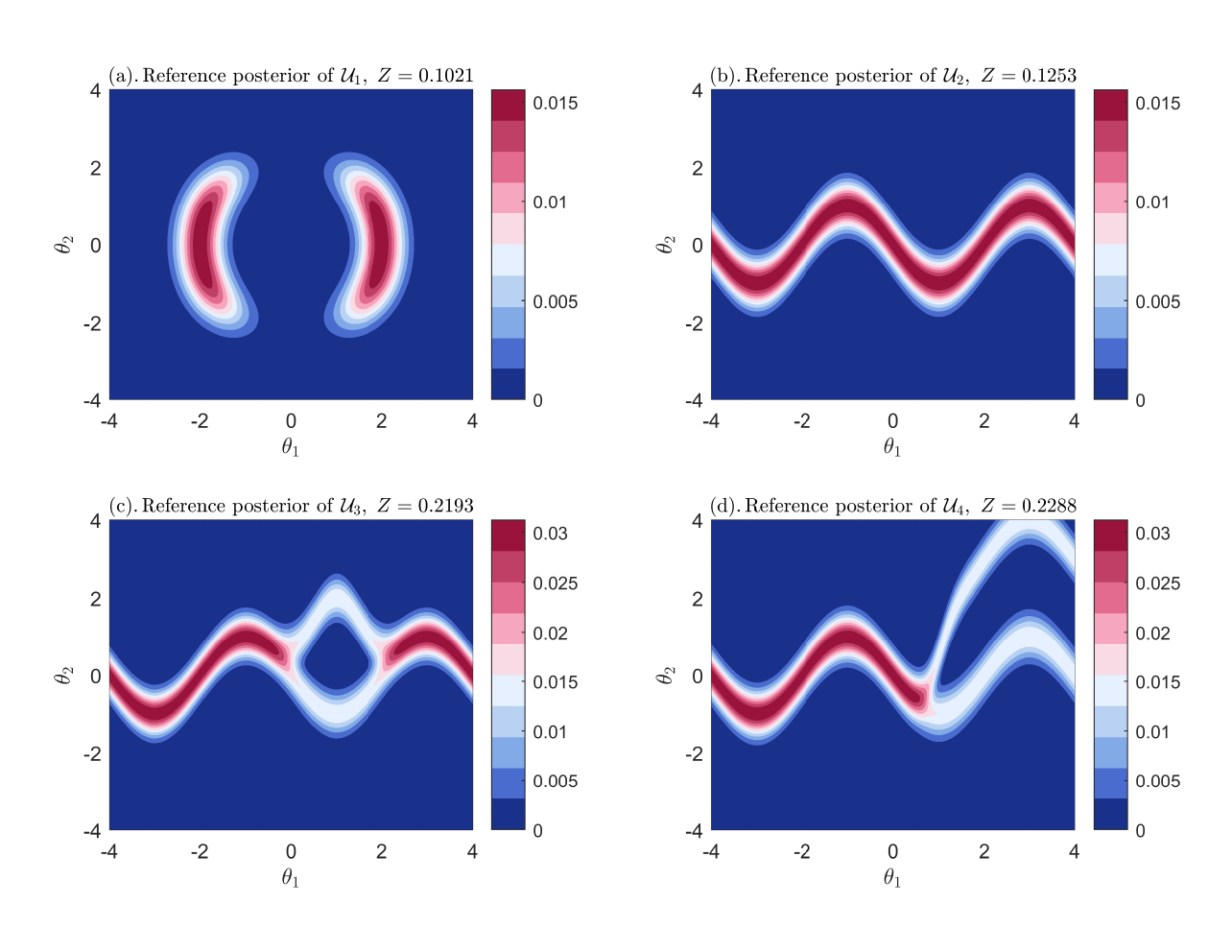}
	\caption{\centering{Reference results of unnormalized posteriors and the associated model evidences of the four 2D examples .}}
	\label{fig:RefPostU}
\end{figure}

The true unnormalized posteriors of these four 2D problems are schematically shown in Figure \ref{fig:RefPostU}, with the corresponding true values of the model evidence given in the respective titles. It is seen that posterior defined by $\mathcal{U}_1$ shows two disconnected modes, and that of $\mathcal{U}_2$ shows nonlinear dependence.  The posteriors defined by $\mathcal{U}_3$ and $\mathcal{U}_4$ show nonlinear dependencies and bifurcating behavior, with each branching mode being weaker than the main mode, making it difficult to be accurately captured. It should be noted that, for all implementations of the BQ and TBQ algorithms in this example, each new training point is searched by maximizing the acquisition function using the Matlab function ``ga'' with the algorithm parameters `UseParallel' as true, `UseVectorized' as true, `PopulationSize' as 50, and all the others parameters as default values. 

\begin{figure}[H]
	\centering
	\setlength{\abovecaptionskip}{-0.2cm}
	\includegraphics[scale=0.65]{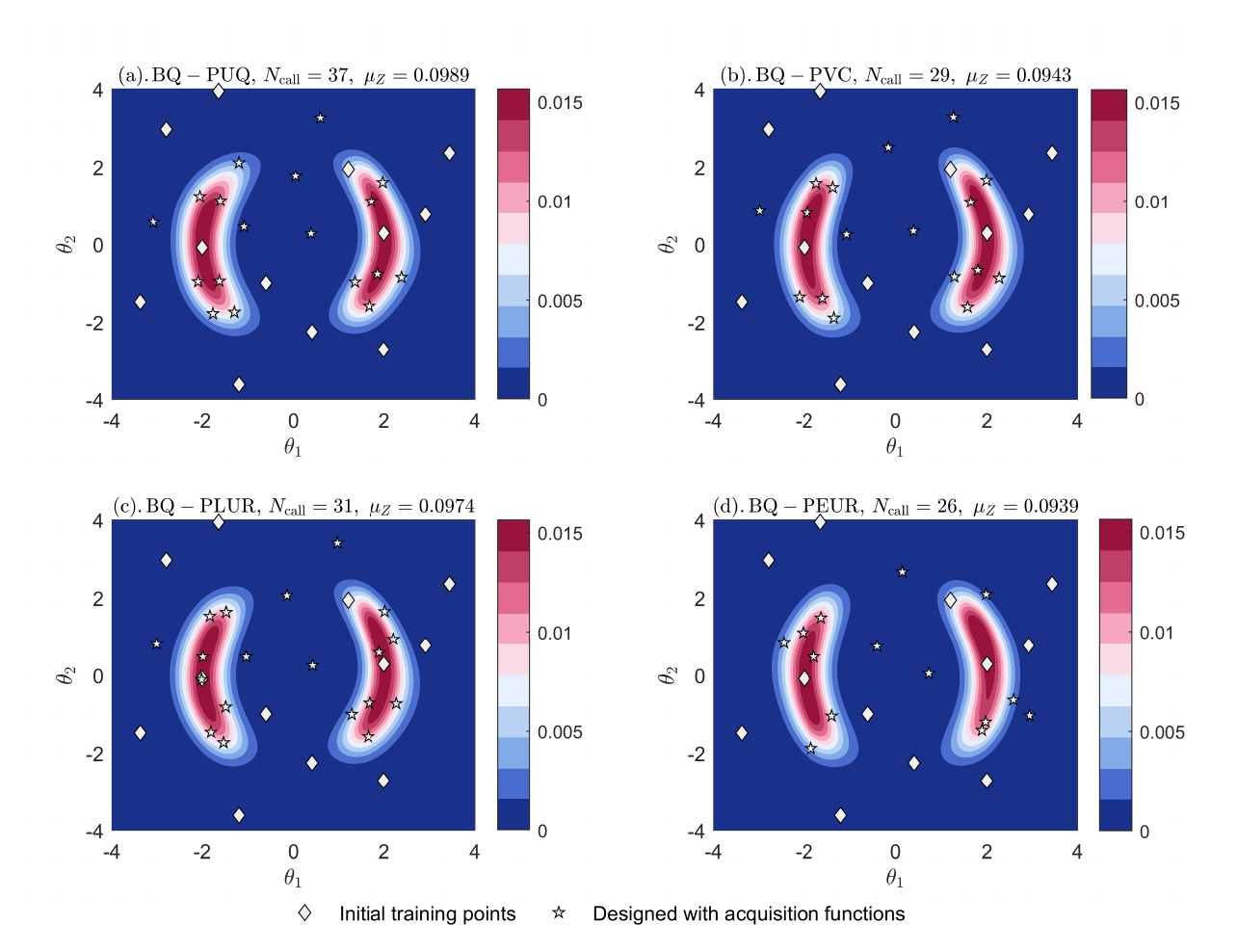}
	\caption{\centering{Results of the posteriors defined by $\mathcal{U}_1$, which are estimated by BQ algorithm driven by the four acquisition functions, together with the training points and the mean estimate of model evidence.}}
	\label{fig:BQU1}
\end{figure}

\begin{figure}[H]
	\centering
	\setlength{\abovecaptionskip}{-0.2cm}
	\includegraphics[scale=0.65]{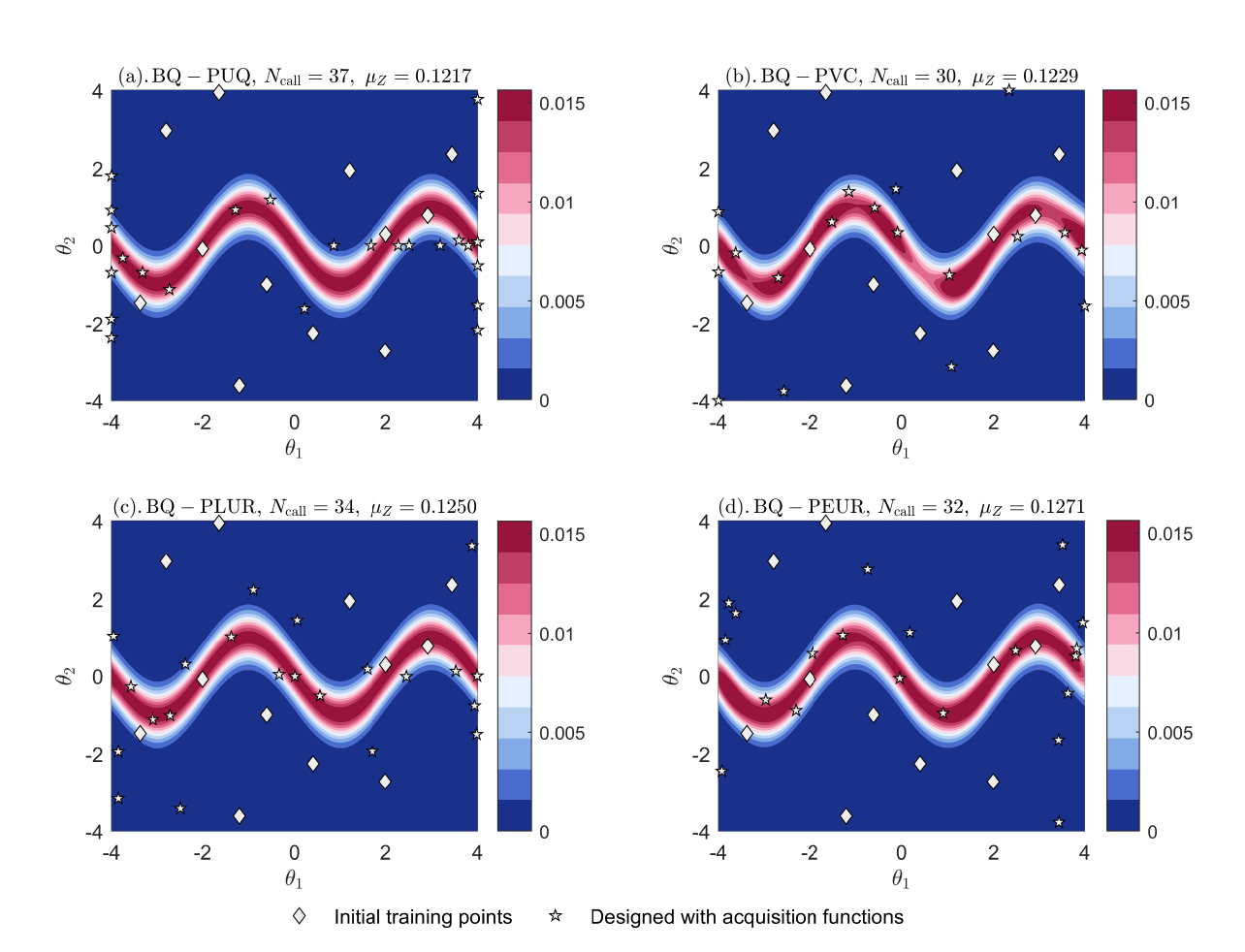}
	\caption{\centering{Results of the posteriors defined by $\mathcal{U}_2$, generated with the BQ algorithm.}}
	\label{fig:BQU2}
\end{figure}

\begin{figure}[H]
	\centering
	\setlength{\abovecaptionskip}{-0.2cm}
	\includegraphics[scale=0.6]{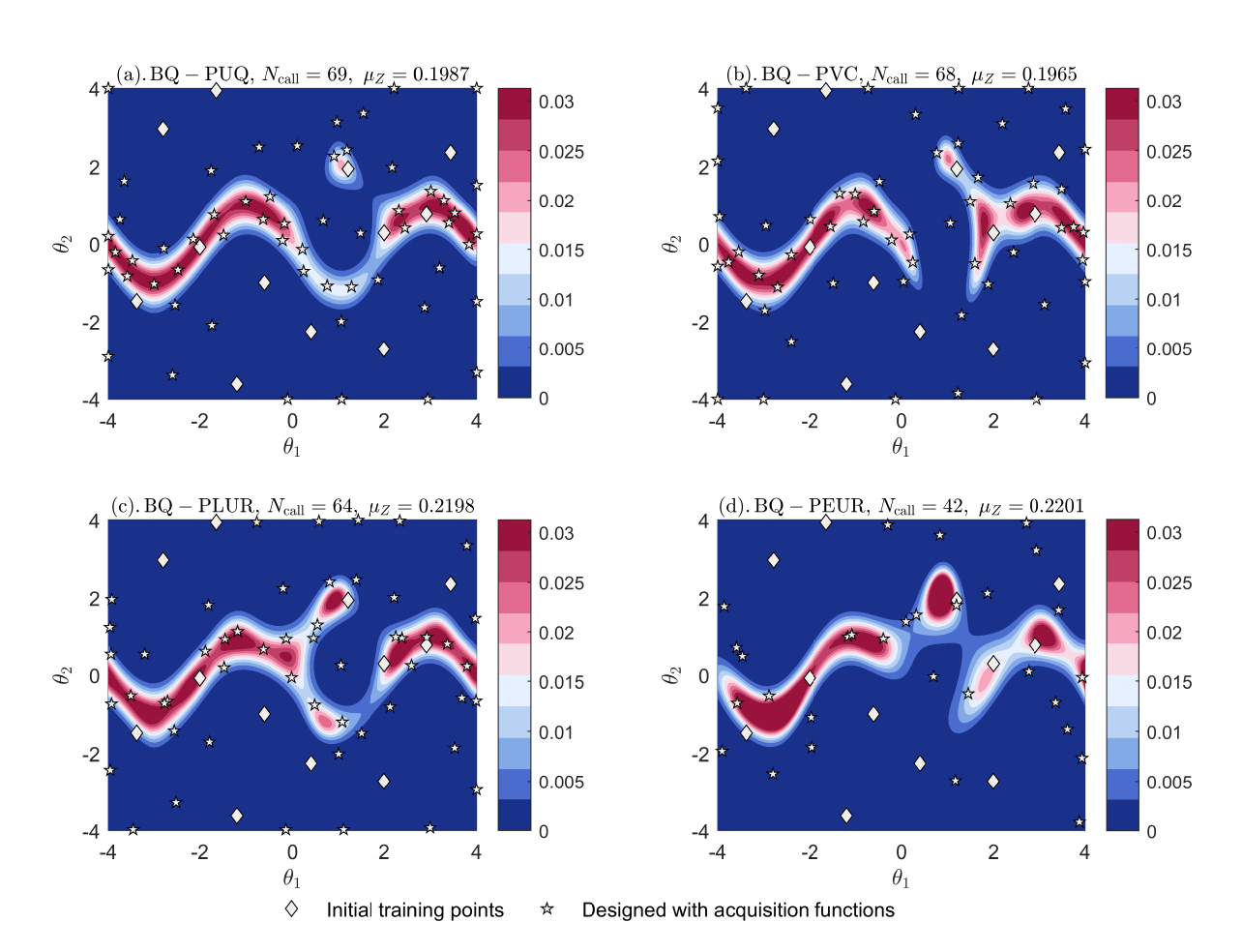}
	\caption{\centering{Results of the posteriors defined by $\mathcal{U}_3$, generated with the BQ algorithm.}}
	\label{fig:BQU3}
\end{figure}

\begin{figure}[H]
	\centering
	\setlength{\abovecaptionskip}{-0.2cm}
	\includegraphics[scale=0.6]{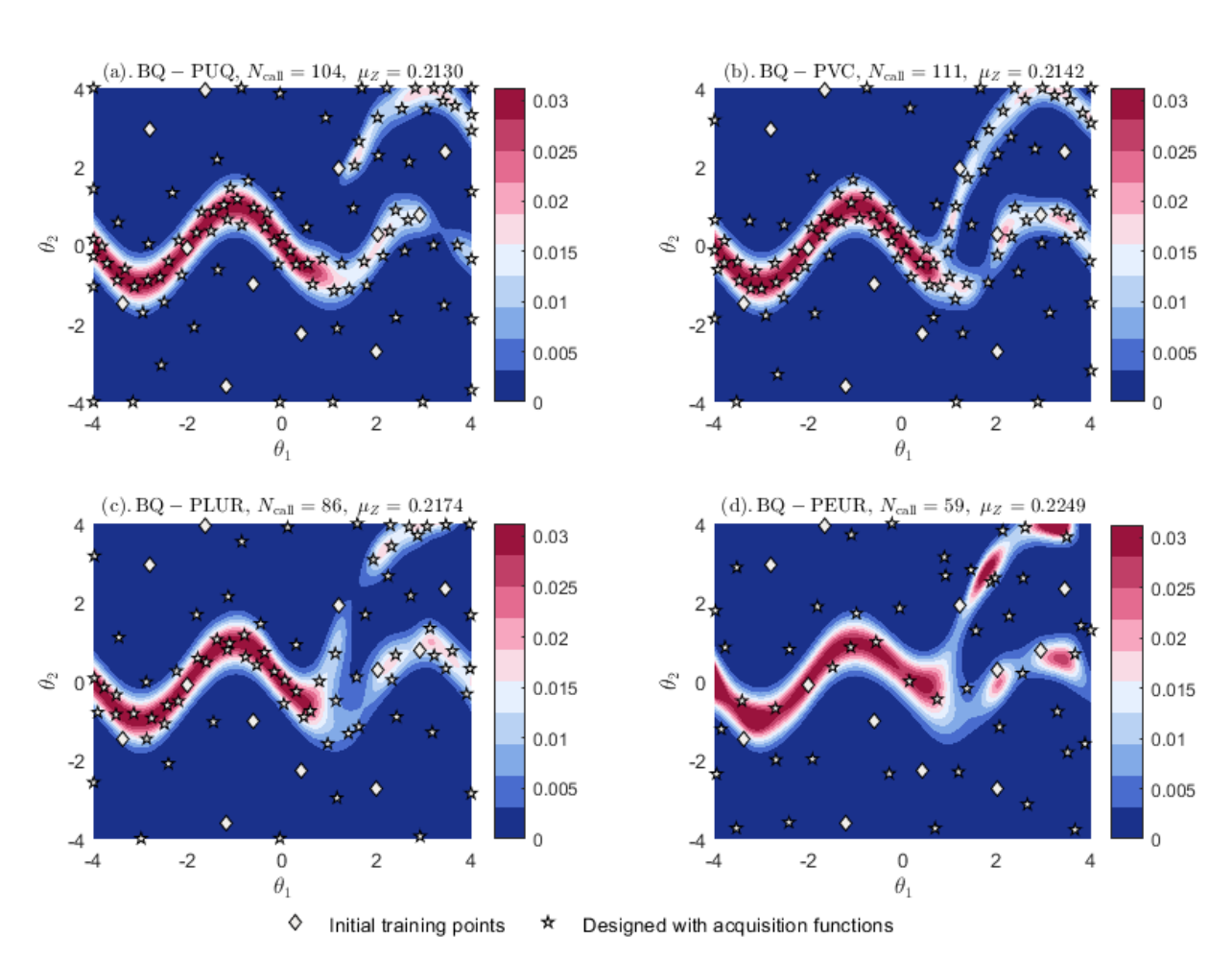}
	\caption{\centering{Results of the posteriors of the forth 2D example, generated with the BQ algorithm.}}
	\label{fig:BQU4}
\end{figure}

Based on the same set of $N_0=12$ initial training points, the BQ algorithm is implemented with all the four acquisition functions for all the four 2D examples, with $\epsilon$ set to be 0.04 for the first and second examples, and 0.02 for the remaining two examples to capture the less important modes. $N_\mathrm{MC}$ is set to be $5\times 10^3$ for all implementations. For the first three examples, the squared exponential kernel is used for training the GP model, while for the last one, the $\frac{5}{2}$-Mat$\acute{\mathrm{e}}$rn kernel is utilized. These parameter settings also apply to the TBQ algorithm. The four unnormalized posteriors estimated by BQ equipped with the four acquisition functions are then reported in Figure \ref{fig:BQU1}-\ref{fig:BQU4} respectively, accompanied with numbers of model calls and the mean estimates of model evidence in the titles. 

For the first example defined with $\mathcal{U}_1$, as shown by Figure \ref{fig:BQU1}, initialized with the same 12 points generated by LHS, all the four acquisition functions produce accurate estimate of both posterior and model evidence. The PUQ function consumes the most energy function calls, followed by PLUR, PVC and then PEUR. This demonstrates the high efficiency and accuracy of the BQ algorithm, equipped with any one of the four proposed acquisition functions, for predicting the multi-modalities of posteriors.         

Results in Figure \ref{fig:BQU2} indicate that, the BQ algorithm equipped with any of the four acquisition functions, has evaluated both the model evidence and posterior, with high accuracy, for the problem defined by $\mathcal{U}_2$. With the same 12 initial training samples and under the same stopping criterion, PUQ requires the most model evaluations (37), followed by PLUR (34), PEUR (32), and PVC (30) in descending order. These results also demonstrate that, regardless of the acquisition function employed, the proposed BQ algorithm can accurately capture the nonlinear dependencies within the posteriors.   

Next, results for the problem formulated with $\mathcal{U}_3$, as reported in Figure \ref{fig:BQU3}, are discussed. As shown, based on the same 12 initial training points, all four acquisition functions successfully and accurately predicted the posteriors in the main modal regions. For the bifurcated secondary modal regions, none of the acquisition functions produced perfect predictions, but they all essentially captured the behavior of these secondary modes. This phenomenon can be explained as follows. The posterior in main modal region exhibits the highest peak values, and its probability mass contributes the most, through the integral over the prior, to the model evidence, and thus, the BQ algorithm inherently tend to focusing on the primary characteristics. When a more stringent convergence criterion is used, such as defining convergence as the maximum STD of the posterior not exceeding a pre-defined threshold, the estimation of the secondary modal regions can be improved, but requires more model calls. This trade-off between accuracy and computational cost remains a decision for users to balance based on their specific purposes. Regardless, all four acquisition functions, especially PLUR and PEUR, successfully predicted the model evidence with acceptable accuracy. As can also be seen from Figure \ref{fig:BQU3}, compared to the first two examples, this example requires more model evaluations, although the convergence threshold is only half of that in the first two examples. This is the additional price for capturing the behavior of secondary modal regions.     

As observed in Figure \ref{fig:BQU4}, a similar phenomenon appears, i.e., to capture the two bifurcated modal regions, each acquisition function required more model evaluations than in the previous three cases. It is also seen that, for this example, the PVC function perfectly captured the behavior of the posterior throughout the support of the prior, but it also consumed the most energy function calls, which is 111. By contrast, the PEUR function required only 59 model calls to capture the behavior of all modal regions with acceptable accuracy. Still, all four acquisition functions produced accurate estimates of model evidence. Thus, the cost of the four acquisition function can be quite different from each other, and the specific choice in practical use depends on user's preference. For example, if the accuracy of model evidence is more preferable, then the PEUR function may be the best choice as it consumes the least energy function calls to reach the same level of accuracy for model evidence.  

To verify the robustness of the BQ algorithm, e.g., with respect to the initial design of training points, it is implemented for ten times for each example, driven by any one of the four acquisition functions. Each implementation employs LHS to randomly generate 12 sample points for initializing the algorithm. The algorithms parameters are exactly the same as the previous settings for each example. Results obtained under the above settings are summarized in Table \ref{table:Result2D_BQ}, where the mean estimates $\mu_Z$, the number of model calls $N_\mathrm{call}$, and the absolute errors to the reference solutions, represent the average of the ten repeated runs, and the CoVs are calculated across the ten mean estimates.  From the results of the first example, it is seen that PUQ, PVC and PLUR consumed a comparable number of model evaluations, thus achieving comparable computational efficiency; while PEUR required less number of function calls. In terms of absolute errors, PVC, PLUR and PEUR produced estimates of comparable accuracy, and is of higher accuracy than PUQ; while with respect to CoVs, PUQ, PVC and PLUR yielded estimates with comparable variation, and is of lower variation than PEUR. One can also analyze the results of the other three examples from the perspectives of efficiency and accuracy respectively, and the conclusions are not present for simplicity. Overall, across all four examples, to achieve the same convergence criteria, PEUR consistently consumed the fewest model evaluations to produce estimates with comparable accuracy.

\newcolumntype{C}{>{\centering\arraybackslash}m{2cm}}
\newcolumntype{D}{>{\raggedright\arraybackslash}m{2.5cm}}
\newcolumntype{E}{>{\centering\arraybackslash}m{1.6cm}}
\newcolumntype{F}{>{\centering\arraybackslash}m{1.6cm}}
\newcolumntype{O}{>{\centering\arraybackslash}m{1.6cm}}
\newcolumntype{G}{>{\centering\arraybackslash}m{2.4cm}}
\newcolumntype{I}{>{\centering\arraybackslash}m{1.6cm}}

\begin{table}[H]
	\caption{Summary and comparison of results of model evidence of the four 2D problems, generated by BQ algorithms driven by the four acquisition functions, where $N_\mathrm{call}$, $\mu_Z$, relative errors and CoVs are computed with the results of ten implementations. }
	\label{table:Result2D_BQ}
	\centering
	\begin{threeparttable}
		\begin{tabular}{C D E F O G I}
			\hline 
			Examples& Methods  & $N_\mathrm{call}$  & $\mu_Z$&$Z$ (Ref.) &  Relative errors    & CoVs \\ 
			\hline
			\multirow{4}{*}{1st}&BQ-PUQ& 31.7  &0.0964&\multirow{4}{*}{0.1021} &0.0608&0.0529\\
			&BQ-PVC&30.6 &0.0980& &0.0455&0.0515\\
			&BQ-PLUR&32.5  &0.0988& &0.0438&0.0473\\
			&BQ-PEUR&25.8   &0.1014& &0.0456&0.0624\\
			\hline
			\multirow{4}{*}{2nd}&BQ-PUQ& 30.7  &0.1260&\multirow{4}{*}{0.1253} &0.0265&0.0443\\
			&BQ-PVC&31.4 &0.1251& &0.0166&0.0214\\
			&BQ-PLUR&31.5  &0.1266& &0.0330&0.0369 \\
			&BQ-PEUR&29.1   &0.1250& &0.0577&0.0727\\
			\hline
			\multirow{4}{*}{3rd}&BQ-PUQ& 58.6  &0.2004&\multirow{4}{*}{0.2193} &0.0863&0.0483\\
			&BQ-PVC&54.7 &0.1978& &0.0981&0.0542\\
			&BQ-PLUR&61.7  &0.2064& &0.0703&0.0620 \\
			&BQ-PEUR&43.1   &0.2088& &0.0531&0.0527\\
			\hline
			\multirow{4}{*}{4th}&BQ-PUQ& 94.0  &0.2126&\multirow{4}{*}{0.2288} &0.0742&0.0444\\
			&BQ-PVC&85.8 &0.2071& &0.0947&0.0507\\
			&BQ-PLUR&92.7  &0.2118& &0.0741&0.0478\\
			&BQ-PEUR&53.2   &0.2268& &0.0535&0.0640\\
			\hline
		\end{tabular}
	\end{threeparttable}
\end{table}   

\begin{figure}[H]
	\centering
	\setlength{\abovecaptionskip}{-0.8cm}
	\includegraphics[scale=0.75]{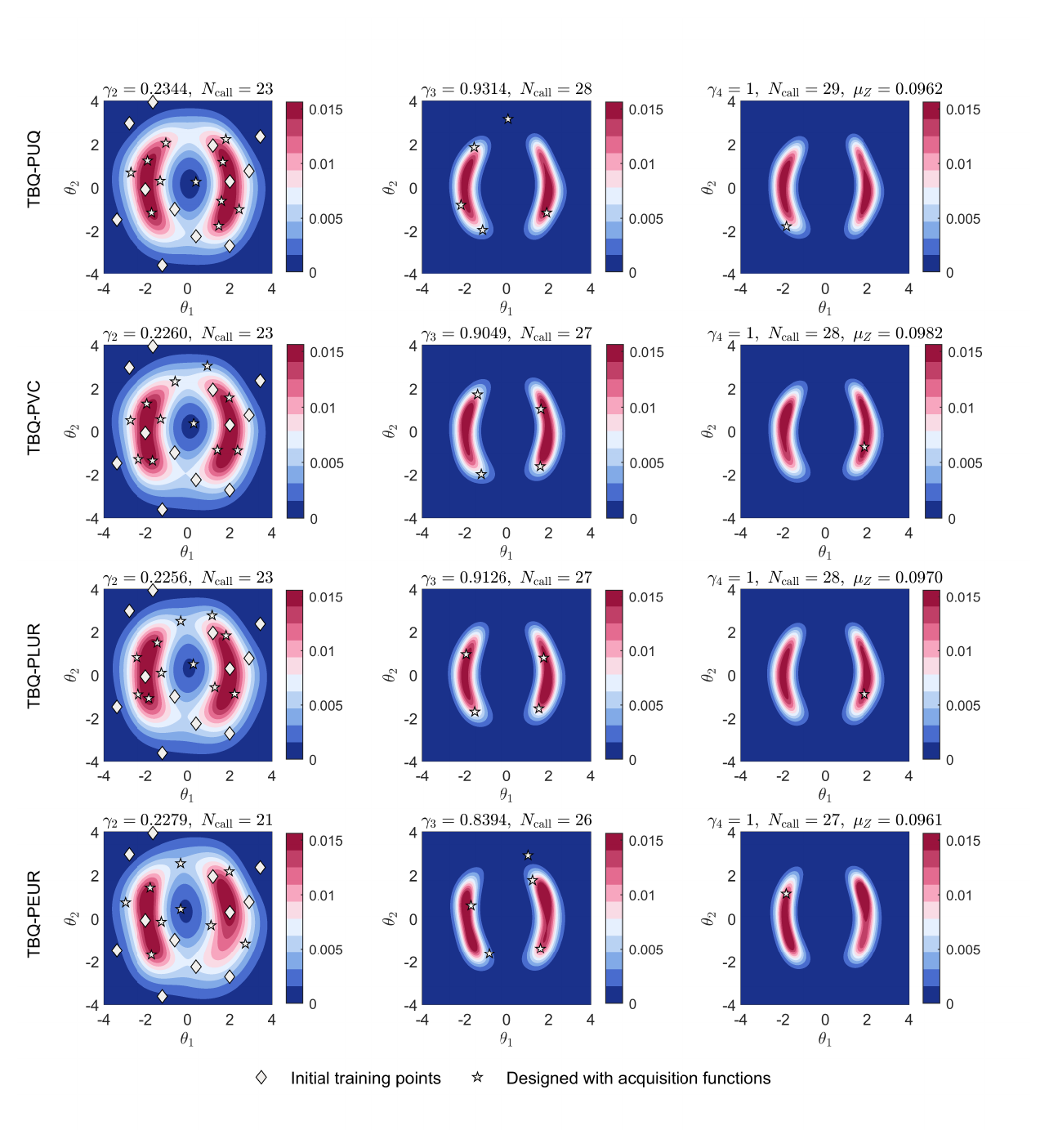}
	\caption{\centering{Results for posterior defined with $\mathcal{U}_1$, generated by the TBQ algorithm driven by PUQ (1st row), PVC (2nd row), PLUR (3nd row) and PEUR (4th row). The newly added training points generated each stage are marked on the corresponding subplots. }}
	\label{fig:TBQU1}
\end{figure}

\begin{figure}[H]
	\centering
	\setlength{\abovecaptionskip}{-0.8cm}
	\includegraphics[scale=0.70]{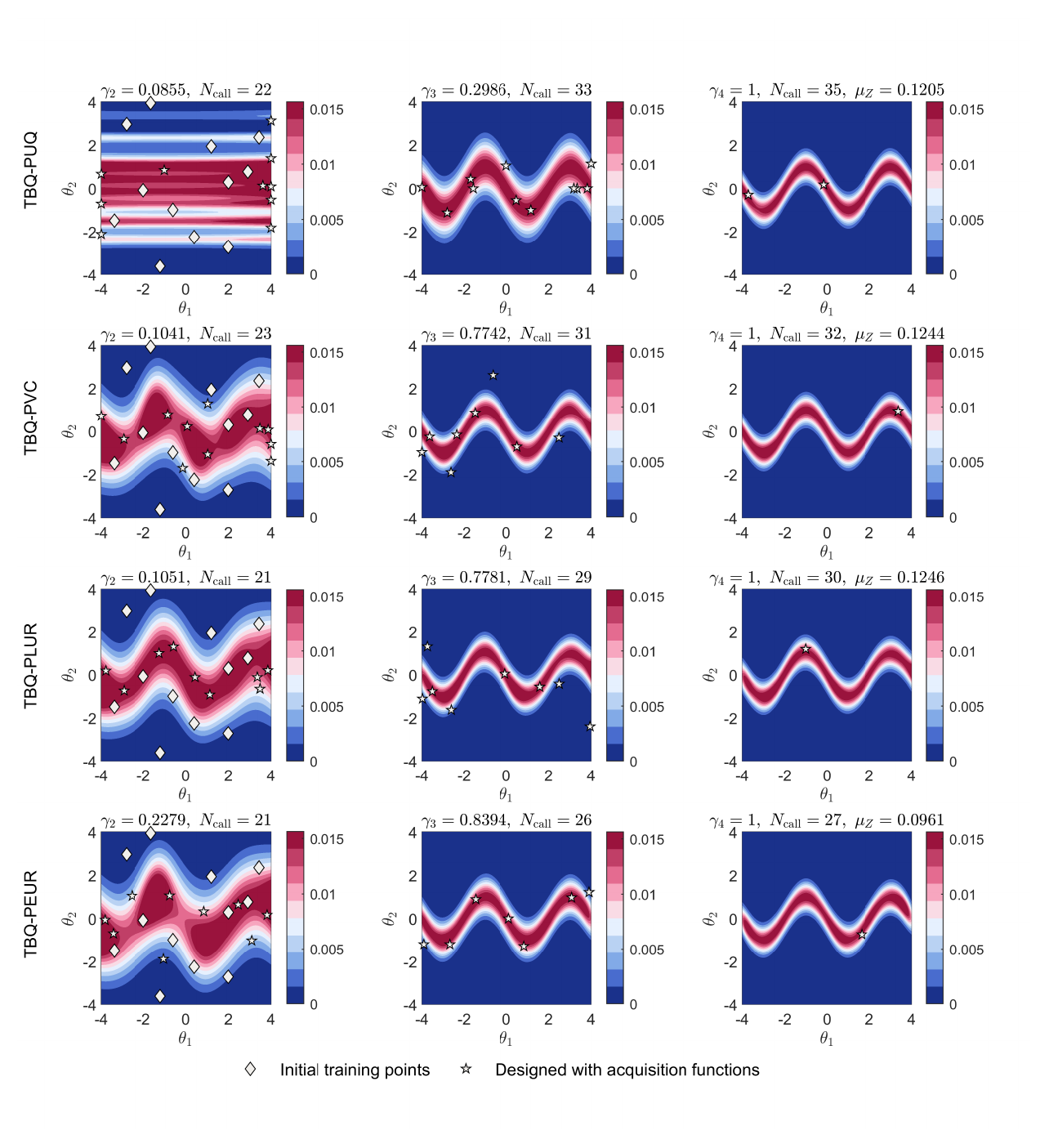}
	\caption{\centering{Results of the posteriors defined by $\mathcal{U}_2$, generated by TBQ equipped with the four proposed acquisition functions.}}
	\label{fig:TBQU2}
\end{figure}

\begin{figure}[H]
	\centering
	\setlength{\abovecaptionskip}{-0.8cm}
	\includegraphics[scale=0.75]{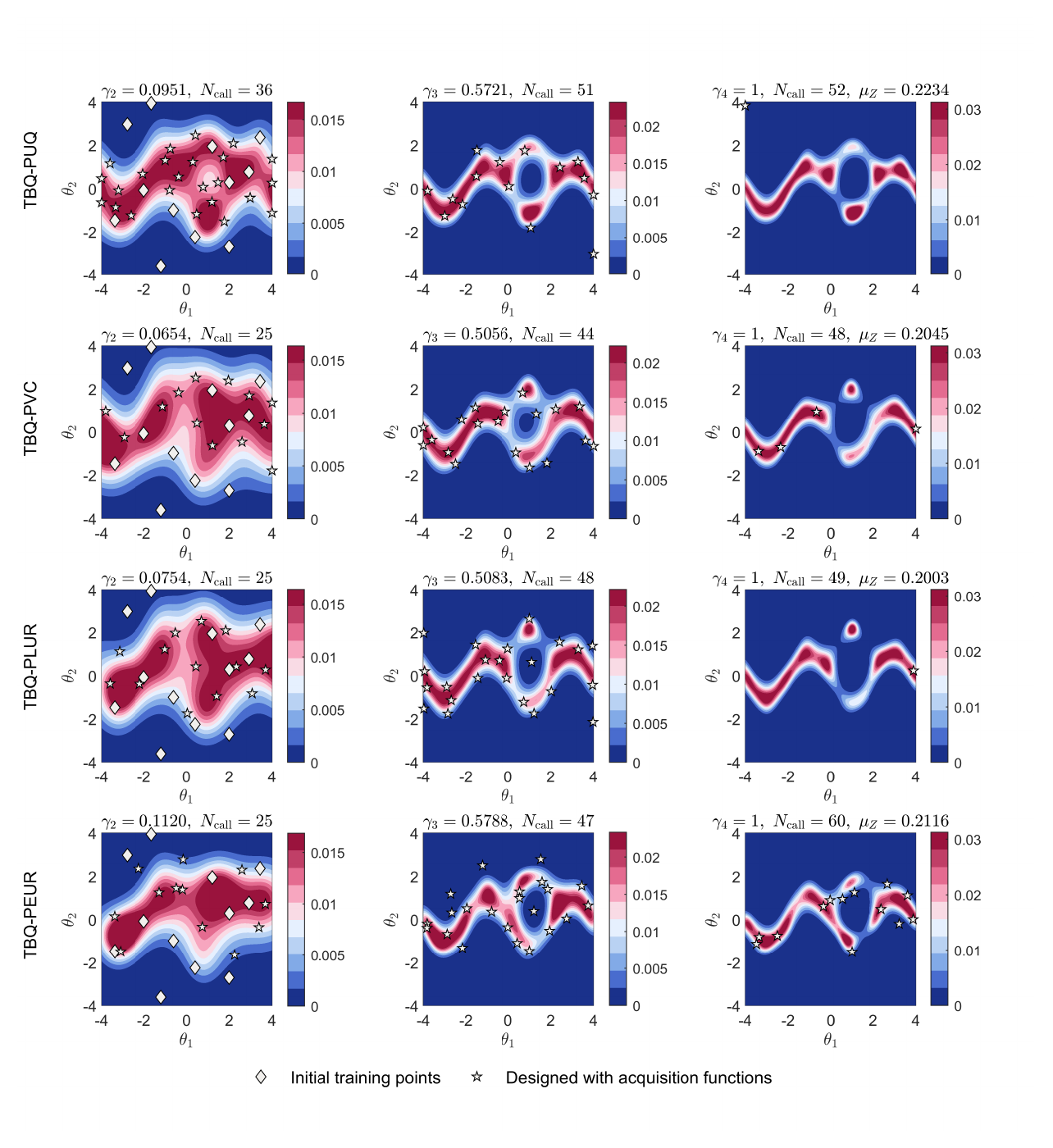}
	\caption{\centering{Results of the posteriors defined by $\mathcal{U}_3$, generated by the TBQ algorithm.}}
	\label{fig:TBQU3}
\end{figure}

\begin{figure}[H]
	\centering
	\setlength{\abovecaptionskip}{-0.9cm}
	\includegraphics[scale=0.75]{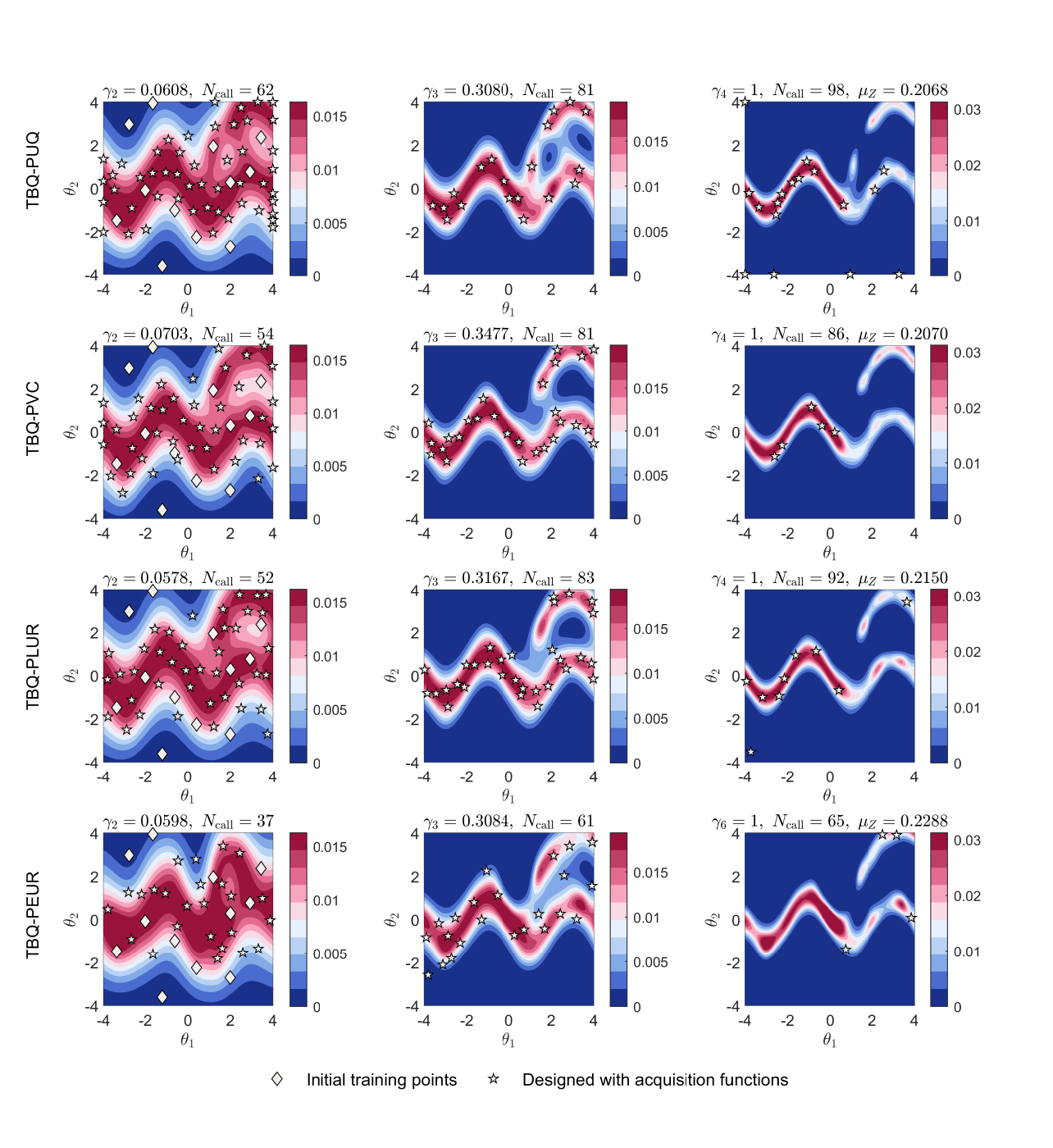}
	\caption{\centering{Results of the posteriors defined by $\mathcal{U}_4$, generated by the TBQ algorithm.}}
	\label{fig:TBQU4}
\end{figure}

Next, the TBQ is implemented for all four examples with each of the four acquisition functions. The algorithm parameters were kept as consistent as possible with those of BQ. For the first two examples, $\epsilon$ is set to be 0.04 for all tempering stages; for the third example, it is set to 0.02; and for the fourth example, it is set to be 0.01 for intermediate stage, and then 0.02 for the last stage. It should be noted that, in general, the TBQ algorithm exhibits high robustness to the values of $\epsilon$ and $\varsigma$ within certain ranges, as has been proved by the first example of Ref. \cite{wei2025bayesian}. However, if the posterior distribution contains multiple unbalanced modes (modes with significant difference in peaks and support bandwidths), each mode contributes differently to the model evidence $Z$. In this case, a smaller value of $\epsilon$ is preferred to precisely capture the behavior of secondary modes. While in case the posterior include only one mode or multiple modes with close peaks, a larger value can be adopted to save computational resources. $\varsigma$ is set to be 1 for the first three examples, and then 0.75 for the fourth example. The length of each Markov chain is set to be 30 for all runs to ensure a safe skip of the burn-in period and to avoid particle degeneracy. With the same 12 initial training samples for producing Figures \ref{fig:BQU1}-\ref{fig:BQU4}, the TBQ algorithm is initialized, and the results of unormalized posteriors for the four examples are sequentially reported in Figure \ref{fig:TBQU1}-\ref{fig:TBQU4}, with the value of tempering parameters, number of accumulated likelihood calls, and the mean estimate of the model evidence listed in the title of each subplot. 

As can be seen from Figure \ref{fig:TBQU1}, regardless of the acquisition function used, the TBQ algorithm automatically generates three tempering stages, with slightly different values of tempering parameters. The posterior of the final stage serves as the estimate of the target posterior. Comparing Figure \ref{fig:TBQU1} with Figure \ref{fig:BQU1}, it is seen that, both the estimation accuracy and computational cost of TBQ are comparable to those of BQ. Similar conclusions can be drawn for the second example by comparing Figure \ref{fig:TBQU2} with Figure \ref{fig:BQU2}. 

Comparing Figure \ref{fig:TBQU3} with Figure \ref{fig:BQU3}, it is seen that, the TBQ algorithms with PUQ, PVC and PLUR as acquisition functions also produced comparable quality of estimates, but with much less model evaluations, compared to those of the BQ algorithm. The TBQ algorithm driven by PEUR, although consumed more model evaluations, provided much better estimation of the posterior especially in the bifurcated secondary modal regions. From Figure \ref{fig:TBQU4}, it is shown that, for the fourth example, the TBQ algorithm, equipped with any of the four acquisition functions, produced three intermediate tempering stages to arrive at accurate estimate of both posterior model evidence.   

Comparing the training points shown in Figures \ref{fig:TBQU1} - \ref{fig:TBQU4} and Figures \ref{fig:BQU1} - \ref{fig:BQU4}, it can be found that TBQ tends to guide training points to concentrate in the modal support regions of each tempering stage, while the BQ algorithm tends to generate training samples more uniformly distributed within the support of prior distribution. This phenomenon is determined by the nature of the two algorithms. The balance between exploration and exploitation in the BQ algorithm is entirely determined by the adopted acquisition function; while that of TBQ can also be realized by the value of the tempering parameters. With smaller value of $\gamma$, the acquisition function tends to have better exploration performance, while with higher value of $\gamma$, it tends to exploit the local identified modes.  This feature makes the TBQ algorithm particularly suitable for handling problems of which the posteriors show great divergence with the priors, especially in higher dimension. For such problems, BQ generally requires more initial samples to ensure that some initial samples lie in the true modal regions.  
\vspace{0.1 cm}

\vspace{0.1 cm}

With identical parameter configurations, and 12 initial training samples randomly generated, the TBQ algorithm equipped with each acquisition function was executed ten times, the results of model evidences are summarized in Table \ref{table:Result2D_TBQ}. Most of these 160 runs produced three tempering stages, while others created four or even five, to approach the target posteriors. Comparing Table \ref{table:Result2D_TBQ} with Table \ref{table:Result2D_BQ}, it can be seen that, the prediction accuracy, in terms of both absolute error and CoVs, has been improved for the first three examples. Comparing the results, e.g., for the fourth example, generated by the two prospective acquisition functions, i.e., PLUR and PEUR, it can be found that PLUR commonly requires more model evaluations to achieve the same stopping criteria, but yields estimates with smaller absolute errors for model evidences (indeed also for posteriors). This is due to the fact that PLUR measures the expected improvements in the predictive accuracy of the posteriors, thus it forces the algorithm to capture as many details of the posteriors, which, in turn, reduces the bias of the estimation for model evidence. However, PEUR measures the expected reduction of the prediction uncertainty of the model evidence, not the bias of the estimate, thus results in a higher rate on reducing the variance of the model evidence. This indicates that, in case the estimation of model evidence is of priority, the PEUR function is preferable.
\vspace{0.1 cm}

Finally, to compare the computational budget of acquisition functions, the CPU time of one TBQ run, and the average time for generating one training point with each acquisition function, are reported in the last two columns of Table \ref{table:Result2D_TBQ}. It is seen that, the PUQ function always consumes the least CPU time, followed by the PVC and PLUR functions, and PEUR requires slightly higher cost. This aligns with the complex levels of the four acquisition functions as reported in the last column of Table \ref{table:SummaryAcqFun}. 
\newcolumntype{C}{>{\centering\arraybackslash}m{1.4cm}}
\newcolumntype{D}{>{\raggedright\arraybackslash}m{1.9cm}}
\newcolumntype{E}{>{\centering\arraybackslash}m{0.8cm}}
\newcolumntype{F}{>{\centering\arraybackslash}m{0.8cm}}
\newcolumntype{O}{>{\centering\arraybackslash}m{1.3cm}}
\newcolumntype{G}{>{\centering\arraybackslash}m{2.3cm}}
\newcolumntype{I}{>{\centering\arraybackslash}m{1.2cm}}
\newcolumntype{J}{>{\centering\arraybackslash}m{2cm}}
\newcolumntype{K}{>{\centering\arraybackslash}m{2cm}}

\begin{table}[H]
	\caption{Summary and comparison of results of model evidence of the four 2D problems, generated by TBQ algorithms, with $N_\mathrm{call}$, $\mu_Z$, relative errors and CoVs being calculated across ten replications to illustrate the robustness of the algorithm. ``Time (Total)'' indicates the total CPU time of one implementation of the TBQ algorithm, and ``Time (Acq.)'' refers to the average time for generating one point using the acquisition function. All CPU times are reported in seconds. }
	\label{table:Result2D_TBQ}
	\centering
	\begin{threeparttable}
		\begin{tabular}{C D E F O G I J K}
			\hline 
			Examples& Methods  & $N_\mathrm{call}$  & $\mu_Z$&$Z$ (Ref.) &  Relative errors    & CoVs&Time (Total)&Time (Acq.)\\ 
			\hline
			\multirow{4}{*}{1st}&TBQ-PUQ& 28.0  & 0.0956&\multirow{4}{*}{0.1021} &0.0628&0.0133&53.8311&0.0594\\
			&TBQ-PVC&28.3 &0.0975& &0.0450&0.0129&61.2212&0.4032\\
			&TBQ-PLUR&28.7  &0.0965& &0.0550&0.0266&65.3000&0.4132 \\
			&TBQ-PEUR&28.3   &0.0993& &0.0307&0.0262&67.0025&0.6049\\
			\hline
			\multirow{4}{*}{2nd}&TBQ-PUQ& 31.9  &0.1246&\multirow{4}{*}{0.1253} &0.0265&0.0089&52.4298&0.0522\\
			&TBQ-PVC&31.0 &0.1238& &0.0184&0.0308&64.8023&0.4083\\
			&TBQ-PLUR&31.4  &0.1255& &0.0116&0.0149&62.5098&0.4002 \\
			&TBQ-PEUR&30.7   &0.1228& &0.0210&0.0175&77.8625&0.6758\\
			\hline
			\multirow{4}{*}{3rd}&TBQ-PUQ& 55.9  &0.2037&\multirow{4}{*}{0.2193} &0.0714&0.0427&66.8861&0.0682\\
			&TBQ-PVC&57.3 &0.2069& &0.0565&0.0313&93.3407&0.5201\\
			&TBQ-PLUR&58.9  &0.2097& &0.0437&0.0374&83.1504&0.4847 \\
			&TBQ-PEUR&53.1   &0.2122& &0.0399&0.0393&82.3796&0.9562\\
			\hline
			\multirow{4}{*}{4th}&TBQ-PUQ& 96.6  &0.2119&\multirow{4}{*}{0.2288} &0.0738&0.0546&90.0506&0.0671\\
			&TBQ-PVC&93.8 &0.2183& &0.0457&0.0422&131.2666&0.6580\\
			&TBQ-PLUR&96.2  &0.2167& &0.0527&0.0410&146.5755&0.6716 \\
			&TBQ-PEUR&70.7   &0.2098& &0.0922&0.0658&119.3403&0.8638\\
			\hline
		\end{tabular}
	\end{threeparttable}
\end{table}   

\subsection{A Ten-dimensional Example}
Next, we test the performance of the TBQ algorithm and the four acquisition functions for higher-dimensional problems where the BQ algorithm is not applicable. A ten-dimensional example is built from the first two energy functions in Eq. \eqref{eq:EnergyFunExample1} with the energy function formulated as:
\begin{equation}
	\begin{split}
		\mathcal{U} \left( \boldsymbol{\theta } \right) &=\frac{1}{2}\left( \frac{\left\| \boldsymbol{\theta }_{1:2} \right\| -2}{0.2} \right) ^2-\log \left( \exp \left( -\frac{1}{2}\left( \frac{\theta _1-2}{0.08} \right) ^2 \right) +\exp \left( -\frac{1}{2}\left( \frac{\theta _2+2}{0.08} \right) ^2 \right) \right) \\
		&+\frac{1}{2}\left( \frac{\theta _4+\sin \left( 0.25\pi \theta _3 \right)}{0.3} \right) ^2 +\frac{1}{2}\left( \boldsymbol{\theta }_{5:6}-\boldsymbol{\mu }_{5:6} \right) ^{\top}\Sigma _{5:6}^{-1}\left( \boldsymbol{\theta }_{5:6}-\boldsymbol{\mu }_{5:6} \right) 
	\end{split}
\end{equation}
, where $\boldsymbol{\theta }=\left( \theta _1,\theta _2,\cdots ,\theta _{10} \right) ^{\top}$, $\boldsymbol{\theta }_{1:2}=\left( \theta _1,\theta _2 \right) ^{\top}$,$\boldsymbol{\theta }_{5:6}=\left( \theta _5,\theta _6 \right) ^{\top}$, $\boldsymbol{\mu }_{5:6}=\left( 0.5,1 \right) ^{\top}$ and $\Sigma _{5:6}=\left( \begin{matrix}
	0.7^2&		0.56\\
	0.56&		1\\
\end{matrix} \right) $. The prior distributions of all parameters are assumed to be independent and uniform with support $\left[ -4,4 \right] $. This example was previously designed in Ref. \cite{wei2025bayesian}. The target posterior exhibits a  number of complex features like multi-modality presented in the marginal posterior of $\boldsymbol{\theta}_{1:2}$, nonlinear dependency of $\boldsymbol{\theta}_{3:4}$, linear dependency of $\boldsymbol{\theta}_{5:6}$, high sharpness of the marginal posterior of $\boldsymbol{\theta}_{1:4}$, as well as negligibility of $\boldsymbol{\theta}_{7:10}$. Meanwhile, the true value of the model evidence is also extremely small (specifically, estimated using the crude MCS to be $1.4680\times 10^{-4}$, with a CoV less than 1\%), making it difficult to be accurately estimated. 
 
 \begin{figure}[H]
 	\centering
 	\setlength{\abovecaptionskip}{-0.1cm}
 	\includegraphics[scale=0.75]{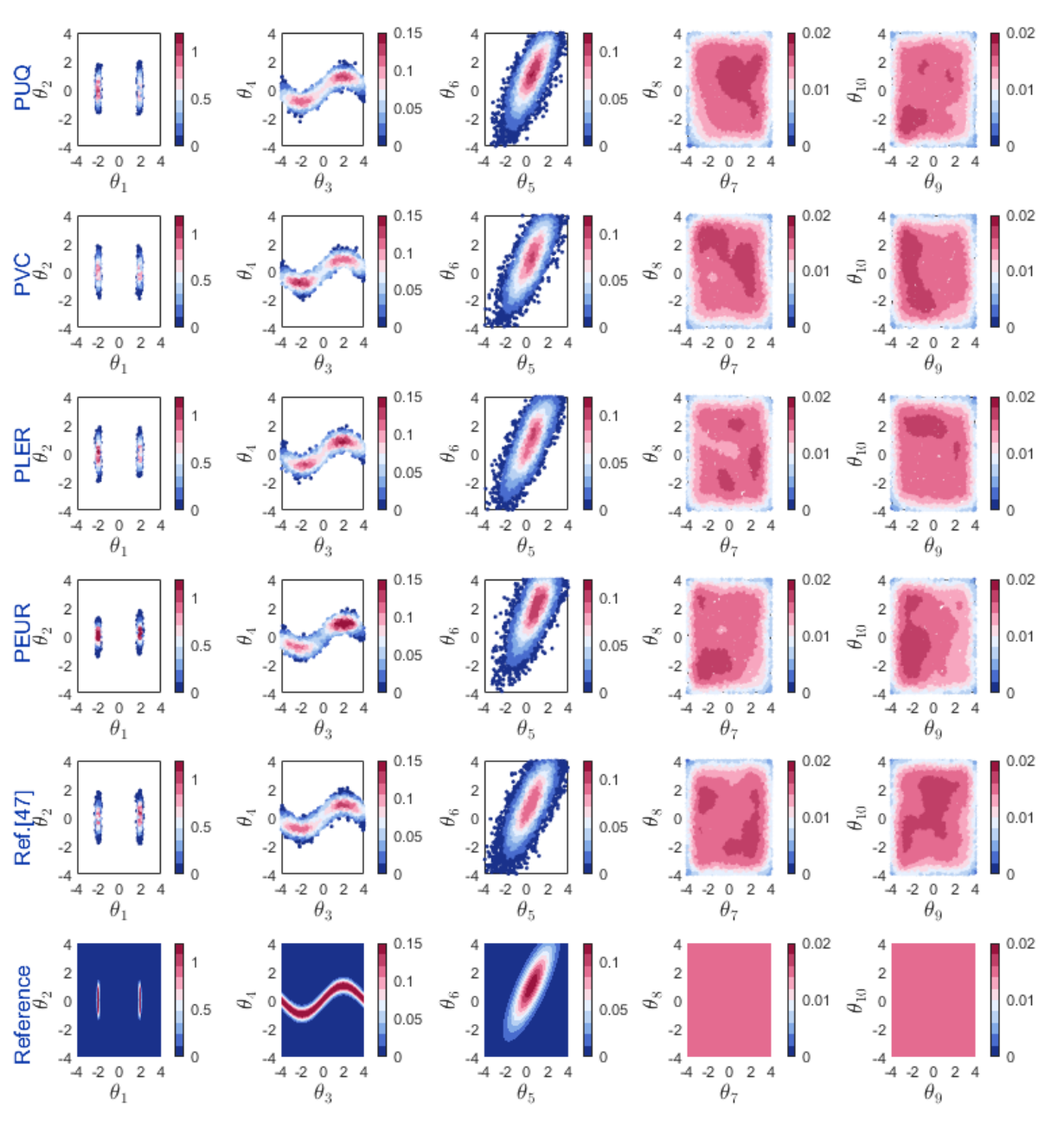}
 	\caption{\centering{Comparison of results of the pairwise marginal posteriors for the 10-dimensional example. Those displayed in the first four rows, each of which is generated with a specific acquisition function as labeled, represent the generated posterior samples of the final tempering stages, and the heatmaps represent the densities of samples. Those shown in the fifth row refer to the results generated by the original TBQ algorithm in Ref. \cite{wei2025bayesian}. Heatmaps in the last row presents the reference pairwise posteriors generated by integrating the other parameters out. }}
 	\label{fig:TenDPost}
 \end{figure}

Due to the high sharpness of the target posterior, the BQ algorithm is not applicable. To implement TBQ with any one of the four developed acquisition functions, as well as the original TBQ algorithm in Ref. \cite{wei2025bayesian}, $\epsilon$ is set as 0.04 for all tempering stages, $\varsigma$ is assumed to be 0.7, and $N_\mathrm{MC}$ is assumed to be $5\times 10^3$. For each iteration of the $j$-th tempering stage, the new training point is selected from the candidate sample pool $\mathcal{T}_\mathrm{MC}^{(j-1)}$ as the one with highest acquisition function value, instead of by maximizing the acquisition functions.  Initialized with the same set of $N_0=20$ initial training samples, the TBQ algorithms equipped with each of the four acquisition functions and the original TBQ algorithm in Ref. \cite{wei2025bayesian}, are implemented, and the results of posteriors are reported in Figure \ref{fig:TenDPost}. Comparing  results in each of the first four row with the reference solution shown in the last row, it can be seen that TBQ with any of the four acquisition functions produces high-quality estimates of the posterior densities with all the complex features being accurately captured. However, each acquisition function consumed different numbers of model calls and produced different numbers of intermediate tempering stages. Specifically, PUQ, PVC, PLER and PEUR consumed 322, 284, 294 and 191 likelihood calls, respectively, produced 11, 10, 11 and 14 intermediate tempering stages respectively. The original TBQ algorithm consumed 326 likelihood calls, while produced 10 tempering stages. The mean estimates of the model evidence generated with these four acquisition functions are 1.4508$\times 10^{-4}$, 1.2412$\times 10^{-4}$, 1.2681$\times 10^{-4}$ and 1.6302$\times 10^{-4}$ respectively, while the one estimated by the original TBQ algorithm is 1.2053$\times 10^{-4}$, with the CoV of each estimate being smaller than the pre-specific threshold  $\epsilon=0.04$. Results in Figure \ref{fig:TenDPost} indicates that all the four acquisitions and the original TBQ algorithm produced accurate mean estimate for the posteriors. The PUQ function consumes close number of likelihood calls, while the other three new acquisition functions consumes notably less number of likelihood function, indicating higher efficiency. 

\begin{table}[H]
	\caption{Comparison of results of model evidence for the 10-dimensional example, where $N_\mathrm{call}$, $\mu_Z$ and CoVs are calculated across ten repeated implementations. }
	\label{table:Result10D}
	\centering
	\begin{threeparttable}
		\begin{tabular}{p{4cm}p{2cm}p{3cm}p{3cm}p{2cm}}
			\hline 
			Acq. functions  & $N_\mathrm{call}$  & $\mu_Z$&$Z$ (Ref.)  & CoVs\\ 
			\hline
			PUQ& 319.00  & $1.3062\times 10^{-4}$&\multirow{4}{*}{$1.4680\times 10^{-4}$} &0.0215\\
			PVC&284.55 &$1.3055\times 10^{-4}$& &0.0146\\
			PLUR&295.30  &$1.3206\times 10^{-4}$& &0.0254 \\
			PEUR&211.70   &$1.3813\times 10^{-4}$& &0.0364\\
			Ref.\cite{wei2025bayesian}&323.10&$1.2964\times 10^{-4}$&&0.0283\\
			\hline
		\end{tabular}
	\end{threeparttable}
\end{table}   

To further verify the robustness, the original TBQ algorithm and the new TBQ equipped with each of the four acquisition functions are implemented for ten times, each of which is initialized with 20 training samples generated randomly using LHS, and the results are summarized in Table \ref{table:Result10D}. The results indicate that all the four acquisition functions and the original TBQ algorithm predicted the model evidence with high accuracy and robustness, and the CoVs of all estimates are less than the pre-specified convergence threshold (e.g., 0.04). In terms of computational efficiency, PUQ consumed the highest number of model calls on average, which is close to the one consumed by the original TBQ algorithm, followed by PLUR and then PVC, while PEUR consumed the smallest number of model calls. This demonstrates that PEUR is the most effective acquisition function for this example. This indicates that for posterior with higher complexities, the PEUR function tend to be more effective than the other threes, and is thus preferable although it consumes more CPU time to generate one new training point. To summary, for this example, the developed PUQ function show similar performance with the original TBQ algorithm for estimating both posterior and model evidence, while the other three new acquisition functions, especially the PEUR function, show much better performance.

\subsection{A three-degree-of-freedom (3-DoF) dynamical system}
\begin{figure}[H]
	\centering
	\setlength{\abovecaptionskip}{-0.0cm}
	\includegraphics[scale=0.3]{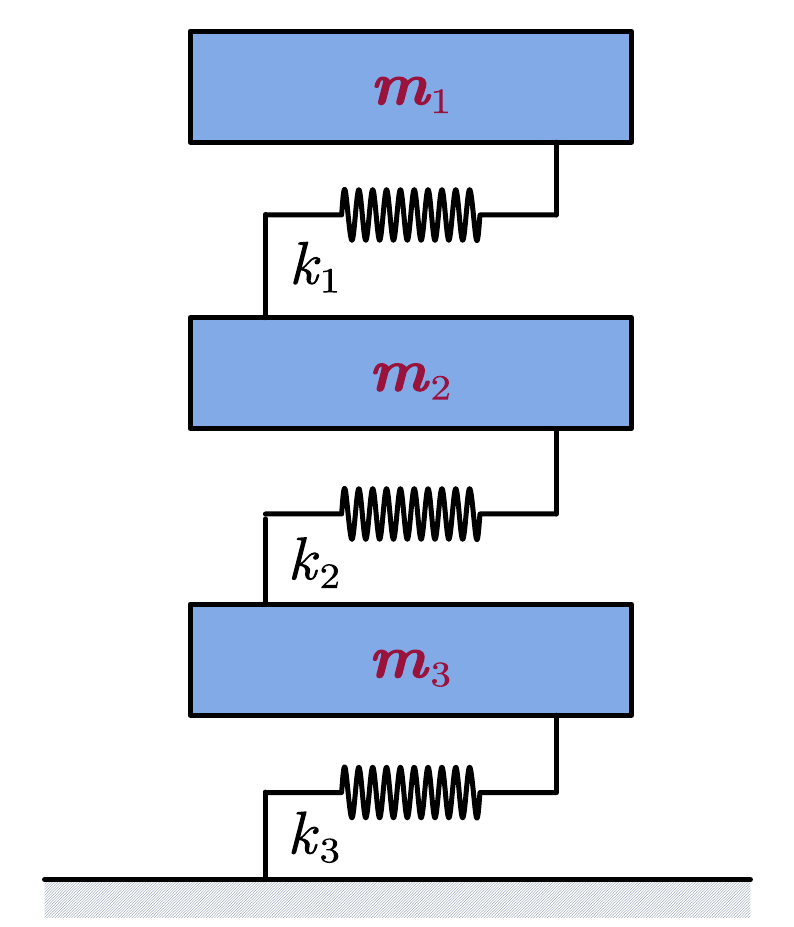}
	\caption{\centering{The 3DoF undamped spring-mass system. }}
	\label{fig:ThreeDoFSystem}
\end{figure}

Consider a 3-DoF undamped spring-mass sistem shown in Figure \ref{fig:ThreeDoFSystem}. The governing equation is $\mathbf{M}\ddot{\boldsymbol{x}}+\mathbf{K}\boldsymbol{x}=0$, with $\mathbf{M}=\mathrm{diag}\left( m_1,m_2,m_3 \right) $ referring to the mass matrix, and 
\begin{equation}
	\mathbf{K}=\left[ \begin{matrix}
		k_1&		-k_1&		0\\
		-k_1&		k_1+k_2&		-k_2\\
		0&		-k_2&		k_2+k_3\\
	\end{matrix} \right] 
\end{equation}
indicating the stiffness matrix. The three mass parameters $m_1\sim m_3$ are assumed to be 1000kg, 1200 kg and 800 kg respectively. The three stiffness parameters are assumed to be $k_i=\bar{k}\theta _i$, with $i=1,2,3$ and $\bar{k}=21.6\times 10^6\,\,\mathrm{N}/\mathrm{m}$ being the nominal stiffness. The three dimensionless parameters $\theta_1\sim \theta_3$ are the deterministic-but-unknown parameters to be calibrated, of which the prior distribution is assumed to be uniform with supports $\theta _1\sim \left[ 0.1,1 \right] $, $\theta _2\sim \left[ 0.5,5 \right] $ and $\theta _3\sim \left[ 0.5,5 \right] $ respectively. The measurement is conducted on the first three natural frequencies to be 11.9191, 26.7954 and 56.5167 respectively, with a Gaussian white measurement noise whose STD is 0.05. The energy function is formulated by Eq. \eqref{eq:Defenergyfunction}.

\begin{figure}[H]
	\centering
	\setlength{\abovecaptionskip}{-0.0cm}
	\includegraphics[scale=0.75]{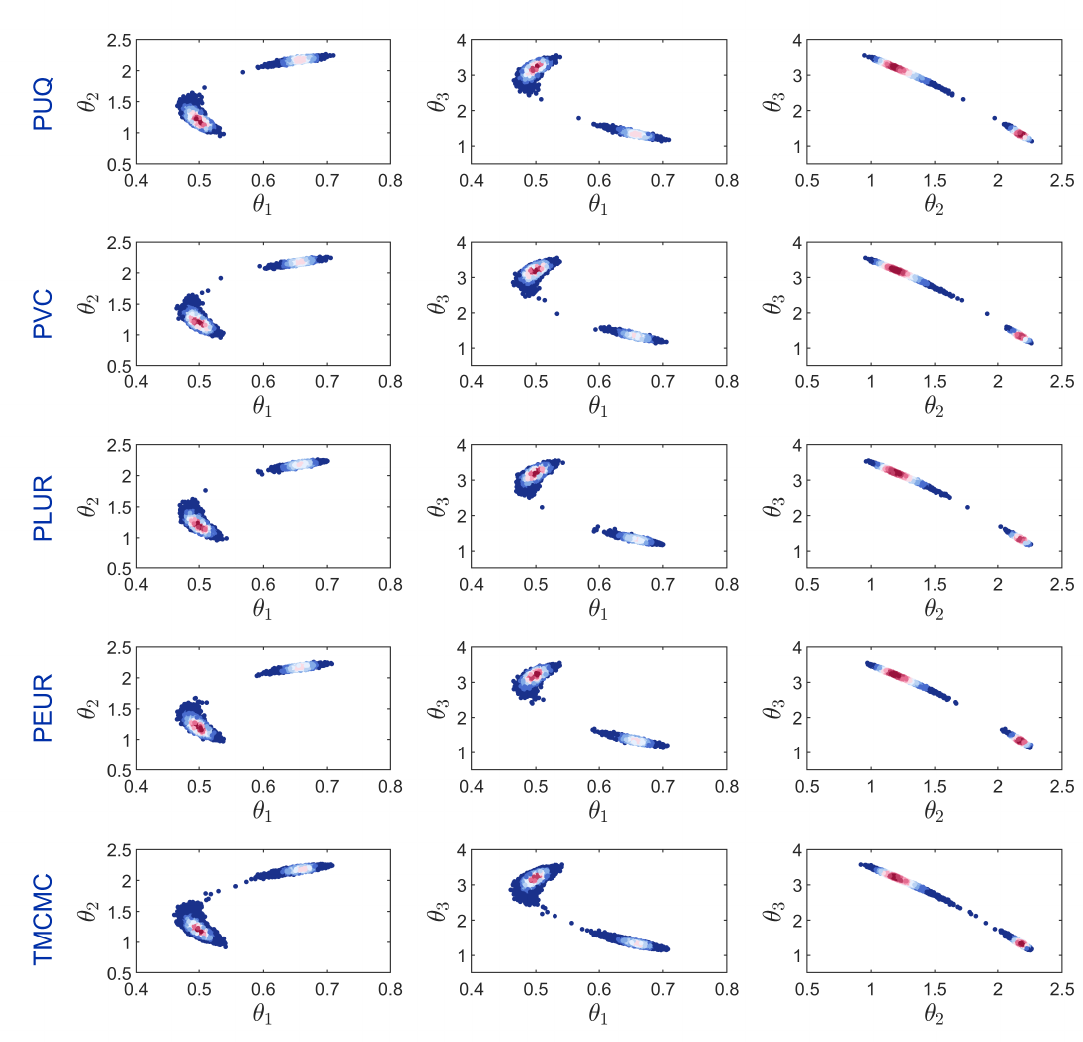}
	\caption{\centering{Comparison of posterior results for the 3DoF system. Note that the plotting range is smaller than the support of the prior distribution to display the posterior samples more clearly, as the posterior has a much smaller support than the prior. }}
	\label{fig:ThreeDOFPost}
\end{figure}

The algorithm parameters of TBQ are set to be $\epsilon=0.02$, $\varsigma=1$, and $N_\mathrm{MC}=10^4$. Each new training point is selected from the candidate sample pool $\mathcal{T}_\mathrm{MC}^{(j-1)}$. Initialized with the same 12 training samples, the TBQ algorithm is implemented with each of the four acquisition functions, and resultant posterior samples of the last tempering stage are reported in the first four rows of Figure \ref{fig:ThreeDOFPost}. The reference solutions displayed in the last row are generated using TMCMC, with the sample size of each tempering stage set to be $5\times 10^4$. It is seen again that all the four acquisition functions produced accurate estimate of the posterior, with the multiple disconnected modes and dependencies among parameters being precisely captured.

To demonstrate the robustness and efficiency, TBQ equipped with each acquisition function and the TMCMC are implemented for ten times, and the results of model evidence are summarized in Table \ref{table:Result3Dof}. The mean estimate by TMCMC can serve as reference solution as the estimator is unbiased and the CoV is smaller than 1\%. It is seen that all the four acquisition functions slightly underestimate the model evidence with small variation, but the accuracy is generally acceptable. It is also seen that, for this example, the PEUR acquisition function is slightly less effective than the other threes. This contradicts the conclusion obtained from the previous examples, indicating that no single acquisition function is the most effective across all problems, and all four developed acquisition functions have their relative merits.  

\begin{table}[H]
	\caption{Comparison of results of model evidence for the 3-DoF dynamic system, with $N_\mathrm{call}$, $\mu_Z$ and CoVs being calculated across ten repeated implementations. }
	\label{table:Result3Dof}
	\centering
	\begin{threeparttable}
		\begin{tabular}{p{5cm}p{3cm}p{4cm}p{3cm}}
			\hline 
			Methods  & $N_\mathrm{call}$  & $\mu_Z$& CoVs\\ 
			\hline
			TBQ-PUQ& 50.3  & $7.9562\times 10^{-4}$ &0.0115\\
			TBQ-PVC&48.9 &$7.8942\times 10^{-4}$& 0.0152\\
			TBQ-PLUR&50.7  &$7.7557\times 10^{-4}$& 0.0142 \\
			TBQ-PEUR&55.9   &$7.8474\times 10^{-4}$& 0.0117\\
			TMCMC&$4.5\times10^5$   &$8.2456\times 10^{-4}$& 0.0057\\
			\hline
		\end{tabular}
	\end{threeparttable}
\end{table}

\subsection{Application to battery module cooling analysis}
\begin{figure}[H]
	\centering
	\setlength{\abovecaptionskip}{-0.0cm}
	\includegraphics[scale=0.7]{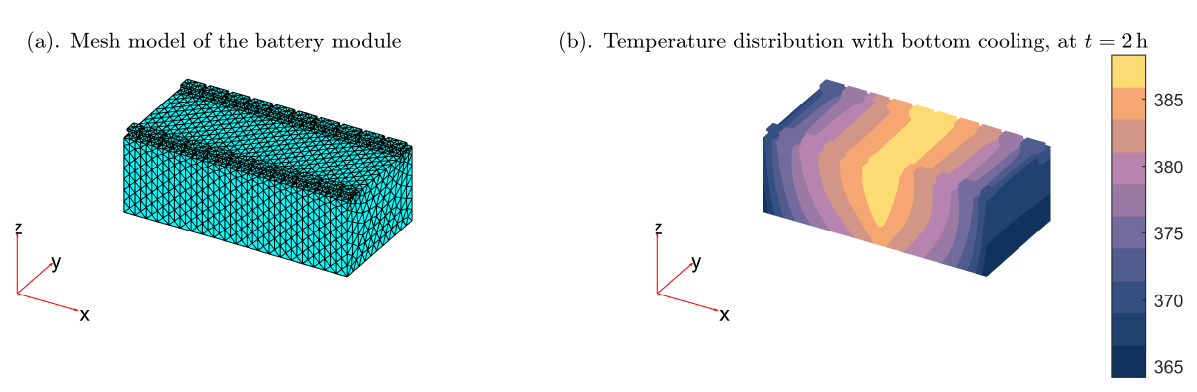}
	\caption{\centering{Battery module cooling model, with (a) indicating the mesh model and (b). being the predicted temperature distribution at $t=2\,\,\mathrm{h}$ with cooling. }}
	\label{fig:BatteryModel}
\end{figure}

Consider the dynamic cooling analysis of a battery module, with the computational model available in the Matlab PDE toolbox. This battery module is composed of 20 cells, with a cooling panel on the bottom face of the module. One can refer to the user document for the geometric information of each cell and the module. The mesh model as well as the predicted temperature distribution at the time instant $t=2\,\mathrm{h}$ are schematically shown in Figure \ref{fig:BatteryModel}. A total number of thirteen model parameters, with information reported in Table \ref{table:Betteryparameter}, are assumed to be unknown and required to be inferred based on measurements. The controllable input $x$ is the ambient temperature. At $x_1=293$ K and $x_2=308$ K, virtual measurements (through noised simulation) are conducted at the top and bottom nodes of the middle cell respectively, as $y_\mathrm{top}(x_1)=367.56$ K, $y_\mathrm{bottom}(x_1)=363.55$ K, $y_\mathrm{top}(x_2)=382.93$ K and $y_\mathrm{bottom}(x_2)=378.53$ K, with the STD of measurement noise being $\sigma_\mathrm{n}=1$ K. The energy function is formulated following Eq. \eqref{eq:Defenergyfunction}.

\begin{table}[H]
	\caption{Summary of model parameters for the battery module cooling analysis, with each parameter admitting a uniform prior distribution. }
	\label{table:Betteryparameter}
	\centering
	\begin{threeparttable}
		\begin{tabular}{p{7cm}p{2cm}p{3cm}p{3cm}}
			\hline 
			Meanings of parameter  & IDs  & True values&Prior supports\\ 
			\hline
			Cell thermal conductivity in plane& $\theta_1$  & 80 W/(K*m) &[80, 90]\\
			Cell thermal conductivity through plane&$\theta_2$&2 W/(K*m)& [1.5, 2.2]\\
			Tab thermal conductivity&$\theta_3$  &386 W/(K*m)& [360, 400] \\
			Connector thermal conductivity&$\theta_4$   &400 W/(K*m)& [380, 410]\\
			Density of cells& $\theta_5$  & 780 kg/$\mathrm{m}^3$ &[760, 800]\\
			Density of tabs&$\theta_6$&2700 kg/$\mathrm{m}^3$& [2500, 3100]\\
			Density of connectors&$\theta_7$  &540 kg/$\mathrm{m}^3$& [520, 550] \\
			Specific heat values of cells&$\theta_8$   &785 J/(Kg*K)& [750, 820]\\
			Specific heat values of tabs&$\theta_9$   &890 J/(Kg*K)& [850, 920]\\
			Specific heat values of connectors&$\theta_{10}$   &840 J/(Kg*K)& [820, 880]\\
			Heat generation of a normal cell&$\theta_{11}$   &15 W& [14, 25]\\
			Heat generation of a faulty cell&$\theta_{12}$   &25 W& [20, 28]\\
			Convection coefficient&$\theta_{13}$   &100 W/($\mathrm{m}^2\cdot \mathrm{K}$)& [80, 102]\\
			\hline
		\end{tabular}
	\end{threeparttable}
\end{table}

\begin{figure}[H]
	\centering
	\setlength{\abovecaptionskip}{-0.0cm}
	\includegraphics[scale=0.8]{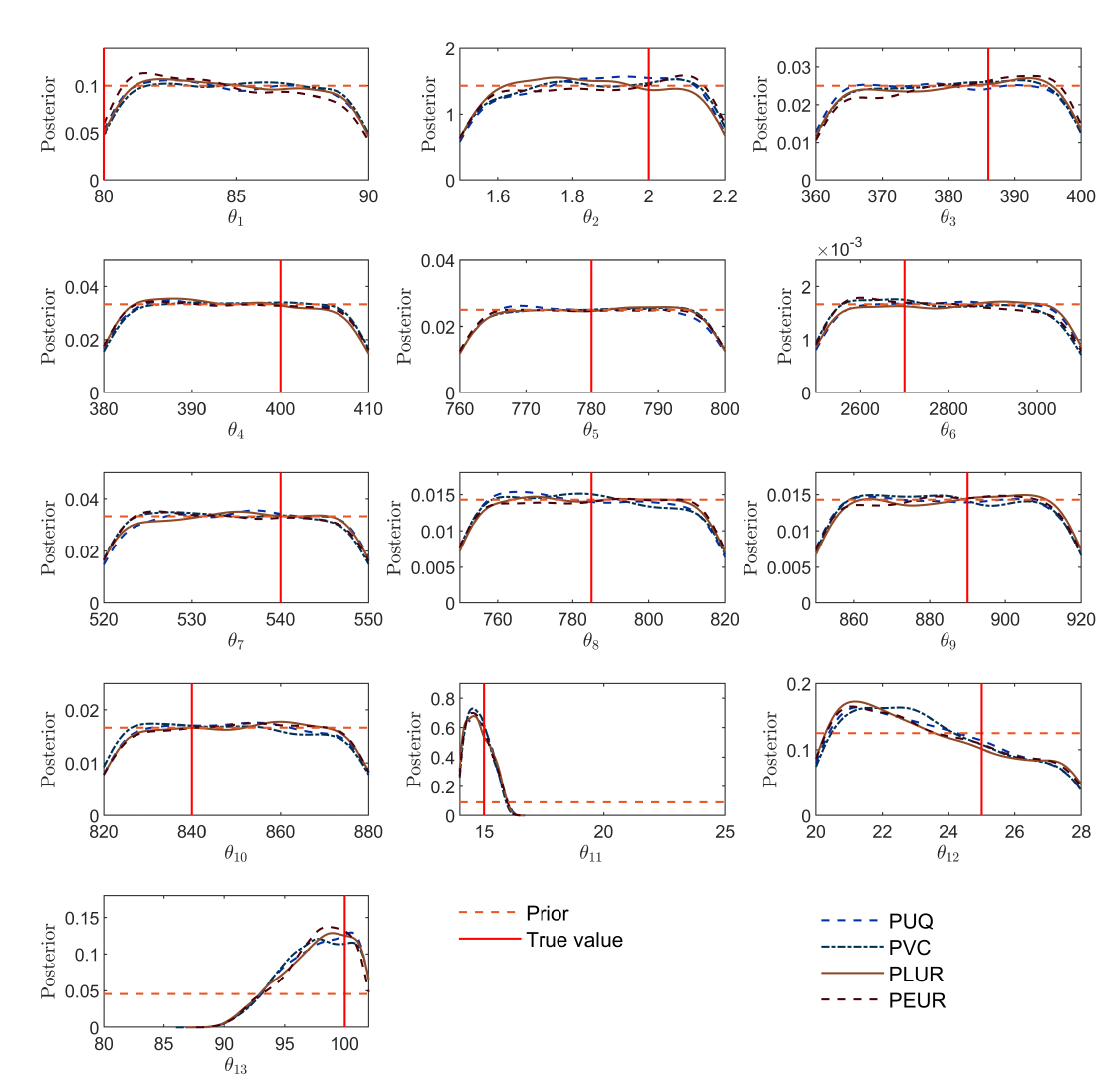}
	\caption{\centering{Results of marginal posterior densities for the battery cooling analysis example. }}
	\label{fig:Battery1DPost}
\end{figure}

\begin{figure}[H]
	\centering
	\setlength{\abovecaptionskip}{-1cm}
	\includegraphics[scale=0.78]{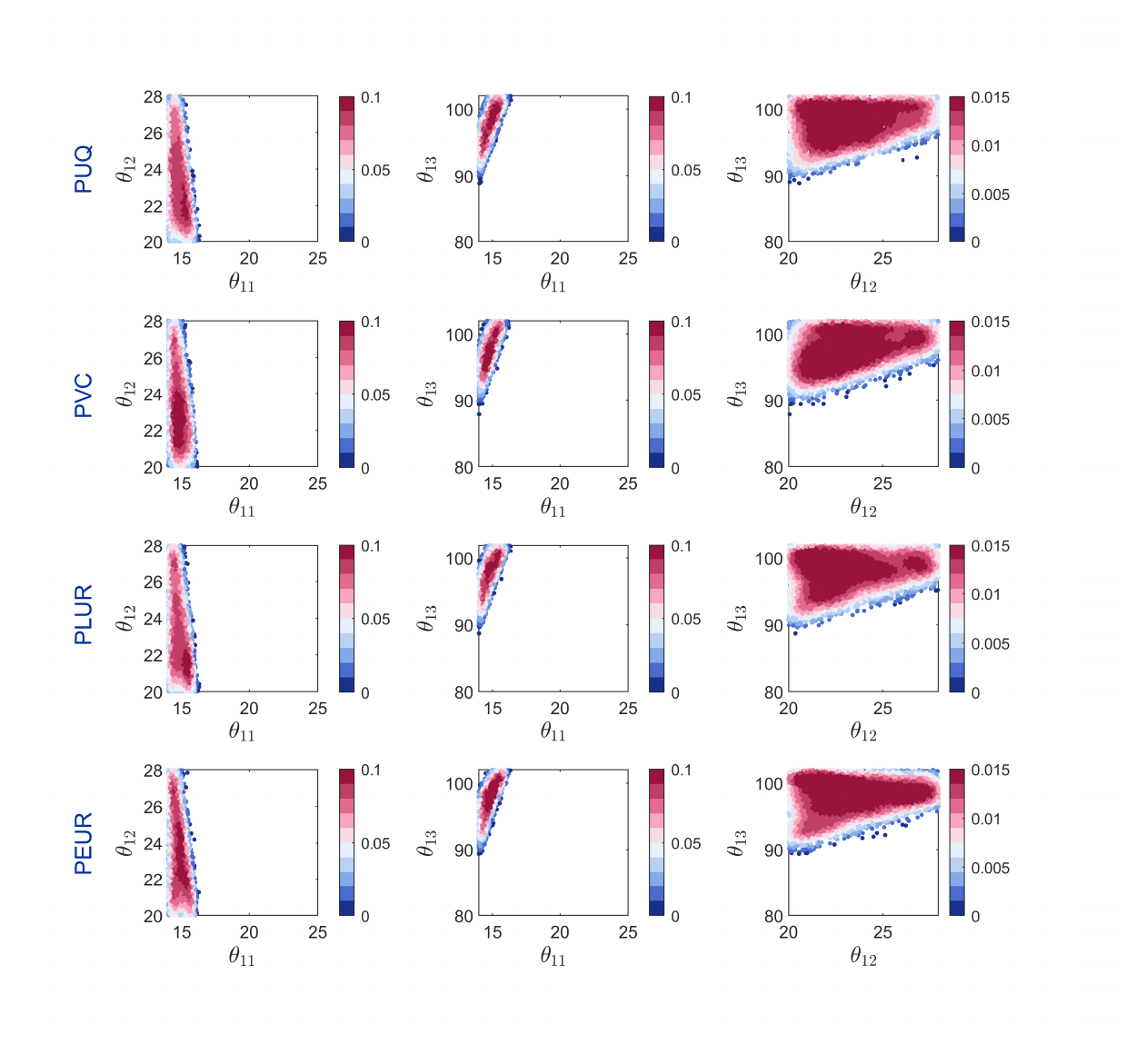}
	\caption{\centering{Results of bivariate marginal posteriors of the three influential parameters for the battery cooling analysis example. }}
	\label{fig:Battery2DPost}
\end{figure}

All the algorithm parameters are set to be the same as those for the 10D example to implement the TBQ algorithm for this example, and each new training point is selected from the candidate sample pool $\mathcal{T}_\mathrm{MC}^{(j-1)}$. Results of univariate marginal posterior densities for all the thirteen parameters, generated with all the four acquisition functions, are reported in Figure \ref{fig:Battery1DPost}, together with the prior density and the true values of parameters for comparison. These marginal densities are estimated using kernel density estimation based on the posterior samples produced using each acquisition function. It is shown that, for each parameter, the marginal posterior densities estimated with all the four acquisition functions show a high degree of consistency, indicating the high accuracy of these estimates. It can also be observed that, except $\theta_{11}\sim \theta_{13}$, the posterior densities of all the other ten parameters show high similarity with their respective prior, indicating that the model response is insensitive to these parameters within their supports. It is seen that, for the two most influential parameters $\theta_{11}$ (heat generation of a normal cell) and $\theta_{13}$ (convection coefficient), the Maximum A Posterior (MAP) estimate, i.e., the location of maximum density, matches well with their true value, indicting a high identifiability; while for the less influential $\theta_{12}$ (heat generation of a faulty cell), the MAP estimate do not match well with the true value, which is caused by the less importance of this parameter and its dependencies with the other two influential parameters. The pairwise joint posterior densities of the three influential parameters are reported in Figure \ref{fig:Battery2DPost}, it is seen again that the results generated by the four acquisition functions are in good agreement, and the three influential parameters show certain degree of linear dependencies. The true value of $\theta_{12}$ lies in the region of high density.

The training details with each acquisition functions and the resultant estimates of model evidence are summarized in Table \ref{table:ResultBattery}, where $N_\mathrm{call}$ refers to the number of calls of the likelihood function. As the stopping conditions are pre-specified as the CoV being smaller than 0.04 for two consecutive times, the post COVs of all estimates are sufficiently small. It is also seen that the mean estimates of the model evidence generated by the four implementations match well with each other. The later three acquisition functions ,i.e., PVC, PLUR and PEUR, consumed almost the same number of model calls, while PUQ consumed more model calls, indicating higher efficiency of PVC, PLUR and PEUR over PUQ.

\begin{table}[H]
	\caption{Results and training details for the battery cooling analysis example, where $N_\mathrm{call}$, $\mu_Z$ and CoVs being calculated from a single run of TBQ equipped with each of the four acquisition functions. }
	\label{table:ResultBattery}
	\centering
	\begin{threeparttable}
		\begin{tabular}{p{3cm}p{2cm}p{5cm}p{3cm}p{1cm}}
			\hline 
			Methods  & $N_\mathrm{call}$&No. of tempering stages  & $\mu_Z$& CoVs\\ 
			\hline
			TBQ-PUQ  & 67  &8& $2.6659\times 10^{-4}$ &0.0176\\
			TBQ-PVC  &53   &8&$2.5617\times 10^{-4}$& 0.0303\\
			TBQ-PLUR &54   &8&$2.3288\times 10^{-4}$& 0.0175 \\
			TBQ-PEUR &53   &9&$2.3212\times 10^{-4}$& 0.0268\\
			\hline
		\end{tabular}
	\end{threeparttable}
\end{table}   

\subsection{Final Remarks}
\begin{figure}[H]
	\centering
	\setlength{\abovecaptionskip}{-0.0cm}
	\includegraphics[scale=0.7]{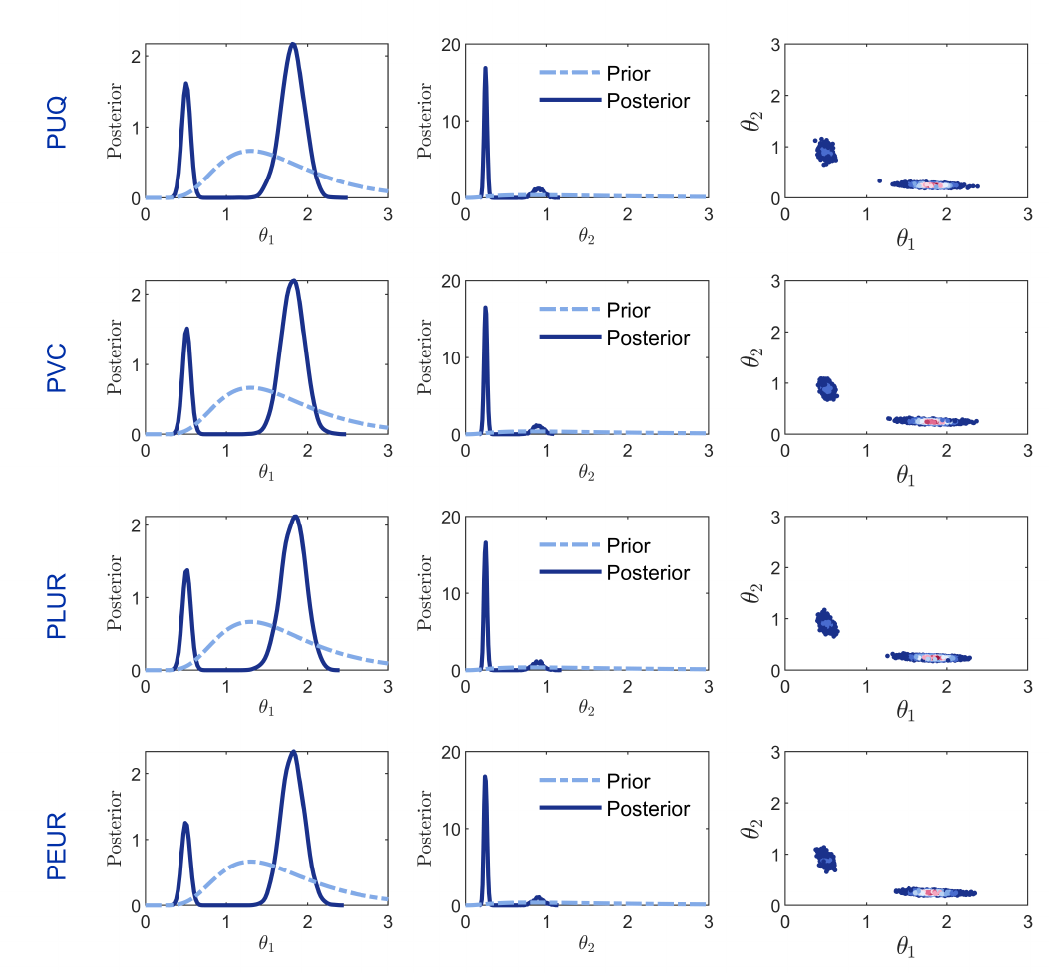}
	\caption{\centering{Results of marginal (first two columns) and joint (last column) posteriors for the 2-DoF system used in Ref. \cite{kitahara2025sequential}. }}
	\label{fig:TwoDoF}
\end{figure}

The challenges associated with Bayesian model updating have spurred increasing research efforts and the proposal of novel methods. Given the space limitations, a comprehensive comparison of all existing methods is not feasible in this paper; therefore, we have made the codes of our methods available on GitHub to enable readers to conduct such comparisons independently (see Section \ref{sec:conclusion} for link). One can simply implement the codes for the examples reported in these literature for comparison. For example, the TBQ algorithm is implemented for the third test example (a 2-DoF system) presented in Ref. \cite{kitahara2025sequential} to compare the performance of TBQ with the Bayesian marginal likelihood inference (BMLI) method developed in this reference. The BMLI method is also based on combination of active leaning GP model with TMCMC, but implemented based on sampling of the GP model, which consumes much more CPU time than TBQ. The results for posteriors generated by one run of TBQ equipped with each acquisition function are reported in Figure \ref{fig:TwoDoF}, and the results of model evidence across ten replications of TBQ are summarized in Table \ref{table:Result2DOF}. It is clearly seen from Figure \ref{fig:TwoDoF} that all the four acquisition functions produced accurate estimation of the bi-modal posterior; and from Table \ref{table:Result2DOF} that TBQ with each acquisition function produced more accurate estimate of model evidence, while consumes smaller number of likelihood calls than the BMLI method reported in Ref. \cite{kitahara2025sequential}. Besides, the CPU time for design of one training point using any of the four acquisition function is much less than the BMLI method as, the acquisition function for BMLI is computed based on sampling the GP model, which can itself be computationally expensive. Except the BMLT method, other developments combining TMCMC and GP model, like the X-TMCMC methods \cite{angelikopoulos2015x},  have also been presented, but the utilized acquisition schemes are different from those developed in this work. Due to space limitations, it is not feasible for us to compare all these related algorithms with TBQ through case studies. Readers may follow a similar procedure as conducted for BMLI to implement the comparison.

The extensive benchmark studies show that two parameters, i.e., the stopping threshold $\epsilon$ and the variation control parameter $\varsigma$, are primarily responsible for the performance of the TBQ algorithm, where the former is used for controlling the prediction uncertainties of posterior and model evidence, and the later one is used for controlling the divergence between intermediate posteriors of consecutive stages. Experiences show that, setting $\epsilon$ as 1$\sim$5\% is usually appropriate, but the specific value should rely on users' tolerance to prediction error. In case the target posterior consists of many modes with unequal peaks, small value of $\epsilon$ encourages capturing more details of less important modes. In the original TBQ algorithm, $\varsigma$ is recommended to be 1$\sim$2, but for the four acquisition functions presented in this work, it is recommended to be 0.5$\sim$1, as higher value causes numerical instability due to the exponential impact on the prediction variance (see Ref. \cite{wei2025bayesian} for explanation of the exponential impact) . Our experience show that setting $\varsigma$ between 0.5 and 1 usually provides guarantee of algorithm robustness. Smaller value for $\varsigma$ also promises numerical stability, but may introduce excessive intermediate tempering stages, which is not necessary.

\begin{table}[H]
	\caption{Results and training details of TBQ for the 2-DoF system utilized in Ref. \cite{kitahara2025sequential}, accompanied with results reported in this reference. The results of each TBQ algorithm are computed across ten replications of the algorithm for indicating robustness. }
	\label{table:Result2DOF}
	\centering
	\begin{threeparttable}
		\begin{tabular}{p{5cm}p{3cm}p{5cm}p{2cm}}
			\hline 
			Methods  & $N_\mathrm{call}$& $\mu_Z$ & CoVs\\ 
			\hline
			TBQ-PUQ  & 36.1  & $1.3816\times 10^{-3}$ &0.0113\\
			TBQ-PVC  &36.9  &$1.3795\times 10^{-3}$& 0.0089\\
			TBQ-PLUR &46.9   &$1.4558\times 10^{-3}$& 0.0105 \\
			TBQ-PEUR &39.4   &$1.4048\times 10^{-3}$& 0.0158\\
			Ref. \cite{kitahara2025sequential}&53&$1.3000\times 10^{-3}$&--\\
			Reference by MC&$10^6$&$1.4190\times 10^{-3}$&$0.0083$\\
			\hline
		\end{tabular}
	\end{threeparttable}
\end{table}

Furthermore, although not explicitly stated, the proposed TBQ algorithm can also be applied to efficient online learning of parameters by leveraging its incremental learning capacity (see e.g., Ref. \cite{rocchetta2018line} for an example of problem description). Specifically, after the posterior $p(\boldsymbol{\theta}|\mathcal{D}_\mathrm{obs})$ and the corresponding model evidence $Z$ have been accurately estimated for the current data $\mathcal{D}_\mathrm{obs}$, given new observations $\mathcal{D}_\mathrm{obs,new}$ received, one can simply set the well-learning posterior as the prior, and reformulate the likelihood as $p(\mathcal{D}_\mathrm{obs,new}|\boldsymbol{\theta})$, to initialize the TBQ algorithm. Then, the algorithm will automatically evaluate the updated posterior $p(\boldsymbol{\theta}|\mathcal{D}_\mathrm{obs},\mathcal{D}_\mathrm{obs, new})$ and the corresponding model evidence with desired accuracy. For simplicity, no more detail will be presented.

Finally, as a supplement to Table \ref{table:SummaryAcqFun}, Table \ref{table:RecommendAcqFun} summarizes key characteristics and scenario-specific recommendations for the four acquisition functions. Generally, for relative simple target posteriors, like the first one reported in Figure \ref{fig:RefPostU} and those with results reported in Figures \ref{fig:ThreeDOFPost} and \ref{fig:TwoDoF}, the PUQ function achieve similar performance in terms of number of likelihood calls to reach the same convergence condition, and thus is recommended as it is computationally the cheapest one among the four. With the increase in complexity, other acquisition functions are recommended; and in case with high complex posteriors, like the third and fourth ones reported in Figure \ref{fig:RefPostU} and the one with results reported in Figure \ref{fig:TenDPost}, the PEUR function is highly recommended as it can substantially save the required number of likelihood function calls. Besides, the four acquisition functions can also be collectively used to enable parallel computing. At each iteration step, each acquisition function can be used to generate one optimal training point, and parallel computing is employed to calculate the likelihood value for each of these four points. This allows for the addition of four training points per iteration, thus accelerating convergence.

\begin{table}[H]
	\caption{Recommendations for acquisition functions.}
	\label{table:RecommendAcqFun}
	\centering
	\fontsize{7}{8}\selectfont 
	 \renewcommand{\arraystretch}{1.5}
	\begin{threeparttable}
		\begin{tabular}{p{1.5cm}p{4.0cm}p{4.0cm}p{5.5cm}}
			\hline 
			Acq. Fun.  & Positive features & Limitations & Recommendations\\ 
			\hline
			PUQ& Admits fully closed-form expressions, and requires the least computer time for design of one training point.  &Commonly requires higher number of likelihood function calls to reach the same convergence criterion, and does not own a prospective view.&Recommended if the likelihood function is cheap to evaluate and/or the complexity of the target posterior is not high, as the first posterior reported in Figure \ref{fig:RefPostU} and the one reported in Figure \ref{fig:TwoDoF}, particularly in case the posterior is given higher priority. \\
			PVC&Commonly requires less training points than PUQ to reach the same error threshold. &Partly in closed form, and consumes more CPU time than PUQ does, and still does not have a prospective view. & Recommended if the likelihood function is moderately expensive to evaluate and if the model evidence is of more concern.\\
			PLUR&Owns a prospective view on learning the posterior, but does not require an integration loop over $y^+$. &The cost to achieve one training point is almost the same as PVC.&Recommended if the likelihood is expensive to evaluate, especially in case the posterior is of special concern.\\
			PEUR&Owns a prospective view on learning the model evidence, commonly requires the smallest number of likelihood function calls, and is independent of the integral over $y^+$.& Consumes more CPU time to design one training point than PVC and PLUR. &Highly recommended if the likelihood function is expensive to estimate and/or the target posterior is highly complex, like last two posteriors reported in Figure \ref{fig:RefPostU} and the one reported in Figure \ref{fig:TenDPost}, particularly when focusing more on model evidence.\\
			\hline
		\end{tabular}
	\end{threeparttable}
\end{table}

\section{Conclusions}\label{sec:conclusion}
This work has presented comprehensive improvements to the BQ and TBQ methods for Bayesian model updating. Specifically, given the logarithm of likelihood being approximated by a GP model, posterior mean prediction and prediction variance were primarily derived in (partly) closed forms for both posteriors and model evidences, which allow for efficient prediction. Then, four new acquisition functions, defined by either investigating the source of prediction uncertainties or evaluating the expected reduction of prediction uncertainties, have been devised. All these four acquisition functions were endowed with mathematically elegant interpretations and efficient computational methods for searching global maxima, and thus are explainable and practical for adaptive design of integration points to achieve desired prediction accuracy for both posteriors and model evidences. The above developments eventually led to the formulation of a new BQ algorithm, which is highly effective and robust for Bayesian inverse problems with multi-modal and highly dependent posteriors. Finally, take it a step further, the above developments have been used for reforming the TBQ algorithm to promote the method more effective for problems with sharply peaked posteriors. Compared with the original TBQ algorithm, both acquisition functions and stopping conditions have been reformulated. 

The effectiveness of the developments have been verified by implementing the proposed BQ and TBQ methods for a number of benchmark studies with diverse features and real-world engineering cases. It is concluded from the results that the proposed methods are capable of estimating multi-modal, nonlinearly dependent and highly sharped posteriors and the associated model evidences with high efficiency and desired accuracy. However, the relative performance of the four acquisition functions is problem-dependent, although for most examples, PEUR consumed the least calls of likelihood function to reach the same convergence condition. With the reformed TBQ algorithm, it is also possible, in case the computational cost (the number of model calls) is strictly limited, to achieve predictions with uncertainties explicitly given.  

Definitely, no single method is universally applicable. Due to the deterioration in the kernels' ability to model spatial correlation, the proposed methods are best suitable for low- to medium-dimensional problems, typically not exceeding 15 dimensions. One approach to address this challenge is to conduct a global sensitivity analysis and fix non-influential parameters, thereby reducing the dimensionality of the target problems. Another route is to develop high-dimension-specific BQ/TBQ algorithms, by e.g., adjusting the kernel parameters using the information of $\gamma_j$ as it indicates the complexity of the intermediate likelihood $p_j(\boldsymbol{\theta}|\mathcal{D}_\mathrm{obs})$. In practical applications, the difference between model predictions and measurements may also be caused by the unknown model bias. In this case, the proposed methods can be combined with the collaborative inference scheme presented in Ref. \cite{hong2026active} for probabilistic inference of both model parameters and bias. All these will be specifically investigated in future research. 

The source Matlab codes for implementing the BQ (Algorithm \ref{Alg: BQ}) and TBQ (Algorithm \ref{Alg:TBQ}) are available via \url{https://github.com/PengfeiWei-NWPU/AcqFunctions-TBQ-BayesModelInference}. Both codes inherit all the four new acquisition functions. 

\appendix

\section{Proof of Eq. \eqref{eq:PLURInterp}}\label{Append:Proof20}
	\footnotesize
 The right-hand term of the first line in Eq. \eqref{eq:PLURInterp} can be expended as:
\begin{equation}\label{eq:ExpectedGainLikMeanExp}
	\begin{split}
		&\int_{\mathbb{R}}{\int_{\mathbb{T}}{\left( \mu _{\mathrm{like},N+1}\left( \boldsymbol{\theta }|\boldsymbol{\theta }^+,y^+ \right) -\mu _{\mathrm{like},N}\left( \boldsymbol{\theta } \right) \right) ^2p\left( \boldsymbol{\theta } \right) f\left( y^+ \right) \mathrm{d}\boldsymbol{\theta }\mathrm{d}y^+}}
		\\
		=&\int_{\mathbb{R}}{\mu _{\mathrm{like},N}^{2}\left( \boldsymbol{\theta } \right) p\left( \boldsymbol{\theta } \right) \mathrm{d}\boldsymbol{\theta }}-2\int_{\mathbb{R}}{\int_{\mathbb{T}}{\mu _{\mathrm{like},N+1}\left( \boldsymbol{\theta }|\boldsymbol{\theta }^+,y^+ \right) \mu _{\mathrm{like},N}\left( \boldsymbol{\theta } \right) p\left( \boldsymbol{\theta } \right) f\left( y^+ \right) \mathrm{d}\boldsymbol{\theta }\mathrm{d}y^+}}
		\\
		&+\int_{\mathbb{R}}{\int_{\mathbb{T}}{\mu _{\mathrm{like},N+1}^{2}\left( \boldsymbol{\theta }|\boldsymbol{\theta }^+,y^+ \right) p\left( \boldsymbol{\theta } \right) f\left( y^+ \right) \mathrm{d}\boldsymbol{\theta }\mathrm{d}y^+}}.
	\end{split}
\end{equation} 
The first term of Eq. \eqref{eq:ExpectedGainLikMeanExp} is free of the integral over $y^+$, and thus substituting Eq. \eqref{eq:PostMeanLikelihood} into this term yields:
\begin{equation}\label{eq:FirstTermLikeMeanGain}
	\int_{\mathbb{R}}{\mu _{\mathrm{like},N}^{2}\left( \boldsymbol{\theta } \right) p\left( \boldsymbol{\theta } \right) \mathrm{d}\boldsymbol{\theta }}=\int_{\mathbb{R}}{\exp \left( 2\mu _{g,N}(\boldsymbol{\theta })+\sigma _{g,N}^{2}(\boldsymbol{\theta }) \right) p\left( \boldsymbol{\theta } \right) \mathrm{d}\boldsymbol{\theta }}.
\end{equation}
Considering $\mu _{\mathrm{like},N+1}\left( \boldsymbol{\theta }|\boldsymbol{\theta }^+,y^+ \right) =\exp \left( \mu _{g,N+1}(\boldsymbol{\theta }|\boldsymbol{\theta }^+,y^+)+\frac{\sigma _{g,N+1}^{2}(\boldsymbol{\theta }|\boldsymbol{\theta }^+)}{2} \right) $ and $y^+\sim \mathcal{N} (\mu _{g,N}(\boldsymbol{\theta }^+),\sigma _{g,N}^{2}(\boldsymbol{\theta }^+))$, the second and third terms on the right-hand side of Eq. \eqref{eq:ExpectedGainLikMeanExp} are derived as:
\begin{equation}\label{eq:SecondTermLikeMeanGain}
	\begin{split}
		&\int_{\mathbb{R}}{\int_{\mathbb{T}}{\mu _{\mathrm{like},N+1}\left( \boldsymbol{\theta }|\boldsymbol{\theta }^+,y^+ \right) \mu _{p,N}\left( \boldsymbol{\theta } \right) p\left( \boldsymbol{\theta } \right) f\left( y^+ \right) \mathrm{d}\boldsymbol{\theta }\mathrm{d}y^+}}
		\\
		=&\int_{\mathbb{R}}{\int_{\mathbb{T}}{\exp \left( 2\mu _{g,N}\left( \boldsymbol{\theta } \right) +\frac{c_{g,N}\left( \boldsymbol{\theta }^+,\boldsymbol{\theta } \right)}{\sigma _{g,N}^{2}\left( \boldsymbol{\theta }^+ \right)}\left( y^+-\mu _{g,N}\left( \boldsymbol{\theta }^+ \right) \right) +\sigma _{g,N}^{2}\left( \boldsymbol{\theta } \right) -\frac{c_{g,N}^{2}\left( \boldsymbol{\theta }^+,\boldsymbol{\theta } \right)}{2\sigma _{g,N}^{2}\left( \boldsymbol{\theta }^+ \right)} \right) p\left( \boldsymbol{\theta } \right) f\left( y^+ \right) \mathrm{d}\boldsymbol{\theta }\mathrm{d}y^+}}
		\\
		=&\int_{\mathbb{T}}{\exp \left( 2\mu _{g,N}\left( \boldsymbol{\theta } \right) +\sigma _{g,N}^{2}\left( \boldsymbol{\theta } \right) \right) \left[ \underset{=1}{\underbrace{\int_{\mathbb{R}}{\exp \left( \frac{c_{g,N}\left( \boldsymbol{\theta }^+,\boldsymbol{\theta } \right)}{\sigma _{g,N}^{2}\left( \boldsymbol{\theta }^+ \right)}\left( y^+-\mu _{g,N}\left( \boldsymbol{\theta }^+ \right) \right) -\frac{c_{g,N}^{2}\left( \boldsymbol{\theta }^+,\boldsymbol{\theta } \right)}{2\sigma _{g,N}^{2}\left( \boldsymbol{\theta }^+ \right)} \right) f\left( y^+ \right) \mathrm{d}y^+}}} \right] p\left( \boldsymbol{\theta } \right) \mathrm{d}\boldsymbol{\theta }}
		\\
		=&\int_{\mathbb{T}}{\exp \left( 2\mu _{g,N}\left( \boldsymbol{\theta } \right) +\sigma _{g,N}^{2}\left( \boldsymbol{\theta } \right) \right) p\left( \boldsymbol{\theta } \right) \mathrm{d}\boldsymbol{\theta }}
	\end{split}
\end{equation} 
, and 
\begin{equation}\label{eq:ThirdTermLikeMeanGain}
	\begin{split}
		&\int_{\mathbb{R}}{\int_{\mathbb{T}}{\mu _{\mathrm{like},N+1}^{2}\left( \boldsymbol{\theta }|\boldsymbol{\theta }^+,y^+ \right) p\left( \boldsymbol{\theta } \right) f\left( y^+ \right) \mathrm{d}\boldsymbol{\theta }\mathrm{d}y^+}}\\
		&=\int_{\mathbb{R}}{\exp \left( 2\mu _{g,N}\left( \boldsymbol{\theta } \right) +\sigma _{g,N}^{2}\left( \boldsymbol{\theta } \right) \right) \left[ \underset{=\exp \left( c_{g,N}^{2}\left( \boldsymbol{\theta }^+,\boldsymbol{\theta } \right) /\sigma _{g,N}^{2}\left( \boldsymbol{\theta }^+ \right) \right)}{\underbrace{\int_{\mathbb{R}}{\exp \left( \frac{2c_{g,N}\left( \boldsymbol{\theta }^+,\boldsymbol{\theta } \right)}{\sigma _{g,N}^{2}\left( \boldsymbol{\theta }^+ \right)}\left( y^+-\mu _{g,N}\left( \boldsymbol{\theta }^+ \right) \right) -\frac{c_{g,N}^{2}\left( \boldsymbol{\theta }^+,\boldsymbol{\theta } \right)}{\sigma _{g,N}^{2}\left( \boldsymbol{\theta }^+ \right)} \right) f\left( y^+ \right) \mathrm{d}y^+}}} \right] p\left( \boldsymbol{\theta } \right) \mathrm{d}\boldsymbol{\theta }}\\
		&=\int_{\mathbb{R}}{\exp \left( 2\mu _{g,N}\left( \boldsymbol{\theta } \right) +\sigma _{g,N}^{2}\left( \boldsymbol{\theta } \right) +\frac{c_{g,N}^{2}\left( \boldsymbol{\theta }^+,\boldsymbol{\theta } \right)}{\sigma _{g,N}^{2}\left( \boldsymbol{\theta }^+ \right)} \right) p\left( \boldsymbol{\theta } \right) \mathrm{d}\boldsymbol{\theta }}.
	\end{split}
\end{equation}
Substituting Eqs.\eqref{eq:FirstTermLikeMeanGain}-\eqref{eq:ThirdTermLikeMeanGain} into Eq. \eqref{eq:ExpectedGainLikMeanExp} and performing the necessary rearrangements, one can obtain:
\begin{equation}
	\begin{split}
		&\int_{\mathbb{R}}{\int_{\mathbb{T}}{\left( \mu _{\mathrm{like},N+1}\left( \boldsymbol{\theta }|\boldsymbol{\theta }^+,y^+ \right) -\mu _{\mathrm{like},N}\left( \boldsymbol{\theta } \right) \right) ^2p\left( \boldsymbol{\theta } \right) f\left( y^+ \right) \mathrm{d}\boldsymbol{\theta }\mathrm{d}y^+}}\\&=\int_{\mathbb{R}}{\exp \left( 2\mu _{g,N}\left( \boldsymbol{\theta } \right) +\sigma _{g,N}^{2}\left( \boldsymbol{\theta } \right) \right) \left( \exp \left( \frac{c_{g,N}^{2}\left( \boldsymbol{\theta }^+,\boldsymbol{\theta } \right)}{\sigma _{g,N}^{2}\left( \boldsymbol{\theta }^+ \right)} \right) -1 \right) p\left( \boldsymbol{\theta } \right) \mathrm{d}\boldsymbol{\theta }}
	\end{split}
\end{equation}
, which is exactly the first line of Eq. \eqref{eq:PLURInterp}.

Next, let's prove the second line of \eqref{eq:PLURInterp}. Based on Eq. \eqref{eq:PostVarLikelihood}, the difference between $\sigma _{\mathrm{like},N}^{2}\left( \boldsymbol{\theta } \right) $ and $\sigma _{\mathrm{like},N+1}^{2}\left( \boldsymbol{\theta }|\boldsymbol{\theta }^+,y^+ \right) $ can be derived as 
\begin{equation}\label{eq:DifferenceVariance}
	\begin{split}
	&\sigma _{\mathrm{like},N}^{2}\left( \boldsymbol{\theta } \right) -\sigma _{\mathrm{like},N+1}^{2}\left( \boldsymbol{\theta }|\boldsymbol{\theta }^+,y^+ \right) 
	\\
	&=\exp \left( 2\mu _{g,N}\left( \boldsymbol{\theta } \right) +\sigma _{g,N}^{2}\left( \boldsymbol{\theta } \right) \right) \left[ \begin{array}{c}
		\left( \exp \left( \sigma _{g,N}^{2}\left( \boldsymbol{\theta } \right) \right) -1 \right) +\left( 1-\exp \left( \sigma _{g,N}^{2}\left( \boldsymbol{\theta } \right) -\frac{c_{g,N}^{2}\left( \boldsymbol{\theta }^+,\boldsymbol{\theta } \right)}{\sigma _{g,N}^{2}\left( \boldsymbol{\theta }^+ \right)} \right) \right)\\
		\times \exp \left( \frac{2c_{g,N}\left( \boldsymbol{\theta }^+,\boldsymbol{\theta } \right)}{\sigma _{g,N}^{2}\left( \boldsymbol{\theta }^+ \right)}\left( y^+-\mu _{g,N}\left( \boldsymbol{\theta }^+ \right) \right) -\frac{c_{g,N}^{2}\left( \boldsymbol{\theta }^+,\boldsymbol{\theta } \right)}{\sigma _{g,N}^{2}\left( \boldsymbol{\theta }^+ \right)} \right)\\
	\end{array} \right] 
	\end{split}
\end{equation}
Considering that the integral of the term $\exp \left( \frac{2c_{g,N}\left( \boldsymbol{\theta }^+,\boldsymbol{\theta } \right)}{\sigma _{g,N}^{2}\left( \boldsymbol{\theta }^+ \right)}\left( y^+-\mu _{g,N}\left( \boldsymbol{\theta }^+ \right) \right) -\frac{c_{g,N}^{2}\left( \boldsymbol{\theta }^+,\boldsymbol{\theta } \right)}{\sigma _{g,N}^{2}\left( \boldsymbol{\theta }^+ \right)} \right) $ in Eq. \eqref{eq:DifferenceVariance} over $f(y^+)$ is equal to $\exp\left(c_{g,N}^{2}\left(\boldsymbol{\theta}^{+},\boldsymbol{\theta}\right)/\sigma_{g,N}^{2}\left(\boldsymbol{\theta}^{+}\right)\right)$, as revealed in Eq. \eqref{eq:ThirdTermLikeMeanGain} the integral of the difference $\sigma _{\mathrm{like},N}^{2}\left( \boldsymbol{\theta } \right) -\sigma _{\mathrm{like},N+1}^{2}\left( \boldsymbol{\theta }|\boldsymbol{\theta }^+,y^+ \right) $ over $f(y^+)$ can be simplified as:
\begin{equation}\label{eq:VarianceREduction}
	\begin{split}
		&\sigma _{\mathrm{like},N}^{2}\left( \boldsymbol{\theta } \right) -\int_{\mathbb{R}}{\sigma _{\mathrm{like},N+1}^{2}\left( \boldsymbol{\theta }|\boldsymbol{\theta }^+,y^+ \right) f\left( y^+ \right) \mathrm{d}y^+}
		\\
		&=\exp \left( 2\mu _{g,N}\left( \boldsymbol{\theta } \right) +\sigma _{g,N}^{2}\left( \boldsymbol{\theta } \right) \right) \left[ \left( \exp \left( \sigma _{g,N}^{2}\left( \boldsymbol{\theta } \right) \right) -1 \right) +\left( 1-\exp \left( \sigma _{g,N}^{2}\left( \boldsymbol{\theta } \right) -\frac{c_{g,N}^{2}\left( \boldsymbol{\theta }^+,\boldsymbol{\theta } \right)}{\sigma _{g,N}^{2}\left( \boldsymbol{\theta }^+ \right)} \right) \right) \exp \left( \frac{c_{g,N}^{2}\left( \boldsymbol{\theta }^+,\boldsymbol{\theta } \right)}{\sigma _{g,N}^{2}\left( \boldsymbol{\theta }^+ \right)} \right) \right] 
		\\
		&=\exp \left( 2\mu _{g,N}\left( \boldsymbol{\theta } \right) +\sigma _{g,N}^{2}\left( \boldsymbol{\theta } \right) \right) \left[ \exp \left( \frac{c_{g,N}^{2}\left( \boldsymbol{\theta }^+,\boldsymbol{\theta } \right)}{\sigma _{g,N}^{2}\left( \boldsymbol{\theta }^+ \right)} \right) -1 \right].
	\end{split}
\end{equation}
The integral of Eq. \eqref{eq:VarianceREduction} over $p\left( \boldsymbol{\theta } \right) $ is exactly the PLUR function. 
\hfill \qedsymbol
\section{Proof of Eq. \eqref{eq:PEURInterp}}\label{Append:Proof22}
Based on Eqs. \eqref{eq:PostMeanUpdatedGP} and \eqref{eq:PostMeanLikelihood}, the difference between $\mu _{Z,N+1}\left( \boldsymbol{\theta }^+,y^+ \right) $ and $\mu _{Z,N}$ can be derived as:
\begin{equation}
	\begin{split}
	&\mu _{Z,N+1}\left( \boldsymbol{\theta }^+,y^+ \right) -\mu _{Z,N}\\
	&=\int_{\mathbb{T}}{\exp \left( \mu _{g,N}\left( \boldsymbol{\theta } \right) +\frac{\sigma _{g,N}^{2}\left( \boldsymbol{\theta } \right)}{2} \right) \left[ \exp \left( \frac{c_{g,N}\left( \boldsymbol{\theta }^+,\boldsymbol{\theta } \right)}{\sigma _{g,N}^{2}\left( \boldsymbol{\theta }^+ \right)}\left( y^+-\mu _{g,N}\left( \boldsymbol{\theta }^+ \right) \right) -\frac{c_{g,N}^{2}\left( \boldsymbol{\theta }^+,\boldsymbol{\theta } \right)}{2\sigma _{g,N}^{2}\left( \boldsymbol{\theta }^+ \right)} \right) -1 \right] p\left( \boldsymbol{\theta } \right) \mathrm{d}\boldsymbol{\theta }}.
	\end{split}.
\end{equation}
As the integral of the exponential term in the square bracket over $f(y^+)$ equal to one, as given in Eq. \eqref{eq:SecondTermLikeMeanGain}, the expectation of $\mu _{Z,N+1}\left( \boldsymbol{\theta }^+,y^+ \right) -\mu _{Z,N}$ over $y^+$ is exactly zero. Thus, the right-hand term of the first line of Eq. \eqref{eq:PEURInterp} is indeed the variance of $\mu _{Z,N+1}\left( \boldsymbol{\theta }^+,y^+ \right) -\mu _{Z,N}$ over $y^+$. It can be further obtained:
\begin{equation}
\left( \mu _{Z,N+1}\left( \boldsymbol{\theta }^+,y^+ \right) -\mu _{Z,N} \right) ^2=\int_{\mathbb{T} \times \mathbb{T}}{A\left( \boldsymbol{\theta },\boldsymbol{\theta }^{\prime} \right) B\left( \boldsymbol{\theta },\boldsymbol{\theta }^{\prime}|\boldsymbol{\theta }^+,y^+ \right) p\left( \boldsymbol{\theta } \right) p\left( \boldsymbol{\theta }^{\prime} \right) \mathrm{d}\boldsymbol{\theta }\mathrm{d}\boldsymbol{\theta }^{\prime}}
\end{equation} 
, where  
\begin{equation}
	A\left( \boldsymbol{\theta },\boldsymbol{\theta }^{\prime} \right) =\exp \left( \mu _{g,N}\left( \boldsymbol{\theta } \right) +\frac{\sigma _{g,N}^{2}\left( \boldsymbol{\theta } \right)}{2}+\mu _{g,N}\left( \boldsymbol{\theta }^{\prime} \right) +\frac{\sigma _{g,N}^{2}\left( \boldsymbol{\theta }^{\prime} \right)}{2} \right) 
\end{equation}
and
\begin{equation}
	B\left( \boldsymbol{\theta },\boldsymbol{\theta }^{\prime}|\boldsymbol{\theta }^+,y^+ \right) =\left[ \exp \left( \begin{array}{c}
		\frac{c_{g,N}\left( \boldsymbol{\theta }^+,\boldsymbol{\theta } \right)}{\sigma _{g,N}^{2}\left( \boldsymbol{\theta }^+ \right)}\left( y^+-\mu _{g,N}\left( \boldsymbol{\theta }^+ \right) \right)\\
		-\frac{c_{g,N}^{2}\left( \boldsymbol{\theta }^+,\boldsymbol{\theta } \right)}{2\sigma _{g,N}^{2}\left( \boldsymbol{\theta }^+ \right)}\\
	\end{array} \right) -1 \right] \left[ \exp \left( \begin{array}{c}
		\frac{c_{g,N}\left( \boldsymbol{\theta }^+,\boldsymbol{\theta }^{\prime} \right)}{\sigma _{g,N}^{2}\left( \boldsymbol{\theta }^+ \right)}\left( y^+-\mu _{g,N}\left( \boldsymbol{\theta }^+ \right) \right)\\
		-\frac{c_{g,N}^{2}\left( \boldsymbol{\theta }^+,\boldsymbol{\theta }^{\prime} \right)}{2\sigma _{g,N}^{2}\left( \boldsymbol{\theta }^+ \right)}\\
	\end{array} \right) -1 \right]. 
\end{equation}
Further, as
\begin{equation}
	\int_{\mathbb{R}}{B\left( \boldsymbol{\theta },\boldsymbol{\theta }^{\prime}|\boldsymbol{\theta }^+,y^+ \right) f\left( y^+ \right)}\mathrm{d}y^+=\exp \left( \frac{c_{g,N}(\boldsymbol{\theta }^+,\boldsymbol{\theta })c_{g,N}(\boldsymbol{\theta }^+,\boldsymbol{\theta }^{\prime})}{\sigma _{g,N}^{2}(\boldsymbol{\theta }^+)} \right) -1
\end{equation}
, the integral of $\left( \mu _{Z,N+1}\left( \boldsymbol{\theta }^+,y^+ \right) -\mu _{Z,N} \right) ^2$ over $f\left( y^+ \right) $ can then be formulated as:
\begin{equation}
	\begin{split}
	&\int_{\mathbb{R}}{\left( \mu _{Z,N+1}\left( \boldsymbol{\theta }^+,y^+ \right) -\mu _{Z,N} \right) ^2f\left( y^+ \right)}\mathrm{d}y^+
	\\
	&=\int_{\mathbb{T} \times \mathbb{T}}{A\left( \boldsymbol{\theta },\boldsymbol{\theta }^{\prime} \right) \left[ \exp \left( \frac{c_{g,N}(\boldsymbol{\theta }^+,\boldsymbol{\theta })c_{g,N}(\boldsymbol{\theta }^+,\boldsymbol{\theta }^{\prime})}{\sigma _{g,N}^{2}(\boldsymbol{\theta }^+)} \right) -1 \right] p\left( \boldsymbol{\theta } \right) p\left( \boldsymbol{\theta }^{\prime} \right) \mathrm{d}\boldsymbol{\theta }\mathrm{d}\boldsymbol{\theta }^{\prime}}
	\end{split}
\end{equation}
, which is exactly the first equality of Eq. \eqref{eq:PEURInterp}. 

Next, consider the second equality of Eq. \eqref{eq:PEURInterp}. The residual variance $\sigma _{Z,N+1}^{2}\left( \boldsymbol{\theta }^+,y^+ \right) $ is formulated as:
\begin{equation}
	\sigma _{Z,N+1}^{2}\left( \boldsymbol{\theta }^+,y^+ \right) =\int_{\mathbb{T} \times \mathbb{T}}{A\left( \boldsymbol{\theta },\boldsymbol{\theta }^{\prime} \right) C\left( \boldsymbol{\theta },\boldsymbol{\theta }^{\prime}|\boldsymbol{\theta }^+,y^+ \right) \left( \exp \left( c_{g,N}\left( \boldsymbol{\theta },\boldsymbol{\theta }^{\prime} \right) -\frac{c_{g,N}\left( \boldsymbol{\theta },\boldsymbol{\theta }^+ \right) c_{g,N}\left( \boldsymbol{\theta }^+,\boldsymbol{\theta }^{\prime} \right)}{\sigma _{g,N}^{2}\left( \boldsymbol{\theta }^+ \right)} \right) -1 \right) p\left( \boldsymbol{\theta } \right) p\left( \boldsymbol{\theta }^{\prime} \right) \mathrm{d}\boldsymbol{\theta }\mathrm{d}\boldsymbol{\theta }^{\prime}}
\end{equation}  
, where 
\begin{equation}
	C\left( \boldsymbol{\theta },\boldsymbol{\theta }^{\prime}|\boldsymbol{\theta }^+,y^+ \right) =\exp \left( \frac{c_{g,N}\left( \boldsymbol{\theta }^+,\boldsymbol{\theta } \right)}{\sigma _{g,N}^{2}\left( \boldsymbol{\theta }^+ \right)}\left( y^+-\mu _{g,N}\left( \boldsymbol{\theta }^+ \right) \right) +\frac{c_{g,N}\left( \boldsymbol{\theta }^+,\boldsymbol{\theta }^{\prime} \right)}{\sigma _{g,N}^{2}\left( \boldsymbol{\theta }^+ \right)}\left( y^+-\mu _{g,N}\left( \boldsymbol{\theta }^+ \right) \right) -\frac{c_{g,N}^{2}\left( \boldsymbol{\theta }^+,\boldsymbol{\theta } \right)}{2\sigma _{g,N}^{2}\left( \boldsymbol{\theta }^+ \right)}-\frac{c_{g,N}^{2}\left( \boldsymbol{\theta }^+,\boldsymbol{\theta }^\prime \right)}{2\sigma _{g,N}^{2}\left( \boldsymbol{\theta }^+ \right)} \right). 
\end{equation}
The integral of $C\left( \boldsymbol{\theta },\boldsymbol{\theta }^{\prime}|\boldsymbol{\theta }^+,y^+ \right) $ over $f(y^+)$ is derived as:
\begin{equation}
	\int_{\mathbb{R}}{C\left( \boldsymbol{\theta },\boldsymbol{\theta }^{\prime}|\boldsymbol{\theta }^+,y^+ \right) f\left( y^+ \right) \mathrm{d}y^+}=\exp \left( \frac{c_{g,N}(\boldsymbol{\theta }^+,\boldsymbol{\theta })c_{g,N}(\boldsymbol{\theta }^+,\boldsymbol{\theta }^{\prime})}{\sigma _{g,N}^{2}(\boldsymbol{\theta }^+)} \right) 
\end{equation}
Then it holds that:
\begin{equation}
	\begin{split}
		&\int_{\mathbb{R}}{\sigma _{Z,N+1}^{2}\left( \boldsymbol{\theta }^+,y^+ \right) f\left( y^+ \right) \mathrm{d}y^+}
		\\
		&=\int_{\mathbb{T} \times \mathbb{T}}{A\left( \boldsymbol{\theta },\boldsymbol{\theta }^{\prime} \right) \left( \exp \left( c_{g,N}\left( \boldsymbol{\theta },\boldsymbol{\theta }^{\prime} \right) \right) -\exp \left( \frac{c_{g,N}(\boldsymbol{\theta }^+,\boldsymbol{\theta })c_{g,N}(\boldsymbol{\theta }^+,\boldsymbol{\theta }^{\prime})}{\sigma _{g,N}^{2}(\boldsymbol{\theta }^+)} \right) \right) p\left( \boldsymbol{\theta } \right) p\left( \boldsymbol{\theta }^{\prime} \right) \mathrm{d}\boldsymbol{\theta }\mathrm{d}\boldsymbol{\theta }^{\prime}}
	\end{split}
\end{equation}

Considering
\begin{equation}
	\sigma _{Z,N}^{2}=\int_{\mathbb{T} \times \mathbb{T}}{A\left( \boldsymbol{\theta },\boldsymbol{\theta }^{\prime} \right) \left( \exp \left( c_{g,N}\left( \boldsymbol{\theta },\boldsymbol{\theta }^{\prime} \right) \right) -1 \right) p\left( \boldsymbol{\theta } \right) p\left( \boldsymbol{\theta }^{\prime} \right) \mathrm{d}\boldsymbol{\theta }\mathrm{d}\boldsymbol{\theta }^{\prime}}
\end{equation}
, it can then be obtained:
\begin{equation}
	\sigma _{Z,N}^{2}-\int_{\mathbb{R}}{\sigma _{Z,N+1}^{2}\left( \boldsymbol{\theta }^+,y^+ \right) f\left( y^+ \right) \mathrm{d}y^+}=\int_{\mathbb{T} \times \mathbb{T}}{A\left( \boldsymbol{\theta },\boldsymbol{\theta }^{\prime} \right) \left[ \exp \left( \frac{c_{g,N}(\boldsymbol{\theta }^+,\boldsymbol{\theta })c_{g,N}(\boldsymbol{\theta }^+,\boldsymbol{\theta }^{\prime})}{\sigma _{g,N}^{2}(\boldsymbol{\theta }^+)} \right) -1 \right] p\left( \boldsymbol{\theta } \right) p\left( \boldsymbol{\theta }^{\prime} \right) \mathrm{d}\boldsymbol{\theta }\mathrm{d}\boldsymbol{\theta }^{\prime}}.
\end{equation} 
The second equality of Eq. \eqref{eq:PEURInterp} is ultimately proved.  
\hfill \qedsymbol
\section*{Acknowledgment}
\normalsize
This work is supported by the National Natural Science Foundation of China under grant number
52475164, the National Key R\&D Program of China with grant number 2023YFB3407103, and the Fundamental and Interdisciplinary Disciplines Breakthrough Plan of the Ministry of Education of China under grant number JYB2025XDXM207.



\end{spacing}

\begin{spacing}{1.3}
 \bibliographystyle{elsarticle-num}

\end{spacing}
\end{document}